\documentclass[aps,twocolumn,prd,showpacs,showkeys,preprintnumbers,superscriptaddress,nobibnotes,floatfix,longbibliography,notitlepage,nofootinbib]{revtex4-2}

\pdfoutput=1
\usepackage{amsmath}
\usepackage{amsfonts}
\usepackage{amssymb}
\usepackage{mathrsfs}
\usepackage{color}
\usepackage{slashed}
\usepackage{graphicx}
\usepackage{xcolor}
\usepackage{microtype}
\usepackage{multirow,makecell}
\usepackage{listings}
\usepackage{upgreek}
\usepackage{bm}
\usepackage[normalem]{ulem}
\usepackage{siunitx}
\usepackage{booktabs}
\usepackage{tabularx}
\usepackage{physics}
\usepackage{adjustbox}
\usepackage{chemformula}
\usepackage[inline]{enumitem}
\usepackage{fontawesome5} 
\usepackage{academicons}
\usepackage[colorlinks=true,linkcolor=purple,citecolor=purple]{hyperref}
\usepackage[capitalise]{cleveref}
\definecolor{orcidlogocol}{HTML}{A6CE39}
\newcommand{\myorcid}[1]{\href{https://orcid.org/#1}{\textcolor{orcidlogocol}{\faOrcid} #1}}
\renewcommand{\phi}{\varphi}
\renewcommand{\epsilon}{\varepsilon}
\def\Z2{\text{Z}_2}

\def\deltaamu{\Delta a_\mu}

\newcounter{CommentCount}
\newcommand{\myparagraph}[1]{\textbf{#1} ---}

\begin{document} 
\preprint{CERN-TH-2023-019, FERMILAB-PUB-23-054-T, IRMP-CP3-23-08}

\title{Semi-Visible Dark Photon Phenomenology at the GeV Scale}

\author{Asli M.  Abdullahi}
\email{asli@fnal.gov}
\thanks{\myorcid{0000-0002-6122-4986}}
\affiliation{Theoretical Physics Department, Fermi National Accelerator Laboratory, Batavia, IL 60510, USA}
\author{Matheus Hostert}
\email{mhostert@pitp.ca}
\thanks{\myorcid{0000-0002-9584-8877}}
\affiliation{Perimeter Institute for Theoretical Physics, Waterloo, ON N2J 2W9, Canada}
\affiliation{School of Physics and Astronomy, University of Minnesota, Minneapolis, MN 55455, USA}
\affiliation{William I. Fine Theoretical Physics Institute, School of Physics and Astronomy, University of Minnesota, Minneapolis, MN 55455, USA}

\author{Daniele Massaro}
\email{daniele.massaro5@unibo.it}
\thanks{\myorcid{0000-0002-1013-3953}}
\affiliation{Dipartimento di Fisica e Astronomia, Universit\`a di Bologna, via Irnerio 46, 40126 Bologna, Italy}
\affiliation{INFN, Sezione di Bologna, viale Berti Pichat 6/2, 40127 Bologna, Italy}

\affiliation{Centre for Cosmology, Particle Physics and Phenomenology (CP3), Universit\'e Catholique de Louvain, B-1348 Louvain-la-Neuve, Belgium}
\author{Silvia Pascoli}
\email{silvia.pascoli@unibo.it}
\thanks{\myorcid{0000-0002-2958-456X}}
\affiliation{Dipartimento di Fisica e Astronomia, Universit\`a di Bologna, via Irnerio 46, 40126 Bologna, Italy}
\affiliation{INFN, Sezione di Bologna, viale Berti Pichat 6/2, 40127 Bologna, Italy}
\affiliation{CERN, Theoretical Physics Department, Geneva, Switzerland}

\begin{abstract}
In rich dark sector models, dark photons heavier than tens of MeV can behave as semi-visible particles: their decays contain both visible and invisible final states. 
We present models containing multiple dark fermions which allow for such decays and inscribe them in the context of inelastic dark matter and heavy neutral leptons scenarios.
Our models represent a generalization of the traditional inelastic dark matter model by means of a charge conjugation symmetry.
We revisit constraints on dark photons from $e^+e^-$ colliders and fixed target experiments, including the effect of analysis vetoes on semi-visible decays, $A^\prime \to \psi_i (\psi_j \to \psi_k \ell^+\ell^-)$.
We find that in some cases the BaBar and NA64 experiments no longer exclude large kinetic mixing, $\epsilon \sim 10^{-2}$, and, specifically, the related explanation of the discrepancy in the muon $(g-2)$. This reopens an interesting window in parameter space for dark photons with exciting discovery prospects.
We point out that a modified missing-energy search at NA64 can target short-lived $A^\prime$ decays and directly probe the newly-open parameter space.

\end{abstract}

\maketitle

{
\hypersetup{linkcolor=black}
\tableofcontents
}
\section{Introduction}

The existence of hidden sectors containing light and feebly-interacting particles offers a promising avenue to address the shortcomings of the Standard Model (SM).
With dynamics and structures of their own, these dark sectors (DS) can contain stable dark matter (DM) particles, new gauge symmetries, new fundamental scales, and additional sources of $C$ and $P$ violation.
While the search for new heavy particles at the LHC continues, this possibility offers a paradigm-shifting framework that is testable and provides fertile ground for model building.
In particular, if the DS contains light dark matter particles, their production typically requires the existence of new mediators that interact with both the DS and SM particles~\cite{Lee:1977ua,Vysotsky:1977pe,Boehm:2003hm,Pospelov:2007mp}. 
These can be new scalars (Higgs portal), heavy neutral leptons (neutrino portal), or new gauge bosons (vector portal).
In the case of heavy neutral leptons, the connection with neutrino masses and leptonic mixing provides additional theoretical motivation. 
The presence of new gauge symmetries with associated Higgs-like breaking mechanisms is also a natural possibility, appearing in many breaking patterns of grand-unified theories.

Targeting experimental searches for these portal mediators can be an efficient way to test the DS framework since they couple to both the SM and DS.
While tremendous experimental progress has been achieved for the three portal cases above (see Refs.~\cite{Agrawal:2021dbo,Gori:2022vri,Batell:2022dpx}), the focus has often been on scenarios with minimal new particle content.
While this is a sensible starting assumption, relaxing the stringent conditions on minimality can help us to uncover rich DS theories~\cite{Harris:2022vnx}.

In this work, we focus our attention on the dark photon $A^\prime$, the vector portal mediator.
Unless explicitly forbidden by new symmetries, kinetic mixing between the dark $U(1)_D$ gauge boson and the SM hypercharge~\cite{Holdom:1985ag}, $\frac{\epsilon}{2 c_{\rm W}} X_{\mu\nu} B^{\mu\nu}$, is expected to be sizeable, providing a clear target for detection. 
The one-loop expectation for kinetic mixing is
\begin{equation}
    \epsilon \sim \frac{g' g_D}{16 \pi^2}\sum_i Q_i^Y Q_i^X \log\left(\frac{M_i^2}{\mu^2}\right) \sim \mathcal{O}\left(10^{-3} - 10^{-2}\right)
\end{equation}
where $M_i$ and $Q_i^{(Y, \, X)}$ are the masses and charges of the heavy new fermions that run in the loop, $\mu$ the renormalization scale, and $g_D$ the gauge coupling, taken to be of the same order as the SM couplings.
So far, a kinetic mixing of this size has not been experimentally observed for dark photons in minimal scenarios. 
Experimental limits have focused primarily on models in which the $A^\prime$ decays to the SM as a fully visible resonance, or decays invisibly to, e.g., DM particles~\cite{BaBar:2014zli,NA482:2015wmo,BaBar:2017tiz,NA64:2017vtt}. 
At first sight, the naive expectation for kinetic mixing looks too strongly constrained to remain a viable possibility, suggesting small dark couplings or a higher-order origin for $\epsilon$~\cite{Gherghetta:2019coi}.
It is, however, possible to avoid laboratory constraints and remain compatible with the naive one-loop estimation of $\epsilon$. 
One possibility is that the dark photon decays \emph{semi-visibly}. 

Semi-visible decays of the dark photon contain both visible and invisible particles in the final state, precluding a full reconstruction of the dark photon mass through its decay products and avoiding constraints from resonance searches.
As we will show, this is a natural prediction of multi-generational DS models, such as inelastic dark matter models~\cite{Tucker-Smith:2001myb}.
Nevertheless, the dark photon may still appear as an invisible particle when produced in an experiment due to the detector geometry and limited experimental resolution. 
The strongest constraints on the GeV-scale invisible dark photon come from the $e^+e^-$ colliders and fixed-target experiments~\cite{Boehm:2003hm,Fayet:2007ua,Essig:2009nc,Gninenko:2013rka,Essig:2013vha,Izaguirre:2014bca,Berlin:2020uwy}, and so we revisit the leading constraints from BaBar~\cite{BaBar:2017tiz} and NA64~\cite{NA64:2016oww,NA64:2017vtt,NA64:2019imj,NA64:2021acr}.

Another motivation for this work is the measurement of the anomalous magnetic moment of the muon, $a_\mu = (g-2)_\mu/2$, at Fermilab (FNAL)~\cite{Abi:2021gix,Albahri:2021ixb}.
Through the same mechanism understood by Schwinger in the early days of Quantum Electrodynamics (QED), 
kinetically-mixed dark photons contribute to $a_\mu$ at the one loop-level with a positive sign, providing an elegant and simple solution to the discrepancy between experiment and theoretical predictions, $\deltaamu$~\cite{Gninenko:2001hx,Pospelov:2008zw}.
This solution is only possible for light mediators, $m_{A^\prime}\lesssim 3$~GeV, and requires large values of kinetic mixing, $\epsilon\sim 10^{-3} - 10^{-2}$, compatible with the naive one-loop expectation.
While this explanation is excluded in fully visible or invisible dark photon models, it remains viable for semi-visible dark photons in the mass region of  $m_{A^\prime}\sim 0.6 - 1$~GeV, as first proposed in Ref.~\cite{Mohlabeng:2019vrz}, but later disputed in Refs.~\cite{Duerr:2019dmv,Duerr:2020muu}. 
We provide a detailed analysis of this option, proposing new models that overcome the problems of the minimal model of Refs.~\cite{Mohlabeng:2019vrz,Duerr:2019dmv,Duerr:2020muu}. 
We also note that dark photons can be semi-visible due if their decays contain dark showers~\cite{Cohen:2015toa,Schwaller:2015gea} (see Ref.~\cite{Bernreuther:2022jlj} for a recent study) and lepton jets~\cite{Falkowski:2010cm,Buschmann:2015awa}.

We focus on DS models with multiple fermions, interpreted as either thermal DM models or seesaw neutrino mass models.
In the DM interpretation, our phenomenological considerations point towards models with dark-photon-mediated DM coannihilations, automatically satisfying strong Cosmic Microwave Background (CMB) constraints on light thermal DM.
The seesaw interpretation, albeit less predictive, has several striking predictions for neutrino experiments~\cite{Abdullahi:2020nyr}.
As we will show, the original proposal for a semi-visible $A^\prime$, based on a minimal inelastic DM (iDM) model~\cite{Mohlabeng:2019vrz}, is strongly constrained by collider, fixed target, and indirect searches and can only explain $\deltaamu$ in a very narrow region of parameter space. 
Our collection of semi-visible $A^\prime$ models constitutes a viable explanation of $\deltaamu$ by adding new heavy neutral fermions (HNFs) with several hundred MeV masses and sizeable mass splittings. 
Due to the conservation of a charge-conjugation symmetry, $C$, many of our scenarios ensure that $A^\prime$ couples only off-diagonally to HNF generations, generalizing the popular iDM scenario.

Testing the allowed parameter space to exclude or discover these models is a tangible task for current collider and fixed-target experiments.
We discuss some strategies to isolate the distinct semi-visible signatures in these experiments. 
In particular, the newly-open parameter space can be explored with displaced vertices and monophoton-like events at Belle-II, such as those studied by the BaBar $e^+e^-$ collider.
At fixed-target experiments, we point out that invisible-$A^\prime$ searches can be adapted to be sensitive to the missing energy in semi-visible dark photon decays.
This strategy is pursued in an accompanying paper to derive new experimental limits using NA64 data~\cite{na64semivisible}.

The paper is organized as follows. 
We introduce the canonical semi-visible dark photon model in \cref{sec:darkphoton} and discuss specific realizations with varying fermionic content. 
The set of relevant constraints in the parameter space is discussed in \cref{sec:model_independent}, and in \cref{sec:recast}, we detail our recasting procedure to obtain the revised constraints from BaBar and NA64.
Our results are then presented in \cref{sec:results}.
We discuss the implications of our results for models of dark matter and heavy neutral leptons in \cref{sec:discussion}, concluding with \cref{sec:conclusions}.

\section{Semi-visible dark photons}
\label{sec:darkphoton}

We are interested in a kinetically-mixed, massive dark photon. 
In general terms, the initial Lagrangian is given by
\begin{align}
    \mathscr{L} &=  \mathscr{L}_{\rm SM} - \frac{\epsilon}{2 c_{\rm W}} F_{\mu \nu} X^{\mu \nu} - \frac{1}{4} X_{\mu\nu}X^{\mu\nu}
    \\\nonumber
    & + g_D X_{\mu} \mathcal{J}^{\mu}_D + \frac{m_{X}^2}{2} X_{\mu} X^{\mu},
\end{align}
where $X_{\mu \nu}$ is the field strength tensor of the dark photon, and $\mathcal{J}^{\mu}_D$ is the dark current containing new fermionic or scalar degrees of freedom.
The origin of the dark photon mass is thus far unspecified~\footnote{While introducing a dark Higgs is compelling from a model building point of view, it also comes with additional assumptions and decreased predictivity.
Therefore, we proceed assuming a St\"uckelberg mass (see Ref.~\cite{Kribs:2022gri} for a recent discussion).}.
After the diagonalization of the gauge kinetic terms, the dark photon mass eigenstate $A^\prime$ with mass $m_{A^\prime} \simeq m_X$ couples to both the SM electromagnetic (EM) current and the SM weak neutral current (NC),
\begin{align}\label{eq:couplings_to_currents}
        \mathscr{L}_{\rm int.} &\supset A^\prime_{\mu} \left( g_D  \mathcal{J}^{\mu}_D - e\epsilon  \mathcal{J}^{\mu}_{\rm EM} - \epsilon  t_{\rm W} \frac{m_{A^\prime}^2}{m_Z^2} \frac{g}{2 c_{\rm W}}\mathcal{J}^{\mu}_{\rm NC}\right) 
        \\\nonumber &+ 
        Z_\mu \left( \frac{g}{2 c_{\rm W}} \mathcal{J}^\mu_{\rm NC} + g_D t_{\rm W} \epsilon \mathcal{J}_{\rm D}^\mu \right) + \mathcal{O}(\epsilon^2),
\end{align}
where $t_{\rm W}\equiv\tan\theta_{\rm W}$, with $\theta_{\rm W}$ the Standard Model Weak angle. 
While the SM photon does not couple to the dark sector, the SM Z boson mass eigenstate can.

Let us now discuss the particle content in the dark sector and how it can render $A^\prime$ semi-visible.
To appear as {\em semi-visible}, a dark photon has to decay predominantly into dark particles that cascade-decay into SM states plus missing energy, avoiding limits on both visible and invisible dark photons, see \cref{sec:recast}. 
A particularly simple choice would be a dark complex scalar $\Phi$, charged under the $U(1)_D$.
The dark current is $\mathcal{J}_D^\mu= \Phi^* i\overset{\leftrightarrow}{\partial^\mu} \Phi$, and the hermiticity of the Lagrangian ensures that the dark photon interactions must always take place between the real components of $\Phi$, $\phi_1$ and $\phi_2$. 
A soft $U(1)_D$-breaking term, $\mu \Phi^2$ can then split the masses of the real scalars, and render the dark photon semi-visible due to the decay cascade $A^\prime \to \phi_1 (\phi_2 \to \phi_1 e^+e^-)$.
For dark fermions, the idea is similar, but a much richer structure can arise due to the half-integer spin.
We will consider models with $n$ fermions, where
\begin{equation}\label{eq:generic_dark_current}
\mathcal{J}_{D}^\mu \equiv \sum_{i,j = 1}^n V_{ij} \overline{\psi_i} \gamma^\mu \psi_j
\end{equation}
with $V_{ij}$ the model-dependent coupling vertices. 
If each of the dark fermions has a dark charge $Q_i$, the vertices are constrained to $\sum_{i,j}^n |V_{ij}|^2 \leq \sum_i^n |Q_i|$.
While we do not pursue this possibility, we note that non-renormalizable interactions between $\psi_i$ could be considered, including electric and magnetic moments for the dark fermions~\cite{Eby:2019mgs,Baryakhtar:2020rwy,Chang:2010en,Feldstein:2010su}. 
Like in the scalar case, the latter has the advantage of being automatically off-diagonal in the $i,j$ fermion indices (if the dark fermions are Majorana particles), and would also generate semi-visible decays.
For an analogous discussion of semi-visible dark sector models, see Ref.~\cite{Hostert:2020gou}.
In what follows, we focus purely on a fermionic dark sector with the couplings in \cref{eq:generic_dark_current}.

If the HNFs mix with neutrinos, they are usually called heavy neutral leptons (HNL) and are typically labeled in the literature as $N_4$, $N_5$, and so on. If the lightest is stable, they may be a DM candidate and would typically be denoted as $\chi_1$, with the heavier particles in the spectrum  $\chi_{j=2,\dots,n}$.
We retain here a more general notation, $\psi_{i=1,\dots,n}$, which encompasses both options.
The heavier states $\psi_{j=2,\dots,n}$, are expected to decay in cascades down to the lightest HNF, emitting two charged particles at each step~\footnote{If the dark photon is heavier than all HNFs, then only three-body decays are allowed. If not forbidden by the $C$-symmetry, decays into three HNFs are assumed to be kinematically forbidden.}.
These models were initially linked to iDM~\cite{Mohlabeng:2019vrz,Duerr:2019dmv}, where the dark photon decays into a heavy and short-lived HNF, and a lighter HNF, $\psi_1$, that is stable and constitutes the DM candidate. 
In this work, we generalize this idea to richer dark sectors.

\begin{figure*}[t]
    \centering
    \includegraphics[width=0.8\textwidth]{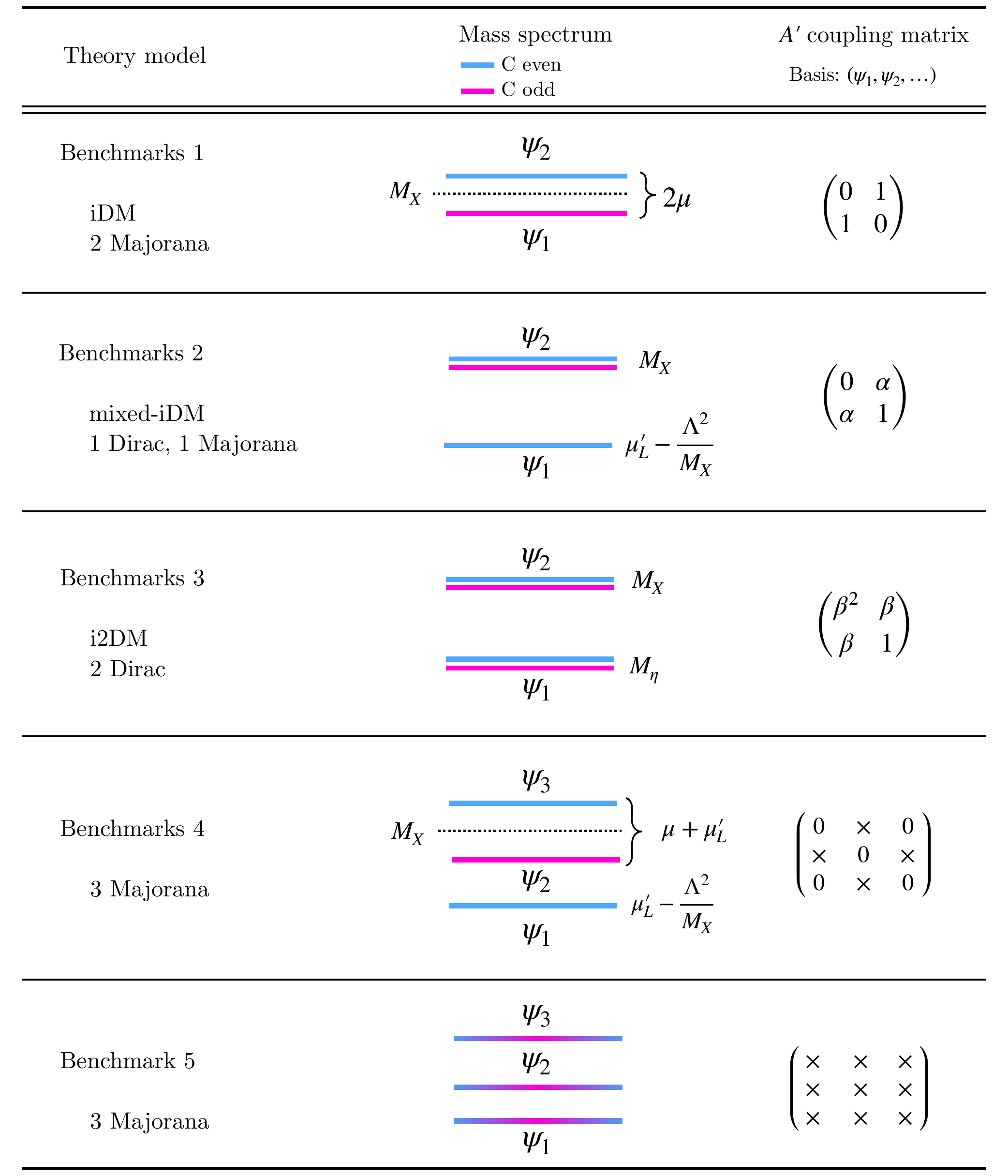}
    \caption{
    A summary of the benchmark models we consider, including the HNF spectra and their interaction vertices with the dark photon, $A^\prime$.
    Blue lines indicate $C$-even states, and pink lines $C$-odd states. 
    Pseudo-Dirac states with negligible mass splitting are denoted by two opposite $C$ lines close together.
    The vertex matrices are defined in the basis $(\psi_1, \psi_2, \dots)$, where the mixing angles are small $\alpha,\beta \ll 1$, and the entries denoted by $\times$ are arbitrary.
    Mixed-color states have no definite $C$ property.
    \label{fig:HNF_diagrams}
    }
\end{figure*}

In what follows, we systematically discuss the fermionic content of the models we study.
We start from the case of two Majorana HNFs, as for iDM. 
Then, following Ref.~\cite{Abdullahi:2020nyr}, we further extend the model to include 3 HNFs, organized into a pseudo-Dirac pair and one Majorana HNF, or 3 Majorana HNFs, depending on the choice of parameters. 
The former option is akin to the iDM model, while the latter is more general and can accommodate the neutrino mass model of Ref.~\cite{Abdullahi:2020nyr}. 
Finally, we present the case of four Majorana HNFs and its limit of two Dirac particles, recently studied in \cite{Filimonova:2022pkj}. 

In all cases, the dark photon decay into two $\psi_1$ needs to be suppressed as this contributes to its invisible branching ratio, which is severely constrained by missing-energy searches. We will show how a $C$-symmetry can satisfy such a requirement.

\begin{widetext}
\subsection{Two heavy neutral fermions (HNFs)}\label{sec:two_hnfs}

The simplest model we consider is that of two Majorana HNFs. 
In the interaction basis, we take $\chi_L$ and $\chi_R$ with charges $Q_L$ and $Q_R$.
The most general Lagrangian for their mass reads
\begin{align}\label{eq:2HNLlagrangian}
    \mathscr{L}_{\chi} &=  \overline{\chi_L}i\left(\slashed{\partial} - i g_D Q_L \slashed{A}^\prime\right)\chi_L
    +
    \overline{\chi_R^c}i\left(\slashed{\partial} + i g_D Q_R \slashed{A}^\prime\right)\chi_R^c
    - \frac{1}{2}\left[ \left( \begin{matrix} \overline{\chi_L^c} & \overline{\chi_R} \end{matrix}\right) \left( \begin{matrix} \mu_L &  m_D \\ m_D & \mu_R \end{matrix} \right) \left( \begin{matrix} \chi_L \\ \chi_R^c \end{matrix}\right) + {\rm h.c.} \right]
\end{align}
where the Majorana masses $\mu_L$ and $\mu_R$ break the $U(1)_D$ softly --- they can be generated by the vacuum expectation value of a dark Higgs $\Phi_2$ with $Q_{\Phi_2} = 2 Q_{L,R}$. 
The diagonalization of the symmetric and complex mass matrix $M$ can be achieved with the usual Takagi diagonalization, $\text{diag}(m_1, m_2) = U^T M U$, where $U = R (\theta) \text{diag}( e^{i\varphi}, 1)$ with $R(\theta)$ the matrix rotation by an angle,
\begin{equation}
    \tan 2\theta = \frac{m_D}{\Delta \mu}.
\end{equation}
We define $\Delta \mu = (\mu_R - \mu_L)/2$ and $\mu = (\mu_L + \mu_R)/2$.
$CP$ conservation is ensured when the Majorana phase is $\varphi = 0$ or $\pi/2$.
In terms of the Majorana mass eigenstates, the dark current is given by
\begin{equation}
\label{eq:iDMcurrent_full}
\mathcal{J}_D^{\mu} = \frac{Q_A - Q_V \cos{2\theta}}{2} \overline{\psi_2} \gamma^\mu \gamma^5\psi_2 
+ \frac{Q_A + Q_V \cos{2\theta}}{2} \overline{\psi_1} \gamma^\mu \gamma^5 \psi_1     
+ iQ_V \sin{2\theta} \sin\varphi   \,\overline{\psi_2} \gamma^\mu \psi_1 
+ Q_V \sin{2\theta} \cos\varphi \, \overline{\psi_2} \gamma^\mu\gamma^5 \psi_1,
\end{equation}
where $Q_{V} \equiv (Q_L + Q_R)/2$ and $Q_{A} \equiv (Q_L - Q_R)/2$.
Gauge anomaly cancellation fixes $Q_L = Q_R$ ($Q_A=0$).\\

\myparagraph{The $C$ symmetry}
For $\Delta \mu \to 0$, the mixing angle $\theta$ is maximal, and the dark photon couples only off-diagonally to the mass eigenstates.
The smallness of the on-diagonal couplings can be understood thanks to a $C$ symmetry.
The $C$ operator $U_c$ acts on Weyl fermions as
\begin{equation}
    U_c \chi_L U_c^{-1}  = \eta_c \psi_R^c, \quad 
    U_c \chi_R U_c^{-1}  = \eta_c \psi_L^c,
\end{equation}
where $\psi_L^c = C \overline{\psi_L}^T$ and we choose the phase factor $\eta_c = +1$, for simplicity.
In the case of a Dirac fermion, the $C$ operation is achieved by charge conjugating the field, $\chi = (\zeta, \, -i\sigma^2\xi^*)^T \mapsto \chi^c = (\xi,\, -i\sigma^2\zeta^*)^T$.
The $C$ symmetry is then respected when the Lagrangian is invariant under the exchange $\xi \leftrightarrow \zeta$ (i.e., $\chi_{L} \leftrightarrow \chi_{R}^c$). 

The left-handed fermions 
\begin{align}\label{eq:Cbasis}
\chi_{+} = \frac{\chi_L + \chi_R^c}{\sqrt{2}}, \quad \chi_{-} =  e^{i\varphi}\left(\frac{\chi_L - \chi_R^c}{\sqrt{2}}\right),
\end{align}
constitute the $C$ eigenbasis, where $\varphi$ is the same Majorana phase as before.
Given that $C(A^\prime_\mu) = -1$, the intrinsic $C$-parity of the fermions can be fixed as $C(\chi_{\pm}) = \pm 1$ without loss of generality.
The Lagrangian in \cref{eq:2HNLlagrangian} in this basis reads
\begin{align}
    \mathscr{L}_{\chi} &= \overline{\chi}_+ i\slashed{\partial} \chi_+ + \overline{\chi}_ - i\slashed{\partial} \chi_-  +  g_D A_{\mu}^\prime 
    \left[ \frac{Q_A}{2} \left( \overline{\chi_+}\gamma^\mu\gamma^5 \chi_+ +\overline{\chi_-}\gamma^\mu\gamma^5 \chi_-\right) + iQ_V  \overline{\chi}_+ \gamma^\mu \chi_- \right]
    \\\nonumber
    &- \left[ \frac{1}{2}\left( \begin{matrix} \overline{\chi_-^c} & \overline{\chi_+^c} \end{matrix}\right) \left( \begin{matrix} m_D - \mu &  i\Delta \mu \\ i\Delta \mu & m_D + \mu \end{matrix} \right) \left( \begin{matrix} \chi_- \\ \chi_+ \end{matrix}\right) + {\rm h.c.} \right],
\end{align}
where we took $\varphi \to \pi/2$ to ensure that the mass terms are positive for $m_D > \mu$. This signals that the two fermions have opposite $CP$ parities.
This basis is identified with the physical basis when $\Delta\mu\to0$.
As expected, $\Delta \mu$ and $Q_A$ are the only parameters that break $C$ in this model. 

In the $C$-symmetric limit, $\chi_\pm$ behaves like the components of a pseudo-Dirac particle with a mass gap $2\mu$. 
We can also conclude that if the interactions with the dark photon are off-diagonal in the $C$-conserving limit, then interactions with a $C$-even dark Higgs in the same limit would be purely diagonal.
In what follows, we assume that if any such scalar degree of freedom is part of the spectrum, it is heavier than $A^\prime_\mu$ and has negligible mixing with the SM Higgs.

Finally, the $C$ symmetry cannot be preserved in the Standard Model due to the different hypercharges of left- and right-handed fermions.
Therefore, we consider the breaking of $C$ to be stronger in the SM than in the DS.
If there are two $C$ symmetries in the theory, one in the SM, $C_{\rm SM}$, and one in the DS, $C_{\rm DS}$, then the conservation of $C_{\rm DS}$, but not $C_{\rm SM}$, would forbid kinetic mixing, $F^{\mu \nu}X_{\mu \nu} \xrightarrow{C_{\rm DS}} - F^{\mu \nu}X_{\mu \nu}$.\\
\end{widetext}

\myparagraph{Inelastic dark matter (iDM)}
In the $C$ symmetric limit and with an anomaly-free charge assignment, $Q_A=0$, we recover the well-known iDM  model~\cite{Tucker-Smith:2001myb,Cui:2009xq}.
Taking $\varphi = \pi/2$ and $Q_V=1$, the dark current is simply
\begin{align}
\label{eq:iDMcurrent}
\mathcal{J}_{\rm iDM}^\mu &= i\overline{\psi_2} \gamma^\mu \psi_1 + \text{h.c.}
\end{align}
This phenomenological model is defined by the following five parameters
\begin{equation}\label{eq:pheno_params}
    m_1, \,\, \Delta_{21} \equiv \frac{m_2 - m_1}{m_1}, \,\, r \equiv \frac{m_1}{m_{A^\prime}}, \,\, \alpha_D \equiv \frac{g_D^2}{4 \pi}, \text{ and } \epsilon.
\end{equation}
Historically, the interest in these models stemmed from explanations of the DAMA observation and the fact that the energy threshold for direct DM detection, as induced by $\Delta_{21}$, varies between Sodium-Iodine and Xenon experiments~\cite{Tucker-Smith:2001myb}. 
This explanation has since been ruled out by other direct detection experiments~\cite{CDMS-II:2010wvq,ZEPLIN-III:2010cnv,XENON100:2011hxw}. 
Still, the interest in iDM has persisted, especially in the context of accelerator experiments~\cite{Izaguirre:2015zva,Berlin:2018jbm,Tsai:2019buq,Kim:2016zjx}, where the co-annihilator can be searched for through its displaced decays.

Self-interactions of $\psi_1$ can only proceed through scalars in this model.
If the scalar mixes with the Higgs, $\lambda |\Phi|^2|H|^2$, direct annihilation to SM fermions can take place.
The mixing with the Higgs induces couplings smaller than the Higgs' Yukawa couplings with SM fermions; annihilations are suppressed by the fermion masses, in addition to the Higgs mixing parameter.
The mixing should, therefore, be small enough for the scalar contribution to the self-annihilation of DM to be sub-dominant to the exponentially suppressed $A^\prime$-mediated coannihilations.
In addition, we assume the dark higgs is heavier than the HNFs, as otherwise secluded annihilation would dominate~\footnote{This scenario was explored in Ref.~\cite{Duerr:2020muu}.
The authors also find that BaBar does not rule out the entire $\deltaamu$ region of preference but did not study constraints from NA64 and Higgs decays.}.

As we will show in \Cref{sec:results}, the explanation of $\deltaamu$ in this model is in tension with invisible dark photon limits.
Each dark photon produced can only be accompanied by a single semi-visible decay, $\psi_2 \to \psi_1 f^+f^-$, with $f$ a SM particle. Even small losses of acceptance in the detector can lead to a missed $e^+e^-$ pair.
With this limitation in mind, we consider new models where multiple unstable fermions accompany dark photon production.

There are two ways to achieve this:
\begin{enumerate*} 
\item in models where $\psi_2$ can be produced in pairs, $A^\prime \to \psi_2 \psi_2$, and 
\item in models of three or more HNFs, where the dark photon couples predominantly to the heaviest and most short-lived states, e.g. $A^\prime \to \psi_3 \psi_2$.
\end{enumerate*}

We explore these possibilities in models with three and four Weyl fermions, always reducing the phenomenological model to, at most, three distinguishable states in the spectrum.
This allows for better compatibility between the $\deltaamu$ anomaly and a dark photon explanation. We will prove this in \Cref{sec:results} with a detailed analysis.

\subsection{Three HNFs} \label{sec:three_hnfs}

We start by extending the two HNF model by a single fully sterile Weyl fermion.
We keep the two dark fermions, $\chi_R$ and $\chi_L$, with the same charges $Q_L = Q_R = 1$, and introduce a new singlet fermion, $\eta_L$. 
This content is a simplified version of the three-portal model of Ref.~\cite{Abdullahi:2020nyr}.

The Lagrangian is given by
\begin{align}\label{eq:3HNFlagrangian}
    \mathscr{L}_{\rm 3-HNF} & = \mathscr{L}_{\chi}  + \overline{\eta_L}i\slashed{\partial}\eta_L  \
    \\\nonumber & - \left[ \frac{\mu_L^\prime}{2}\overline{\eta_L^c}\eta_L
    +  \Lambda_L\overline{\eta_L^c} \chi_L + \Lambda_R \overline{\eta_L^c}  \chi_R^c  + \text{ h.c.}\right].
\end{align}
The mixing terms break the $U(1)_D$ and can be generated by the vacuum expectation value of a scalar particle $\Phi_1$, which carries charge $Q_{\Phi_1} = 1$. 
In that case, $\Lambda_{L,R} \equiv Y_{L,R} \,v_{\Phi_1}/\sqrt{2}$ where $Y_{L,R}$ are the Yukawa couplings. 
Another dark Higgs $\Phi_2$ with $Q_{\Phi_2} = 2$ could generate the Majorana masses of the two dark fermions after symmetry breaking.

Because $\eta_L$ is completely neutral, it can couple to the SM lepton doublets via the Yukawa coupling $\overline{L}\tilde{H}\eta_L^c$.
While these terms play an essential role in the mass generation of light neutrinos, they only give a small correction to the HNF masses. The neutrino Yukawa coupling is constrained to be small and will have no impact on the collider and fixed-target phenomenology we discuss.
We will consider the impact of this coupling on neutrino mass generation in \cref{sec:HNL_theory}.
These terms allow the lightest HNF to decay into SM neutrinos.
To ensure the stability of the dark matter candidate, we forbid these terms by charging all DS fermions, $\chi_L, \chi_R, \eta_L$, under a dark parity, e.g. $Z_2$ symmetry. This dark parity can also be attributed to the conservation of lepton number if $L(\eta_L) = 0$, which would forbid the neutrino Yukawa coupling~\cite{Ma:2015xla}.

We use the left-handed dark fermion basis $\chi_+$ and $\chi_-$ introduced in \cref{eq:Cbasis}, and set the Majorana phases to be such that $CP$ is conserved and the mass terms are positive when $M_X > \mu$.
In that case, the DS fermion mass matrix is
\begin{widetext}
\begin{align}
\label{eq:mass3HNF}
    -\mathscr{L_{\rm 3-HNF}} &\supset 
    \frac{1}{2}\left( \begin{matrix} \overline{\eta_L^c} 
    & 
    \overline{\chi_-^c} 
    & 
    \overline{\chi_+^c} \end{matrix}\right)   
    \left( 
    \begin{matrix} 
                    \mu_L^\prime & \Delta \Lambda &\Lambda  
                    \\ \Delta\Lambda & M_X - \mu & \Delta \mu 
                     \\ \Lambda & \Delta \mu & M_X + \mu \end{matrix} 
    \right)
    \left( \begin{matrix} \eta_L \\ \chi_- \\ \chi_+ \end{matrix}\right) + {\rm h.c.},
\end{align}
\end{widetext}
where $\Lambda = (\Lambda_L + \Lambda_R)/\sqrt{2}$ and $\Delta \Lambda = (\Lambda_R - \Lambda_L)/\sqrt{2}$.
Imposing the $C$ symmetry in the $\chi$ sector~\footnote{Just like in the SM, the $C$ symmetry is broken in this model due to the odd number of Weyl fermions. This would indicate that only the $U(1)_D$-charged sector respects $C$.}, we recover the limit where $\Delta \mu = \Delta \Lambda = 0$. 
We find an analogous situation to the two HNF cases, with the difference that $\chi_+$ can now mix with a sterile state. 
Indeed, since C$(\eta_L) = +1$, $C$-conservation implies that only the $C$-even fermion can mix with $\eta_L$.
As we will see, the spectrum can consist of one Dirac and one Majorana particle or three Majorana states.


The $C$-odd state $\chi_- \equiv \psi_2$ decouples, and $\eta$ and $\chi_+$ mix. 
A single rotation in the $C$-even sector leads to the mass basis,
\begin{align}
\psi_1 &= c_\alpha \eta + s_\alpha \chi_+, & m_1 = \mu_L^\prime - M \frac{\sin^2 \alpha}{\cos{2\alpha}}
&
\\
\psi_2 &= \chi_-, & m_2 = M_X - \mu_L^\prime&
\\
\psi_3 &= -s_\alpha \eta + c_\alpha \chi_+, & m_3 = \mu_L^\prime + M \frac{\cos^2 \alpha}{\cos{2\alpha}}&
\end{align}
with $\tan 2 \alpha = 2 \Lambda /M$ and $M = M_X + \mu - \mu_L^\prime$.
If $\tan{2\alpha} \ll 1$, a seesaw mechanism is in place, and $\psi_2$ and $\psi_3$ form a pseudo-Dirac pair.
The other possibility, $\tan 2 \alpha \gg 1$, does not preserve the pseudo-Dirac limit.
The splittings in the model are then given by $m_3 - m_1 \sim M + 2\Lambda^2/M$ and $m_3 - m_2 \sim \Lambda^2/M + 2\mu$.

In the $C$ symmetric case, the dark current is also fully off-diagonal and is given by
\begin{equation}
\mathcal{J}_{\rm 3-HNF}^{\mu} \supset s_\alpha\overline{\psi_2} \gamma^\mu \psi_1 + c_\alpha \overline{\psi_2} \gamma^\mu \psi_3 + \text{ h.c.}
\end{equation}
In the limit of small $\alpha$, the dark photon interacts more strongly with the pseudo-Dirac pair. 
When the $\psi_1$ HNF is a dark matter particle, its relic abundance is set exclusively through coannihilation with the heavier pseudo-Dirac partner.
Similarly to the minimal iDM model, this mechanism evades constraints from the CMB and direct detection experiments.
Below, we highlight the two types of phenomenological models that can be derived from \cref{eq:3HNFlagrangian} above.

\myparagraph{Mixed inelastic dark matter (mixed-iDM)}
The first phenomenological scenario we can consider is the limit where two of the Weyl fermions make up a mostly-dark, pseudo-Dirac particle, while the third Weyl fermion remains a Majorana particle, mostly in the direction of the sterile state.
In this case, the lighter, mostly-sterile Majorana fermion would constitute dark matter, while the mostly-dark pseudo-Dirac fermion plays the role of the co-annihilator.
In this case, the self-annihilation of dark matter via $A^\prime$ interactions is forbidden by the $C$ symmetry, and not constrained by CMB limits.
The model is a trivial extension of the iDM model and invokes the same $C$ symmetry used there.

Considering the Majorana state $\psi_1$ and a \mbox{(pseudo-)Dirac} state $\Psi_2$, the dark current is
\begin{equation}
    \mathcal{J}_{\rm mixed-iDM}^{\mu} \supset s_\alpha\overline{\Psi_2} \gamma^\mu \psi_1 + c_\alpha \overline{\Psi_2} \gamma^\mu \Psi_2 + \text{h.c.},
\end{equation}
and so this model is fully specified by \cref{eq:pheno_params} and $\alpha$.
To make use of the model above, one must guarantee the coherence of the Majorana states $\psi_2$ and $\psi_3$ in the pseudo-Dirac state $\Psi_2$, so that we are justified in treating them as a single Dirac particle in the phenomenological work.
Note that $\Delta_{32}$ will only play a minor role in the dark matter hypothesis since the relevant splitting for coannihilations is the one between $\psi_1$, the dark matter candidate, and $\Psi_2$, its interaction partner.
We can express $M_X$ and $\Lambda$ in terms of $\tan 2 \alpha$, which controls the decay rate $\Psi_2 \rightarrow \psi_1$, and the splitting $\Delta_{21} \equiv (m_2 - m_1)/m_1$, which has an important impact on the coannihilation rate for dark matter.
For $\mu < M_X$, we find
\begin{align}
m_1 &\simeq \mu_L^\prime - \frac{1}{4} \mu \Delta_{21} \tan^2 2 \alpha ~, \\
m_2 &= \mu_L^\prime (1 + \Delta_{21}) ~,  \\
m_3 &\simeq \mu_L^\prime (1 + \Delta_{21}) + \frac{1}{4} \mu \Delta_{21} \tan^2 2 \alpha ~.
\end{align}
We notice that $\Delta_{32} = \frac{1}{4} \frac{\Delta_{21}}{1+ \Delta_{21}} \tan^2 2 \alpha$ and is small as far as the condition  $\tan^2 2 \alpha \ll 1$ holds.
For $\mu_L^\prime = \mu = 0$, the splitting of the pseudo-Dirac pair is $\Delta_{32} \propto \alpha^2$, so that in the limit of $\alpha \to 0$, we recover an exact Dirac state.
In summary, provided the mixing angle is small, the $\Delta_{32}$ splitting is negligible. 
The decay rate for three-body decays like $\psi_3 \to \psi_2 + \dots$ are suppressed by \mbox{$\Delta_{32}^5 \propto \alpha^{10}$}, and so, can be safely neglected for the mixing angles considered here.

\myparagraph{Three Majorana fermions} 
Relaxing the condition on $\alpha \ll 1$ and the $C$ symmetry in the dark sector, it is possible to split the mass eigenstates away from a heavy Pseudo-Dirac pair and therefore have three hierarchical Majorana HNFs. 
This structure enhances the semi-visible decay rates of $\psi_3$ and $\psi_2$ while suppressing the on-diagonal terms in the dark sector current. 
The benchmark points exemplify this in Ref.~\cite{Abdullahi:2020nyr}. 
This case is of interest for providing both a viable inelastic DM model, compatible with CMB bounds, or, alternatively, a heavy neutral lepton interpretation with interesting phenomenological consequences, e.g. Ref.~\cite{Abdullahi:2020nyr}.

We can also obtain some useful approximate formulas. 
For the $CP$ conserving case, a mild hierarchy can be obtained for large values of $\tan 2 \alpha$. 
For $0<  \mu \simeq M_X$, we have $(M_X-\mu)/M_X \tan^2 2 \alpha \simeq 2 \Delta_{21}/(1+ \Delta_{21})$ and $\Delta_{32} \simeq \Delta_{21}/(1+ \Delta_{21})$, implying a mildly hierarchical HFN spectrum.  
Moving away from the $C$ symmetric limit, a sizeable $\Delta \Lambda$ can lead to a stronger hierarchy of masses, as, for instance, in benchmark BP5.

Depending on the mass hierarchy, it would also be possible for the heavy states to decay into multiple lighter HNFs, in particular into 3 $\psi_1$. In the case under consideration, such decays are kinematically forbidden. We avoid this possibility as these decay channels could easily dominate and enhance the invisible branching ratio of the dark photon if $\psi_1$ is stable or long-lived, as is typical in these models.

\subsection{Four HNFs} \label{sec:four_hnfs}

If we further enlarge the fermionic sector, it is possible to recover a 2-Dirac fermion picture. Two families of HNF exist: one neutral under all gauge symmetries, $\eta$, and one charged under the dark gauge symmetry, $\chi$. 
Our Lagrangian in this case reads
\begin{widetext}
\begin{equation}
    \mathscr{L} = \mathscr{L}_{\chi} + \overline{\eta}i\slashed{\partial}\eta - M_\eta \overline{\eta} \eta - \left[ \frac{\mu^\prime_R}{2} \overline{\eta_R} \eta_R^c + \Lambda_L^\prime \overline{\eta_R}\chi_L 
    +\Lambda_R^\prime \overline{\eta_R}\chi_R^c + 
    \frac{\mu_L^\prime}{2}\overline{\eta_L^c}\eta_L
    +  \Lambda_L\overline{\eta_L^c}   \chi_L + \Lambda_R \overline{\eta_L^c}  \chi_R^c
    + \text{h.c.}\right],
\end{equation}
where again we have omitted potential Yukawa couplings between SM neutrinos and the sterile fermions, $\eta_L$ and $\eta_R$ (cf. \cref{sec:three_hnfs}).
In the $C$ symmetric limit, one can show that $\Lambda_L = \Lambda_R^\prime$ and $\Lambda_L = \Lambda_R^\prime$.
In this limit, for an appropriate choice of Majorana phases, the mass matrix in the $C$ eigenbasis is,
\begin{align}
\label{eq:mass4HNF}
    -\mathscr{L_{\rm 4-HNF}} &\supset 
    \frac{1}{2}\left( \begin{matrix} \overline{\eta_-^c} & \overline{\eta_+^c} & \overline{\chi_-^c}  \overline{\chi_+^c} \end{matrix}\right)   
    \left( 
    \begin{matrix} 
    M_\eta - \mu^\prime & 0 & \Lambda_- & 0 \\    
    0 & M_\eta + \mu^\prime & 0 & \Lambda_+ \\
    \Lambda_- & 0 &  M_X - \mu & 0 \\
    0 & \Lambda_+ & 0 & M_X+\mu 
    \end{matrix} 
\right)
    \left( \begin{matrix} \eta_- \\ \eta_+ \\ \chi_- \\ \chi_+ \end{matrix}\right) + {\rm h.c.},
\end{align}
where $\Lambda_\pm \equiv (\Lambda_L^\prime + \Lambda_R)/2 \pm (\Lambda_L + \Lambda_R^\prime)/2$.
The $C$-even and $C$-odd sectors decouple.
\end{widetext}
We introduce two commuting rotations defined by the mixing angles $\tan 2\beta_{\pm} = 2\Lambda_\pm/\Delta_\pm$, where $\Delta_\pm = \pm (M_X -M_\eta) + \mu - \mu^\prime$.
The spectrum is then given by
\begin{align}\label{eq:spectrum-4HNF}
    \psi_1 &= c_{\beta_-}\eta_- + s_{\beta_-}\chi_-, &m_1 = M_\eta - \mu^\prime + \Delta_- \frac{\sin^2\beta_-}{\cos{2\beta_-}} \nonumber
    &\\\nonumber
    \psi_2 &= c_{\beta_+}\eta_+ + s_{\beta_+}\chi_+, &m_2 = M_\eta +\mu^\prime  - \Delta_+ \frac{\sin^2\beta_+}{\cos 2\beta_+}
    &\\ \nonumber
    \psi_3 &= s_{\beta_-} \chi_- - c_{\beta_-}\eta_-, &m_3 = M_X - \mu - \Delta_- \frac{\sin^2\beta_-}{\cos{2\beta_-}}& \\ \nonumber
    \psi_4 &= s_{\beta_+}\eta_+ - c_{\beta_+} \chi_+, &m_4 = M_X + \mu  + \Delta_+ \frac{\sin^2\beta_+}{\cos 2\beta_+} 
\end{align} 
When $\mu, \mu^\prime, \Lambda \ll M_X, M_\eta$, the spectrum is composed of two pseudo-Dirac particles, split by the $U(1)_D$-breaking terms.

\begin{table*}
    \centering
        \begin{tabular*}{\textwidth}{@{\extracolsep{\fill}}|cc|cccc|cccccc|l|}
            \hline
             \multirow{2}{*}{BP} & \multirow{2}{*}{model} & \multirow{2}{*}{$r$} & \multirow{2}{*}{$\Delta_{21}$} & \multirow{2}{*}{$\Delta_{32}$} & \multirow{2}{*}{$\alpha_D$} & \multicolumn{1}{c}{$V_{11}$} & $V_{21}$& $V_{22}$ & $V_{31}$ & $V_{32}$ & $V_{33}$ & \multirow{2}{*}{Comment}\\
             & & & & & &  \multicolumn{6}{c|}{$/10^{-2}$} & \\
            \hline
            \hline
            1a & iDM & $1/3$ & $0.5$ &  $-$ & 0.5 & $0$ & $1$ & $0$ & $0$ & $0$ & $0$ & - \\
            1b & iDM & $1/3$ & $0.4$ &  $-$ & 0.1 & $0$ & $1$ & $0$ & $0$ & $0$ & $0$ & same as \cite{Mohlabeng:2019vrz} \\
            \hline \hline
            2a & mixed-iDM & $1/3$ & $0.3$ &  $-$ & 0.5 & $0$ & $s_\alpha c_\alpha$ & $c_\alpha^2$ & $-$ & $-$ & $-$ & $\alpha = 8^\circ $\\
            2b & mixed-iDM & $1/3$ & $0.3$ &  $-$ & 0.5 & $0$ & $s_\alpha c_\alpha$ & $c_\alpha^2$ & $-$ & $-$ & $-$ & $\alpha = 4^\circ $\\
            \hline \hline
            3a & i2DM & $1/3$ & $0.4$ & $-$ & 0.5 & $s_\beta^2$ & $s_\beta c_\beta$ & $c_\beta^2$ & $-$ & $-$ & $-$ & $\beta = 8.6^\circ$ \\
            3b & i2DM & $1/3$ & $0.4$ & $-$ & 0.5 & $s_\beta^2$ & $s_\beta c_\beta$ & $c_\beta^2$ & $-$ & $-$ & $-$ & $\beta = 4.6^\circ$ \\
            3c & i2DM & $1/3$ & $0.4$ & $-$ & 0.5 & $s_\beta^2$ & $s_\beta c_\beta$ & $c_\beta^2$ & $-$ & $-$ & $-$ & $\beta = 2.3^\circ$ \\
            3d & i2DM & $1/3$ & $0.4$ & $-$ & 0.5 & $s_\beta^2$ & $s_\beta c_\beta$ & $c_\beta^2$ & $-$ & $-$ & $-$ & $\beta = 1.1^\circ$ \\
            \hline \hline
            4a & 3 HNFs & $0.11$ & $2.44$ & $0.54$ & $0.3$ & $0$ & $3.9$ & $0$ & $0$ & $99$ & $0$ & same as \cite{Abdullahi:2020nyr} \\
            4b & 3 HNFs & $0.16$ & $2.44$ & $0.54$ & $0.3$ & $0$ & $3.9$ & $0$ & $0$ & $99$ & $0$ & same as \cite{Abdullahi:2020nyr} \\
            4c & 3 HNFs & $0.15$ & $0.85$ &  $0.77$ & $0.3$ & $0$ & $0.10$ & $0$ & $0$ & $99$ & $0$ & same as \cite{Abdullahi:2020nyr} \\
            \hline\hline
            5 & 3 HNFs & $0.16$ & $0.573$ & $0.586$ & $0.3$ & $0.40$ & $7.8$ & $8.3$ & $2.8$ & $98$ & $69$ & same as \cite{Abdullahi:2020nyr} \\
            \hline
        \end{tabular*}
    \caption{The benchmark models for semi-visible dark photons used in this work.
    In the second column, we specify the type of model considered.
    Here, $r=m_1/m_{A^\prime}$ and $\Delta_{ij}=(m_i - m_j)/m_{j}$. The dark photon coupling vertices $V_{ij}$ are defined in \cref{eq:generic_dark_current}.
    \label{tab:BPtable}}
\end{table*}

One more limit of interest is considering the $U(1)_D$ to be exclusively broken by one unit, such that $\mu = \mu^\prime = 0$ and $\Delta_+ = -\Delta_- \equiv \Delta$. In that case, the Dirac pairs are split by 
\begin{equation}
    \Delta_{43} \sim \Delta_{12} \sim \Delta (\beta_+^2 - \beta_-^2),
\end{equation}
which is small for small mixing angles and vanishes when $\beta_+ = \beta_-$.
This last regime is the limit where the two pairs compose exact Dirac fermions, achieved in two cases: 
i) $\Lambda_+ = - \Lambda_- $ ($\Lambda_L^\prime = \Lambda_R = 0$), or ii) $\Lambda_+ = \Lambda_- $ ($\Lambda_L = \Lambda_R^\prime = 0$).

In terms of the mass eigenstates, the dark current takes the simple form,
\begin{align}
    \mathcal{J}_X^\mu &= c_{\beta_+} c_{\beta_-} \overline{\psi_4} \gamma^\mu  \psi_3 + s_{\beta_+} c_{\beta_-} \overline{\psi_4} \gamma^\mu \psi_1 
    \\\nonumber
    &+ s_{\beta_-} c_{\beta_+} \overline{\psi_3} \gamma^\mu \psi_2 + s_{\beta_+} s_{\beta_-} \overline{\psi_2} \gamma^\mu \psi_1 + \text{h.c.},
\end{align}
where only interactions between $C$-odd and $C$-even states are allowed, and where the heaviest pseudo-Dirac pair couples most strongly to the dark photon.
Decays of the type $\psi_{4,2} \to \psi_{3,1} + \dots$ are suppressed with respect to the dominant $\psi_{4,3} \to \psi_{1,2} + \dots$  by factors of $(\beta_+^2-\beta_-^2)^5 \beta_\pm^2$.
In fact, for the typical mixing angles we consider, the particle $\psi_2$ is semi-stable, as $m_2 - m_1 < 2 m_e$.

In summary, in the $C$ symmetric limit, we find a pair of pseudo-Dirac particles, each split by a small gap, proportional to $\Delta (\beta_+^2-\beta_-^2)$, where $\beta_{+}$ and $\beta_-$ are the mixing angles in the $C$-even and $C$-odd sectors, respectively.
The individual splittings are only relevant for large mixings, and vanish exactly in the limit $\beta_+ = \beta_-$.



\myparagraph{Inelastic Dirac dark matter (i2DM)} 
In the exact Dirac limit, a simple phenomenological model arises with only two particles in the spectrum. 
A light, mostly-neutral Dirac fermion $\Psi_1$, constituting a dark matter candidate, and a heavier, mostly-dark Dirac fermion $\Psi_2$.
In terms of these degrees of freedom, the dark current is given by
\begin{equation}
    \mathcal{J}_{\rm i2DM}^\mu = s_\beta^2 \overline{\Psi_1} \gamma^\mu \Psi_1 + s_\beta c_\beta \left( \overline{\Psi_2} \gamma^\mu \Psi_1 + \text{h.c.} \right)+ c_\beta^2 \overline{\Psi_2} \Psi_2.
\end{equation}
As expected, $\Psi_2$ is coupled more strongly to the dark photon, and, therefore, will aid in the depletion of $\Psi_1$ particles in the freeze-out mechanism through its coannihilations.
The mixing-suppressed self-interactions of the dark matter particle weaken the CMB limits on $\Psi_1 \Psi_1 \to e^+ e^-$.
The relic density is typically set by the self-annihilation of the heavy partner, $\Psi_2 \Psi_2 \to f^+f^-$, and coscattering, $\Psi_2 \Psi_2 \leftrightarrow \Psi_{2,1} \Psi_1$ and $\Psi_2 f \leftrightarrow \Psi_{1} f$~\cite{Filimonova:2022pkj}.
This model differs from the mixed-iDM scenario mostly in that the off-diagonal interactions between dark matter and its co-annihilator are suppressed with respect to the self-interactions of the co-annihilator, $\Psi_2$.
The phenomenology is fully determined by the parameters in \cref{eq:pheno_params}, in addition to $\beta$.
Similar ideas of a sterile dark matter particle co-annihilating with heavier dark partners have been explored before in the context of a toy model with two Majorana fermions~\cite{DAgnolo:2018wcn}.

With regards to the accelerator phenomenology, the branching ratios of the dark photon to the lighter fermions will be hierarchical, approximately following a proportion of $(1: \beta^2:\beta^4)$ for decays into $(\Psi_2 \Psi_2, \Psi_2\Psi_1, \Psi_1\Psi_1)$ final states. 
The dominance of $A^\prime \to \Psi_2 \Psi_2$ decays guarantees a large number of events with two semi-visible particles, further relaxing constraints on kinetic mixing coming from invisible dark photon searches.
The presence of more visible final states enhances the prospects for discovery.

\subsection{Mixing with light neutrinos}
\label{sec:HNL_theory}

So far we have considered a secluded sector that feebly interacts with the SM only via the vector (and possibly scalar) portal. Generically, in the presence of sterile fermions, Yukawa couplings with both the SM leptonic doublet and the DS are allowed, and, after symmetry breaking, will lead to mixing between neutrinos and HNFs. Conventionally, the HNFs are called HNLs in this scenario.
 
The HNLs are unstable, as they can always decay to e.g. neutrinos, and the lightest particle in the spectrum cannot constitute DM.
In order to recover a stable candidate for DM, it is necessary to advocate an additional symmetry which distinguishes the HNLs from the light neutrinos and forbids Yukawa couplings with the leptonic doublets. 
The simplest such symmetry is a $Z_2$ and would guarantee the stability of the lightest HNL.

The HNL scenario is most easily realized in models with three or more HNFs, in which the neutral fermions $\eta$ are free to mix with the SM neutrinos, in the absence of any additional symmetries.
For the model in \cref{sec:three_hnfs}, one can add the following Yukawa interaction
\begin{equation}
    \mathscr{L} = \mathscr{L}_{3{\rm -HNF}}  - \sum_{\alpha=e,\mu,\tau} \left(y_\alpha \overline{L_{\alpha}}\widetilde{H} \eta_L^c + \text{h.c.}\right),
\end{equation}
where $L_\alpha$ is the $SU(2)$ leptonic doublet of the SM, and $\widetilde{H}= i\sigma_2 H^*$ the conjugate of the Higgs doublet.
Similar terms involving $\eta_R$ could be added to the four HNF models of \cref{sec:four_hnfs}. 
After EW and $U(1)_D$ symmetry breaking, the HNFs mix amongst themselves and we can justifiably call them HNLs, $N_i \equiv \psi_i$.
With the addition of a dark scalar $\Phi$ with a dark charge $Q_\Phi = 1$, the Lagrangian above is identical to the model discussed in Ref.~\cite{Abdullahi:2020nyr}.

The mixing of active SM neutrinos and HNLs is constrained to be small by direct laboratory searches.
For the values of kinetic mixing and $A^\prime$ mass considered in this paper, the lightest HNL, $N_4$, will decay via $N_4 \to \nu \ell^+\ell^-$ or $N_4 \to \nu \pi^+\pi^-$. 
Due to the suppression of the small mixing with the neutrinos, $N_4$ is usually long-lived ($c\tau_{N_4} \gg 1$~m) and constitutes missing energy at $e^+e^-$ colliders and fixed-target experiments. 
The experimental consequences of the mixing with neutrinos is discussed in \cref{sec:HNL_pheno}.
For a review of the phenomenology of HNLs, see Ref.~\cite{Abdullahi:2022jlv}.

A key consequence of this setup is light neutrino mass generation.
In fact, a GeV-scale seesaw mechanism as the origin of the observed neutrino masses and mixing angles has been extensively studied in the literature~\cite{Minkowski:1977sc,Mohapatra:1979ia,GellMann:1980vs,Yanagida:1979as,Lazarides:1980nt,Mohapatra:1980yp,Schechter:1980gr,Cheng:1980qt,Foot:1988aq}. For a review of this topic, see Ref.~\cite{Agrawal:2021dbo} and references therein.

It is also interesting to consider the role of lepton number in these scenarios. Charging $\eta_L$ so that the Yukawa coupling is lepton number conserving implies that the Majorana mass term $\mu^\prime_L$ breaks it by two units. In the dark sector, the charge assignment of $\chi_L$ and $\chi_R$ is arbitrary and, depending on the specific choice, lepton number will be broken by $\Lambda_{L,R}$, $M_X$ and $\mu_{L,R}$, or by $\Lambda_{L}$ and $\mu_{L,R}$, for $L(\chi_L)=L(\chi_R^c)=0$, $L(\chi_L)=L(\chi_R^c)=1$ or $L(\chi_L)=-L(\chi_R^c)=1$, respectively. 
Both $\Lambda_{L,R}$ and $\mu_{L,R}$ terms also break $U(1)_D$ by one and two units, respectively, and can arise once multiple scalars, carrying $U(1)_D$ charges, develop a vacuum expectation value. In the most minimal case of one scalar $\Phi$ with dark charge $Q_\Phi = 1$, either the resulting $\Lambda_L$, or $\Lambda_R$, term breaks the lepton number explicitly by 2 units.  We leave further theoretical considerations to future work. We also note that the C-symmetry introduced earlier is not compatible with $U(1)_L$ in this minimal realization if $\eta_L$ is charged under $U(1)_L$.

Light neutrino masses need to depend on all the $U(1)_L$-breaking parameters. As an interesting example, let us consider the case in which the $\chi_{L,R}$ do not carry lepton number and only one scalar is included in the theory, so that $\mu_L=\mu_R=0$. We assume that lepton number violating terms are small, implying $\Lambda_{L,R}\ll M_X$  after $U(1)_D$ breaking. Another choice for the charges consists in having all new fields neutral. In this case, the lepton number violating term is the Yukawa coupling itself, explaining naturally its smallness. For negligible $\mu^\prime_L$ and  $\Lambda_{L,R}\ll M_X$, we have $m_1 \simeq \Lambda^2/M$, $m_2\simeq m_3= M$ and light neutrino masses arise
\begin{equation}
    m_\nu \simeq \frac{y^2 v_H^2}{\Lambda^2} M .
\end{equation}
Note that this is a one-generation estimate, but a full flavor analysis is needed to determine the values of the three light masses and mixing parameters.

It should be pointed out that additional contributions can also come from loops, especially if $\mu^\prime_L$ is large, and they can be significant owing to the fact that the scale of symmetry breaking in the dark sector is smaller than the electroweak scale~\cite{Ballett:2019cqp}.

The case with four HNFs is even richer in possibilities owing to the enlarged fermionic sector. In this case, it is possible to add to Yukawa interactions with the SM
\begin{equation}
    \mathscr{L} = \mathscr{L}_{4{\rm -HNF}} - \sum_{\alpha= e,\mu,\tau} \left(y_\alpha \overline{L_{\alpha}}\widetilde{H} \eta_R + y_\alpha^\prime \overline{L_{\alpha}}\widetilde{H} \eta_L^c + \text{h.c.}\right).
\end{equation}
One option is not to charge either of $\eta_{L,R}$ implying that both Yukawa coupings are suppressed being lepton number violating. On the contrary, if $L(\eta_L)=L(\eta_R)=1$, $M_\eta$ is allowed while $L$-conservation implies $y_\alpha^\prime$ to be very small. Depending on the lepton charge assignment of the $\chi_{L,R}$ fields, the different terms in the full Lagrangian are $L$-violating, in addition to being $U(1)_D$ violating, and will be either forbidden or can be taken to be naturally small, if the L symmetry is just approximate.
A full analysis of the different cases is beyond the scope of the current discussion. We highlight one specific case which is of particular interest: the case in which $L(\chi_L)=L(\chi_R)=1$~\footnote{The case $L(\chi_L)=L(\chi_R)=-1$ is equivalent.}. This choice is compatible with the $C$-symmetry discussed earlier in order to avoid diagonal dark photon vertices and forbids the terms $\Lambda^\prime_R$ and $\Lambda_L$. We notice that in this case, the lightest neutrino mass is zero as it is protected by the accidental lepton number symmetry. Small neutrino masses can then be controlled by lepton number breaking terms either introduced directly in the Lagrangian as technically natural, as in standard extended seesaw models, or induced by additional scalars which take a $U(1)_L$-breaking vev.

\section{Model-independent limits} \label{sec:model_independent}

Our region of interest for dark photons includes $10\text{ MeV} < m_{A^\prime} < 10$~GeV and kinetic mixing of order $10^{-4} < \varepsilon < 0.1$.
In this region, it has long been known that colliders, fixed-target, and beam dump experiments provide the best limits on dark photons~\cite{Pospelov:2008zw,Batell:2009yf,Batell:2009di,Essig:2013vha,Izaguirre:2014bca}.
For a compilation of constraints on dark photons, see Refs.~\cite{Fabbrichesi:2020wbt,Agrawal:2021dbo,Batell:2022dpx}.
A discussion of the phenomenology at colliders is given in Ref.~\cite{Graham:2021ggy}.

Semi-visible dark photons decay into visible particles and missing energy, modifying both bounds on visible and invisible $A^\prime$ models. 
The dominant branching ratio (BR) of semi-visible $A^\prime$ is into the HNFs, which subsequently produce both visible and invisible particles.
This BR cannot be reconstructed as a visible resonance due to the missing energy, and it also does not satisfy the criteria of missing energy searches when the visible products are picked up by the detector.
We leave a detailed discussion of the reinterpretation of searches for missing energy to \cref{sec:recast}.

\myparagraph{Visible resonance searches}
In principle, resonance searches at $e^+e^-$ colliders~\cite{KLOE-2:2012lii,BaBar:2014zli} and the LHC~\cite{LHCb:2017trq} can still constrain the direct decays of $A^\prime$ into SM particles, $A^\prime \to \ell^+\ell^-, \, \pi^+\pi^-, \pi^+\pi^- \pi^0$.
These BRs, while still present, are typically much smaller than the BRs into HNFs, as they are of the order of $\epsilon^2 \alpha/\alpha_D$.
In addition, when $\alpha_D$ is large, the dark photon will be a much wider resonance, somewhat decreasing the effectiveness of the bump hunt method.
We do not show the rescaled limits from visible searches in our plots, as they are typically much weaker than the model-independent constraints discussed below.
We come back to the importance of visible searches in \cref{sec:discussion}.

\begin{figure*}
    \centering
    \includegraphics[width=\textwidth]{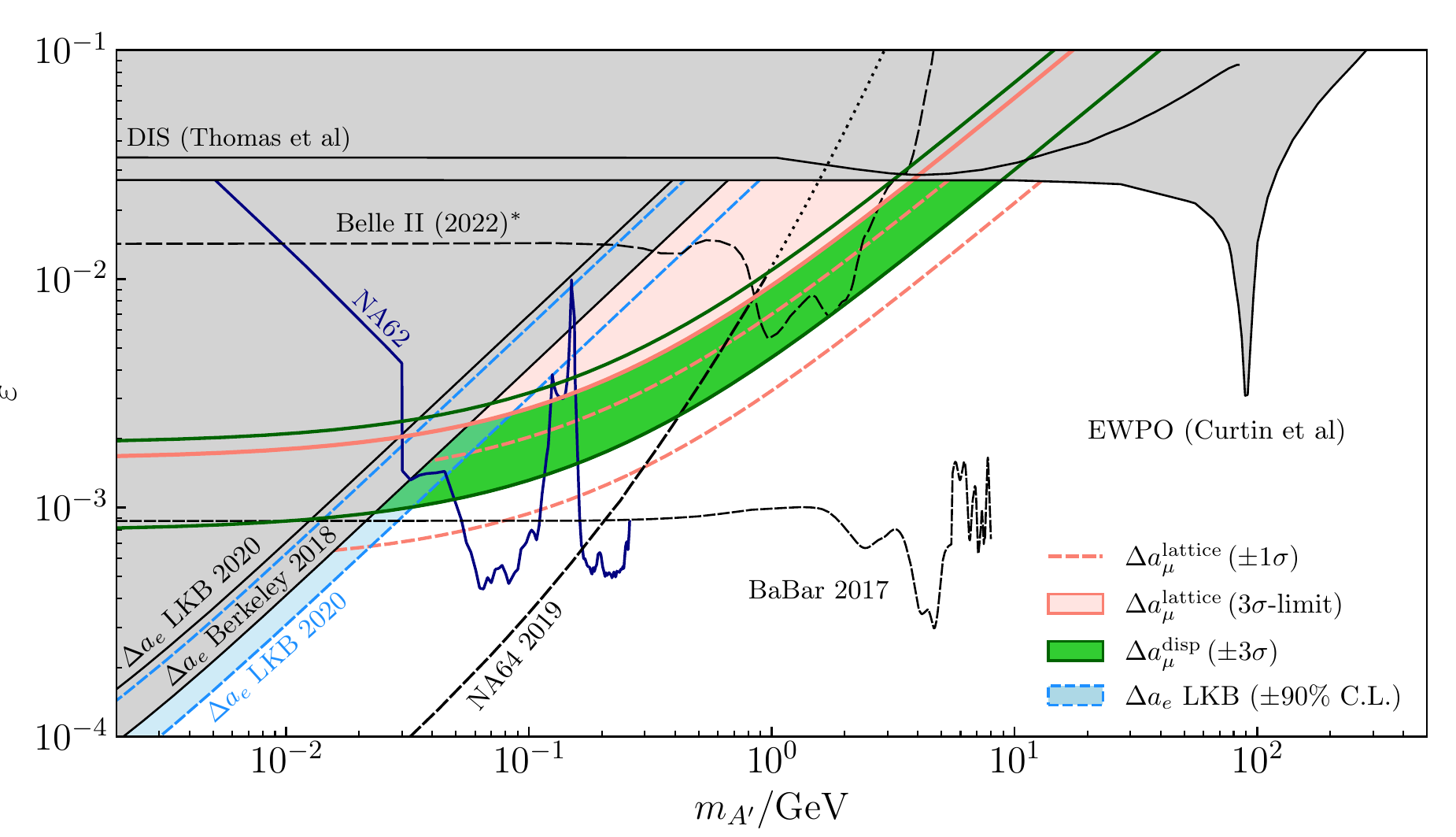}
    \caption{Model independent limits on kinetic mixing $\epsilon$ alongside limits on fully invisible dark photons.
    The limits shown as gray regions are independent of the decay channels of the dark photon.
    Navy colors indicate constraints from meson decays, $\pi^0\to \gamma A^\prime$ and $K^+\to \pi^+ A^\prime$.
    Both assume $A^\prime$ is invisible.
    In dashed curves, we show the limits from BaBar~\cite{BaBar:2017tiz}, NA64~\cite{NA64:2019imj}, and our recast of Belle-II~\cite{Belle-II:2022yaw}.
    To obtain the latter, we neglect the interference between initial and final state radiation of $A^\prime$ (see text).
    In green, we show the preferred region to explain $\deltaamu^{\rm disp}$, at $3\sigma$, and in solid orange, we show the constraints obtained from the lattice results for $\deltaamu^{\rm lattice}$.
    In dashed orange, we also show the region preferred by $\deltaamu^{\rm lattice}$ at $1\sigma$.
    \label{fig:model_independent}
    }
\end{figure*}

Constraints on kinetic mixing that are independent of the BRs of $A^\prime$ can be obtained from processes that are sensitive to the exchange of virtual dark photons.
Barring fine-tuning from other new-physics contributions to these observables, the derived limits on kinetic mixing can be regarded as model-independent.
We show these constraints in \cref{fig:model_independent}, comparing them with the limits on fully invisible dark photons, shown in thin purple and dotted black lines.

\myparagraph{Deep-inelastic scattering}
A dark photon contributes to the deep-inelastic-scattering (DIS) of charged leptons on nuclei via t-channel exchange, impacting the extracted values of Parton distribution functions (PDF)~\cite{Kribs:2020vyk,Thomas:2021lub,Thomas:2022qhj,McCullough:2022hzr}.
The authors of Ref.~\cite{Kribs:2020vyk} set a limit for the first time, fixing the PDF to the best-fit values of the HERA measurement, finding $\epsilon \lesssim 0.015$ for $m_{A^\prime} \lesssim 2$~GeV at $95\%$~C.L.
These limits are slightly relaxed when accounting for the effect of new physics on the extraction of PDFs, as discussed in Refs.~\cite{Carrazza:2019sec,Thomas:2021lub}.
In this work, we use the limits of Ref.~\cite{Thomas:2021lub}, where $\epsilon \lesssim 0.034$ for $m_{A^\prime}<1$~GeV at $95\%$~C.L.

\myparagraph{Electroweak precision observables} 
Among the Electroweak precision observables (EWPO) modified by kinetic mixing, the most important is $M_Z^2 \sim M_{Z^0}^2 - \epsilon^2 M_{A^\prime}$ and the corresponding shift in the mass of the $W$ boson. 
In Ref.~\cite{Curtin:2014cca}, the global fit to EWPO uses the W-mass measurements of LEP, finding $\epsilon^2_{\rm EWPO} < 7.3 \times 10^{-4}$ at $95\%$~C.L for $M_{A^\prime} \ll 10$~GeV. 
Since then, new $W$ mass measurements have been performed by ATLAS~\cite{ATLAS:2017rzl} and LHCb~\cite{LHCb:2021bjt}. 
In addition, a recent analysis by the CDF collaboration reported a significant deviation from the previous measurements~\cite{CDF:2022hxs}.
In view of these discrepancies with the SM, and the fact that light dark photons decrease $M_W$ in the EW fit, we proceed by showing the limits from Ref.~\cite{Curtin:2014cca} with the caveat that limits may turn to regions of preference, depending on future developments with the $W$-mass measurement.

\myparagraph{Meson decays}
Direct production of the dark photon in meson decays provides robust constraints on various dark photon models~\cite{Pospelov:2008zw}.
For invisible dark photons, there are searches for $\pi^0\to \gamma A^\prime$~\cite{NA62:2019meo} and $K\to \pi A^\prime$, with $A^\prime$ invisible~\cite{BNL-E949:2009dza}.
We update the latter using the latest NA62 measurement of $K\to \pi \nu\overline{\nu}$~\cite{NA62:2020xlg}, including also the dedicated search for $\pi^0\to{\rm inv}$ in Ref.~\cite{NA62:2020pwi}.
Since these limits assume the new vector to be invisible, they would be modified in the semi-visible models of interest, especially in hermetic detectors like NA62.
In the rest of the paper, we follow the aggressive strategy of showing these limits without modifications in all our plots.

\myparagraph{Electron $(g-2)$} 
Precision measurements of the electron anomalous magnetic moment provide model-independent limits on $\epsilon$ due to the exchange of virtual dark photons.
The dark photon contribution, in this case, comes with a negative sign and acts to decrease $a_e \equiv (g-2)_e/2$. 
The two most recent measurements of $a_e$ include the one in 2008~\cite{Hanneke:2008tm} and a recent update with 2.2 times more precision in 2022~\cite{Fan:2022eto}.
To use these results to constrain new physics, it is necessary to compare them with high-precision SM predictions~\cite{Aoyama:2019ryr}.
The predictions, however, are not robust due to the inconsistencies in the experimental determination of the fine structure constant, $\alpha$.
A group at Berkeley~\cite{Parker:2018vye} measures the fine structure constant using cesium-133 atoms to 120 parts per million.
Another technique employed by a group in Paris, referred here to as LKM, measures $\alpha$ to 81 parts per million~\cite{Morel:2020dww}.
These measurements are in disagreement at more than $5~\sigma$, indicating that more experimental progress is needed for a meaningful constraint to be derived.
Hereafter, we follow the conservative approach of quoting the most stringent limits, provided by Ref.~\cite{Parker:2018vye}, to set 
In \cref{fig:model_independent}, we show both limits, as well as the region of preference that would explain the measurement of Ref.~\cite{Morel:2020dww}.
In general, these constraints exclude the $\deltaamu$-explanation for dark photon masses below $m_{A^\prime} \sim 30$~MeV (see below).

\myparagraph{Muon $(g-2)$}
In the case of the muon anomalous magnetic moment, the theoretical and experimental uncertainties on $a_\mu \equiv (g-2)_\mu/2$ are much larger than the uncertainty in $\alpha$.
Therefore, it is not subject to the ambiguities discussed above and can still be more sensitive to new physics due to the $a_\mu/a_e \sim m_\mu^2/m_e^2$ enhancement.
The theoretical predictions for $a_\mu$ in the SM (see Refs.~\cite{Jegerlehner:2009ry,Miller:2012opa,Aoyama:2020ynm} for a review) have differed from the experimental measurement at E821 at the Brookhaven National Laboratory (BNL) with a significance of 3.7$\sigma$~\cite{Muong-2:2004fok}.
The E989 experiment at FNAL \cite{Grange:2015fou}, running since 2018, has now confirmed the central value measured at BNL, reporting
$a_\mu^{\rm FNAL} = 116\,592\,040\,(54) \times 10^{-11}$,
where the error in parenthesis is a sum in quadrature of systematical and statistical errors. 
This result combined with the BNL measurements,
    $a_\mu^{\rm BNL} =116\,592\,920\,(63) \times 10^{-11}$,
provides an experimental average
\begin{equation}
    a_\mu^{\rm comb} = 116\,592\,061(41) \times 10^{-11}.
\end{equation}
Eventually, with the five data-taking stages at FNAL completed, the FNAL measurement can achieve 20 times more statistics than the BNL experiment~\cite{Grange:2015fou} and improve the precision on $a_\mu$ by a factor of 4. 
Eventually, this value can also be tested by next-generation experiments, such as at the J-PARC muon facility~\cite{Abe:2019thb}, which plans to achieve a similar precision to the BNL measurement using a complementary technique with lower muon momenta, $p_\mu \sim 300$~MeV.

The SM predictions are obtained by combining QED corrections up to 5 loops~\cite{Aoyama:2012wk,Aoyama:2019ryr}, electroweak ($a_\mu^{\rm EW}$) corrections up to 2 loops~\cite{Czarnecki:2002nt,Gnendiger:2013pva}, and hadronic contributions including the hadronic vacuum polarization ($a_\mu^{\rm HVP}$)~
\cite{Kurz:2014wya,Keshavarzi:2019abf,Davier:2019can,Hoferichter:2019mqg,Colangelo:2018mtw,Keshavarzi:2018mgv} and hadronic light-by-light scattering ($a_\mu^{\rm HLbL}$) diagrams~
\cite{Melnikov:2003xd,Masjuan:2017tvw,Colangelo:2017fiz,Hoferichter:2018kwz,Gerardin:2019vio,Bijnens:2019ghy,Colangelo:2019uex,Pauk:2014rta,Danilkin:2016hnh,Jegerlehner:2017gek,Knecht:2018sci,Eichmann:2019bqf,Roig:2019reh,Blum:2019ugy,Colangelo:2014qya}, with $a_\mu^{\rm HVP}$ dominating the uncertainty in the overall prediction and $a_\mu^{\rm HLbL}$ well below the value needed to explain the discrepancy.

Using a data-driven dispersive calculation for the hadronic contributions $a_\mu^{\rm HVP}$ and $a_\mu^{\rm HLbL}$, the authors of Ref.~\cite{Aoyama:2020ynm} converge on the prediction
\begin{equation}\label{eq:SMprediction}
a_\mu^{\rm disp} = 116\,591\,810\,(43) \times 10^{-11}.
\end{equation}
If the dispersive value holds, the tension between experiment and theory reaches $4.2\sigma$, where $\deltaamu^{\rm disp} = a_\mu^{\rm disp} -a_\mu^{\rm comb}$ is
\begin{equation}\label{eq:discrepancy}
\deltaamu^{\rm disp} = 251(59) \times 10^{-11}.
\end{equation}
A strong ongoing effort aims at reducing the dominant source of uncertainties in these hadronic contributions, but so far, it has not been demonstrated that hadronic uncertainties alone can reconcile the discrepancy~\cite{Passera:2008jk,Crivellin:2020zul,deRafael:2020uif}. 
The agreement between EW precision measurements and the $e^+e^-\to$ hadrons data corroborates this hypothesis. 
A deviation in $\sigma(e^+e^-\to\text{hadrons}, s)$ to explain the observed value of $\deltaamu$ was shown to be ruled out for $e^+e^-$ center-of-mass energies of $\sqrt{s} > 0.7$~GeV~\cite{Keshavarzi:2020bfy}, implying that any missed contributions ought to be mostly concentrated in the $\pi^+\pi^-$ region~\cite{Colangelo:2020lcg}. 
It has also been suggested that new physics could be hiding in this data~\cite{Darme:2021huc,DiLuzio:2021uty,Crivellin:2022gfu,Darme:2022yal}.

The discrepancy of \cref{eq:discrepancy} is still not conclusive, however.
In particular, a calculation of $a_\mu^{\rm HVP}$ on the lattice by the BMW collaboration~\cite{Borsanyi:2020mff} finds a $2.1\sigma$ significant disagreement with the value obtained using the data-driven dispersive method of Ref.~\cite{Aoyama:2020ynm}.
Using the BMW result for $a_\mu^{\rm HVP}$ reduces the disagreement between theory and the experiment average, $a_\mu^{\rm comb}$, down to $1.5\sigma$,  
\begin{equation}
\deltaamu^{\rm lattice} = 107(69) \times 10^{-11}.
\end{equation}
This lattice result has been increasingly scrutinized in the search for additional systematic uncertainties that are specific to discretization and finite-volume effects of the lattice.
Consistency checks of the BMW results have been performed by other collaborations using ``Euclidean window observables", namely, observables calculated in Euclidean time windows that enhance or suppress specific systematic uncertainties~\cite{Ce:2022kxy}.
One of these observables isolates contributions to $a_\mu^{\rm HVP}$ into short, $a_\mu^{\rm SD}$, and intermediate, $a_\mu^W$, time-distance pieces.
The data-driven dispersive method~\cite{Colangelo:2022vok} and lattice calculations of $a_\mu^{\rm SD}$ are in good agreement.
However, a $3.8\sigma$ significant tension exists between the intermediate time-distance observable, $a_\mu^{\rm W}$, and all lattice results~\cite{Borsanyi:2020mff,Ce:2022kxy,Alexandrou:2022amy,Blum:2023qou,Alexandrou:2022kbk}, suggesting the disagreement with $e^+e^- \to$~Hadrons data is, in fact, largest in the energy region of $\sqrt{s} \sim 1-3$~GeV~\cite{Colangelo:2022vok}.
While the nature of this discrepancy is not identified, in this study, we proceed to entertain BSM explanations to both $\deltaamu^{\rm disp}$ and $\deltaamu^{\rm lattice}$.
The $3\sigma$ preference region for $\deltaamu^{\rm disp}$ is shown as a green band in \cref{fig:model_independent}, alongside the $1\sigma$ preference band and the $3\sigma$ exclusion limit from $\deltaamu^{\rm lattice}$.
We summarize other new-physics explanations to $\deltaamu^{\rm disp}$ in \cref{app:amu_explanations}.

\myparagraph{Belle II} 
As of now, no dedicated search for invisible dark photons has been released by the Belle II  $e^+e^-$ collider.
Nevertheless, the collaboration has performed a search for $L_\mu - L_\tau$ gauge bosons, focusing on final state radiation (FSR) of dark photons, $e^+e^- \to \mu^+\mu^- Z^\prime_{\mu-\tau}$.
The search is based on the missing energy carried by the gauge boson.
A similar signature can take place for dark photons, with the difference that $A^\prime$ can be emitted as either initial state radiation (ISR) or FSR.
In the analogous QED processes, $e^+e^- \to \mu^+\mu^- \gamma$, the interference between ISR and FSR can lead to significant charge asymmetries~\cite{BaBar:2015onb}, so the differential event rate will be different for a dark photon.
This interference in QED vanishes, however, if integrated over the total phase space.
While a dedicated study of the efficiencies is needed to recast the Belle-II limit on kinetic mixing, we provide a simple estimate by neglecting the interference term and simply adding the ISR and FSR pieces together.
The estimated limit is shown in \cref{fig:model_independent} with an asterisk to emphasize the approximation in the recast method.
The semi-visible decay of the dark photon can also lead to additional energy in the final state, and relax this constraint.
Therefore, we show it as a dashed line in \cref{fig:model_independent}.
Because of the similar geometries of BaBar and Belle II, we do not include this limit in our recast analyses: a strong relaxation of BaBar would lead to a similar effect in Belle II, which is, in addition, a more hermetic detector.


\section{Reinterpreting constraints on invisible \texorpdfstring{$A^\prime$}{A'}}\label{sec:recast}

Searches for invisibly decaying dark photons at $e^+e^-$ colliders and fixed-target experiments provide the strongest constraints on models of semi-visible dark photons. 
At collider experiments, the dark photon is produced directly alongside initial state radiation (ISR) in the $e^+e^- \to \gamma A^\prime$ process. 
Prompt decays of $A^\prime$ to a pair of HNFs, $A^\prime \to \psi_i \psi_j$, in which the HNFs are:
\begin{enumerate}
    \item long-lived and decay outside the detector, or
    \item short-lived and decay inside to $\psi_i \to \psi_j \, \ell^+ \ell^-$, with final state leptons pairs whose energies fall below detector thresholds,
\end{enumerate} would appear identical to a monophoton signature accompanied by missing $\slashed{E}_T$. The strongest bound of this kind is obtained by the BaBar experiment, excluding explanations of $(g-2)_\mu$ with invisibly decaying dark photons~\cite{BaBar:2017tiz}.
Dark photons can also be produced in bremsstrahlung at fixed-target experiments such as NA64.~\cite{NA64:2017vtt,NA64:2019imj,NA64:2021acr}. 
In this type of experiment, the dark photon signature constitutes a large amount of missing energy in the primary electron beam that scatters on the target.
Below we discuss the reinterpretation of these two leading constraints on the parameter space of the models discussed in \cref{sec:darkphoton}.
In the following, the vector $\bm{x}$ represents the set of free model parameters, $\bm{x} = (\epsilon,\, m_{A^\prime},\, \Delta_{21},\, \dots)$, that have been varied in the analysis performed.

\subsection{BaBar monophoton search} 
\begin{figure*}[t]
\centering
    \includegraphics[width=\textwidth]{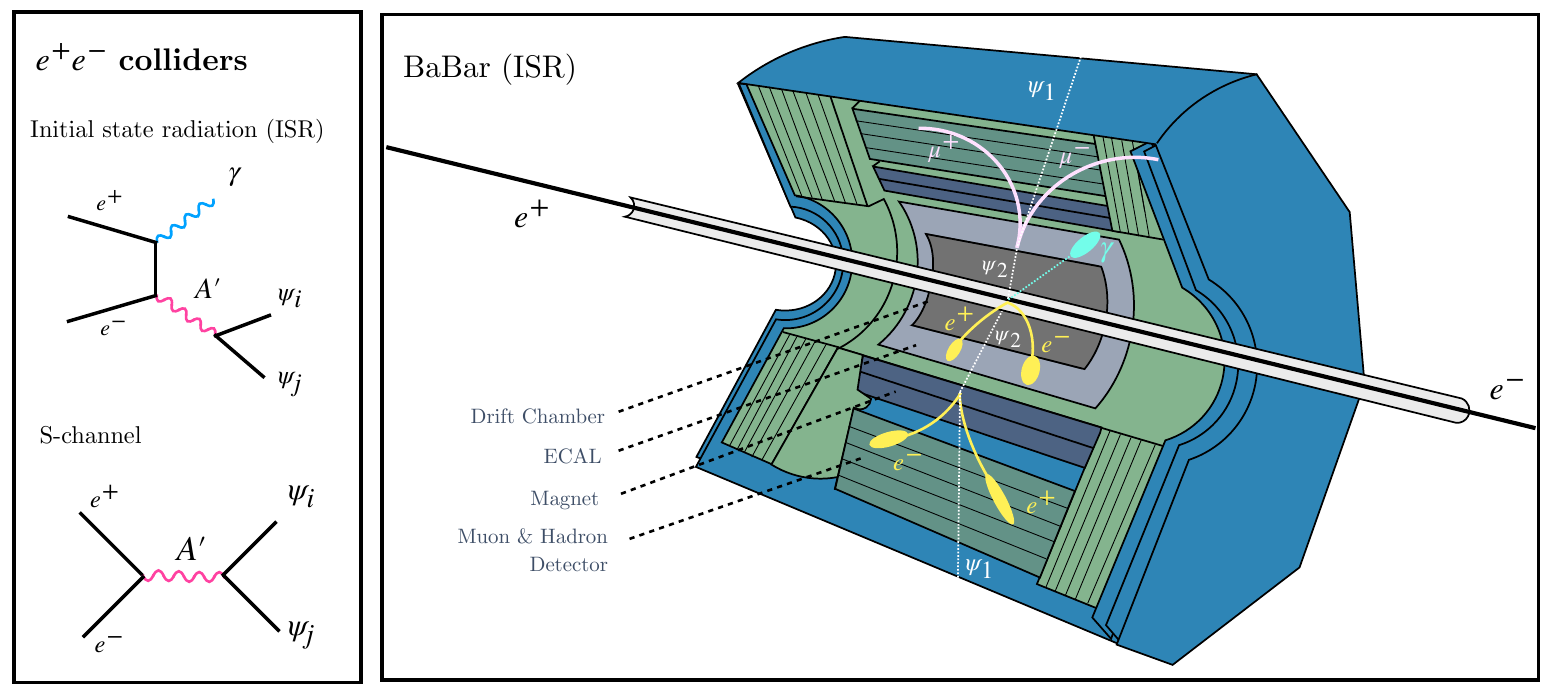}
    \caption{
    The signatures of semi-visible dark photons at $e^+e^-$ colliders. 
    On the right, an inner view of the BaBar detector with the displaced, semi-visible decay of the HNFs into charged lepton pairs.
    In this example, the decay cascade is \mbox{$A^\prime \to (\psi_2\to \psi_1 \mu^+\mu^-) (\psi_3\to \psi_2 e^+e^- \to \psi_1 e^+e^-e^+e^-)$}, where $\psi_3$ decayed promptly.
    }
    \label{fig:BaBardiagram}
\end{figure*}

Based on the PEP-II asymmetric $e^+e^-$ collider at the SLAC National Accelerator Laboratory, the BaBar experiment searched for single photons (monophotons) accompanied by missing energy and momentum in the process $e^+e^- \to \gamma A^\prime$. 
The search was conducted in the $53$~fb$^{-1}$ dataset collected between 2007-2008 at the center-of-mass (COM) energies $\Upsilon(2S), \Upsilon(3S)$ and $\Upsilon(4S)$. 
The components of the BaBar detector relevant for our analysis are a charged-particle-tracking system provided by a five-layer, double-sided silicon vertex tracker (SVT) and $40$-layer drift chamber (DCH); an electromagnetic calorimeter (EMC) of $6580$ CsI(Tl) crystals. 
These systems are all contained within a $1.5$~T superconducting solenoid magnet. 
Beyond the superconducting coil is located an instrumented flux return (IFR) barrel that provides muon and neutral hadron identification.
An illustration of the detector and the HNF signatures is shown in \cref{fig:BaBardiagram}.

\myparagraph{Event generation}
Using our own MC event generator, we simulate the production of dark photons at BaBar in the process $e^+e^- \to \gamma A^\prime$ at a COM energy $\sqrt{s} \approx 10$~GeV. 
We boost and rotate to the laboratory frame, considering that the $e^+e^-$ collision frame is itself already boosted with respect to the laboratory by $\beta_z \approx 0.5$. In the laboratory frame, we take the z-direction to be aligned with the direction of $e^+e^-$ collision. 
The experiment employs a primary selection cut on the photon COM angle to suppress SM backgrounds, $|\cos\theta_\gamma^*| < 0.6 $. 
To a good approximation, the $A^\prime$ decay to HNF pairs is prompt upon production. 
However, the HNF can travel before decaying, modeled by random sampling an exponential distribution according to its decay width. 
We simulate the decays $A^\prime \to e^+e^-, \mu^+\mu^-$, and $\pi^+\pi^-$ according to their differential decay rates, taking into account the differences in the decay kinematics of Majorana and Dirac particles.

As the original analysis searched for single photons and vetoed additional activity in the detector above a certain energy, we introduced a set of veto criteria to dispense those events that would not have passed the event selection.
We show kinematical distributions of $e^+$ and $e^-$ decay products in \cref{app:distributions}, also showing the impact of the analysis selection criteria discussed next.

\begin{figure*}[th]
\centering
\includegraphics[width=0.49\textwidth]{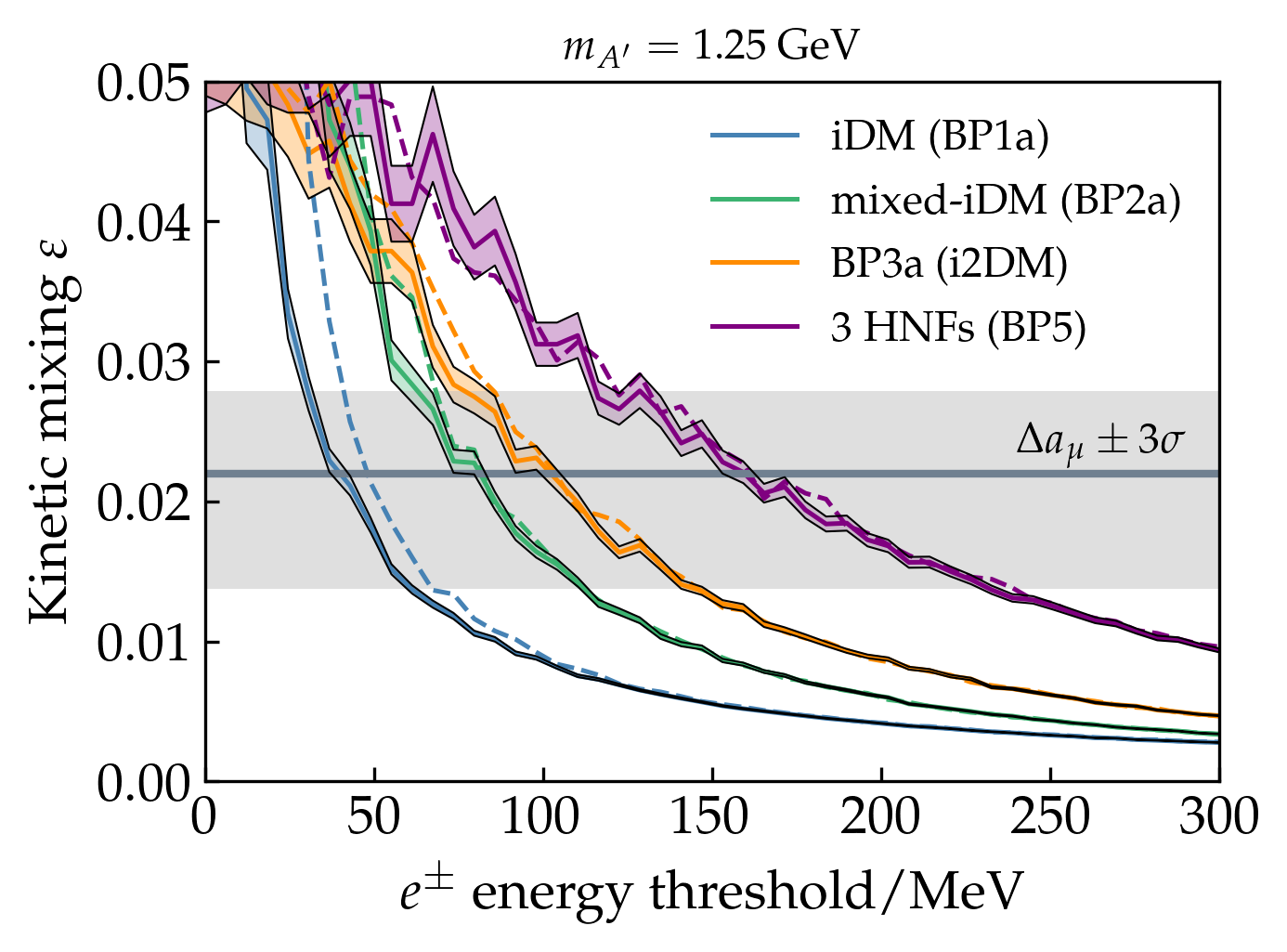}
\includegraphics[width=0.49\textwidth]{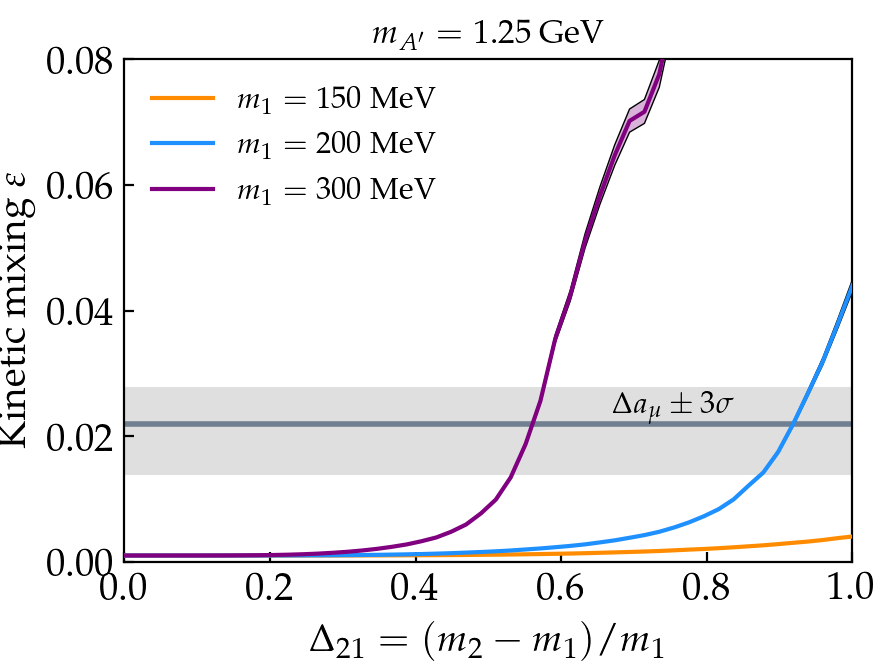}
    \caption{
    The BaBar limit on the kinetic mixing parameter, $\epsilon$. 
    On the left panel, we show the limit as a function of our analysis's individual $e^\pm$ energy threshold. 
    In solid (dashed) lines, we use an analysis with (without) a cut on the angle between the leptons and the pipeline.
    On the right, we show the limit as a function of the mass splitting for BP1b. 
    \label{fig:thresholds}
    }
\end{figure*}

\myparagraph{Monophoton selection}
An $e^+e^- \to \gamma A^\prime$ event passing the initial monophoton selection is vetoed if, anywhere along the decay chain, a charged particle is produced in an instrumented region of the detector, i.e., within the SVT, DCH, EMC, or IFR regions, and satisfies both of the following conditions:
\begin{enumerate}
    \item For $e^+e^-$ pairs with angular separation $\theta_\text{sep.} > 10^{\circ}$, the energy of each electron exceeds BaBar's energy threshold to resolve charged particle tracks, $E_{\pm} > 100$~MeV. For overlapping $e^+e^-$ pairs with $\theta_\text{sep.} < 10^{\circ}$, we require $(E_{+} + E_{-}) > 100$~MeV.
    \item The polar angles, $\theta_{\text{pol.}}$, of the electrons individually, or as a pair, are sufficiently wide that the electrons do not escape along the beam pipeline, $ 17^{\circ} < \theta_{\text{pol.}} < 142^{\circ}$.
\end{enumerate}
The criteria above amount to a statement that all HNF decays that occur inside the detector and produce charged lepton final states that leave visible tracks are vetoed.

The threshold is assumed to be a step function with $100\%$ detection efficiency above $E_{\rm threshold}$ and $0\%$ below it.
Realistically, the final state leptons can escape detection even if their energy is large, as leptons can escape between the active materials in the detector.
This effect requires a more detailed description of the geometry and particle propagation model, which is beyond the capabilities of our simulation.
Nevertheless, we expect this will not significantly change our conclusions, as the leptons are always produced in pairs and follow bent trajectories due to the magnetic field, especially at low energies.

We show the impact of varying energy thresholds used in our analysis on the left panel of \cref{fig:thresholds}.
We do this for a few benchmark points, demonstrating the dependence on the threshold assumptions and confirming that the dominant source of invisible events at BaBar comes from soft leptons.
The effect of omitting the pipeline is small, as seen in the comparison between the solid lines (considering the pipeline effect) and dashed lines (neglecting it).
In addition, in varying the mass splitting, we vary the total energy emitted in SM particles, observing a strong effect on the relaxation of constraints for larger $\Delta_{21}$.
The band in each curve represents the uncertainty associated with Monte-Carlo statistics.

\myparagraph{Recasting the bound}
To derive their bound, BaBar assumed an invisibly decaying dark photon 
\begin{equation}\label{eq:isr_xsecn_bbr1}
   \sigma^{\text{BaBar}}_{e^+e^- \to \gamma + \text{inv.}}\left(\bm{x}\right) = 
    \sigma_{e^+e^- \to \gamma A^\prime}\left(\bm{x}\right) \times \mathcal{B}\left(A^\prime \to \text{inv.} \right),
\end{equation}
with $\mathcal{B}\left(A^\prime \to \text{inv.}\right) = 1$. To re-interpret the bound for a semi-visible dark photon, we introduce a factor $\text{P}^{\text{inv.}}$ that accounts for the probability that decays of semi-visible dark photons produced alongside ISR appear invisible and contribute to the monophoton dataset:
\begin{equation}\label{eq:isr_xsecn_bbr2}
    \sigma^{\text{BaBar}}_{e^+e^- \to \gamma + \text{inv.}}\left(\bm{x}\right) = \sigma_{e^+e^- \to \gamma X}\left(\bm{x}\right)\times \text{P}^{\text{inv.}}\left(\bm{x}\right),
\end{equation}
where $X$ is the semi-visible dark photon. From \cref{eq:isr_xsecn_bbr1} and \cref{eq:isr_xsecn_bbr2}, we may obtain the relation
\begin{equation}\label{eq:recast_eqn}
    \epsilon^{\text{BaBar}} =  \epsilon \, g_D \sqrt{\text{P}^{\text{inv.}}\left(\bm{x}\right)},
\end{equation}
where $\epsilon^{\text{BaBar}}$ is the bound on the kinetic mixing parameter obtained by BaBar. We may define the function $\text{P}^{\text{inv.}}$ as follows
\begin{equation}
     \text{P}^{\text{inv.}}\left(\bm{x}\right) = 1 - \text{P}^{\text{veto}}\left(\bm{x}\right) = 1 - \frac{N^{\text{veto}}\left(\bm{x}\right)}{N\left(\bm{x}\right)},
\end{equation}
with $N$ being the total number of events that pass the initial monophoton selection, and $N^{\text{veto}}$ the subset of events that are vetoed according to the criteria set out above. To recast BaBar's monophoton limit, we solve Eq.~\ref{eq:recast_eqn} for $\epsilon$ at each value of $\left(\bm{x}\right)$. The function $\text{P}^{\text{inv.}}$ contains all model dependencies, including the HNF masses and any mixing parameters. The results of our recast are given in \cref{sec:results} for all benchmark points of interest.

\myparagraph{Pseudo-monophotons}
It was noticed that BaBar has a mild excess of mono-photon events~\cite{Abdullahi:2020nyr}. 
If one of the two HNFs decays with a lifetime of a few cm, a significant fraction of them will decay within the EMC of BaBar, mimicking a monophoton signature.
These events have a relatively broad spectrum in missing energy $M_\textrm{miss}^2 \equiv s - 2 E_\gamma \sqrt{s} $ where a mild excess has been observed in the region $24~\mathrm{GeV}^2 < M_\textup{miss}^2 < 50~\mathrm{GeV}^2$.
This explanation can fit the events well, explaining the $\sim 2.5\sigma$ excess~\cite{Abdullahi:2020nyr}.
 
\subsection{NA64 dark photon searches}
\begin{figure*}[t]
    \includegraphics[width=\textwidth]{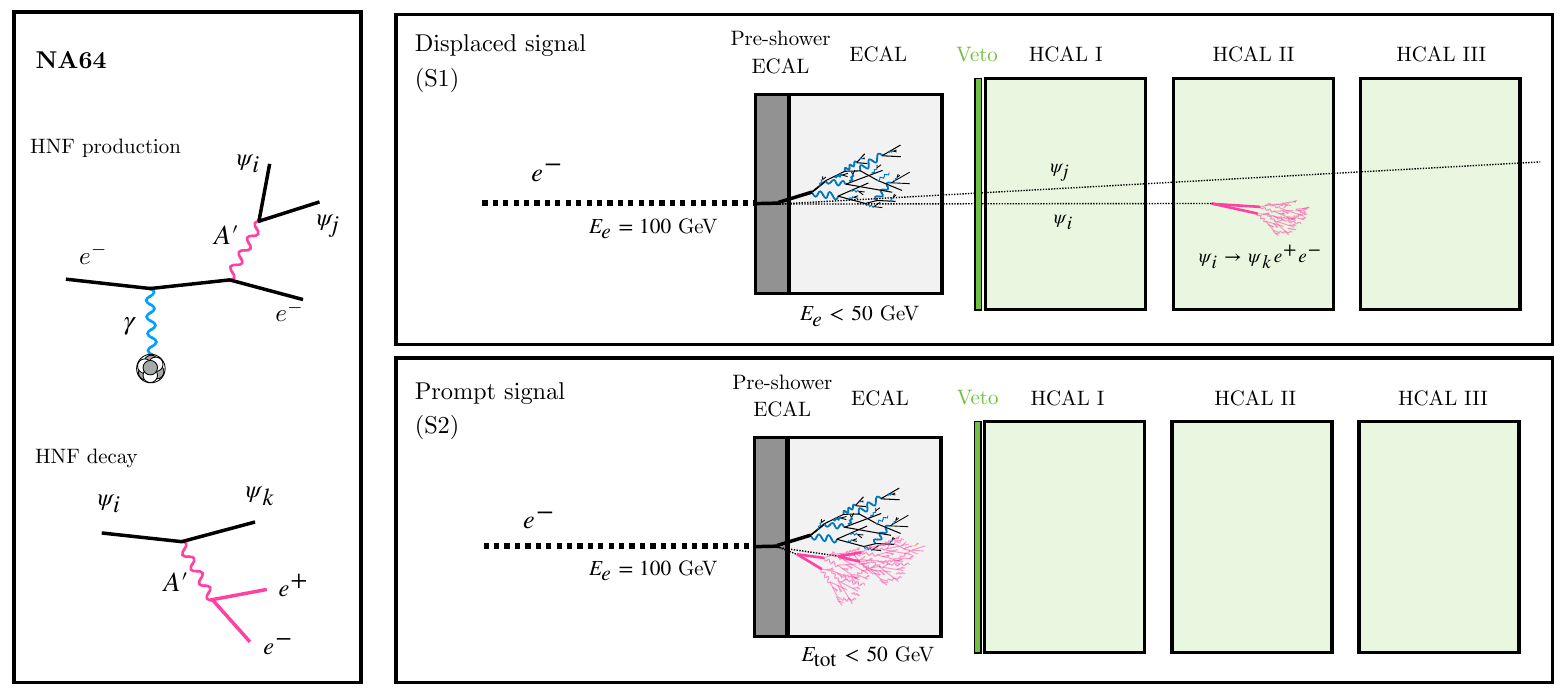}
    \caption{The NA64 setup and signatures considered in this work. Left panel: production and decay processes of the heavy neutral fermions (HNF). Top right panel: the first kind of signature with a displaced vertex already considered in~\cite{NA64:2021acr}. 
    Bottom right panel: the prompt decay signature that NA64 can use to constrain regions of parameter space where the HNFs are promptly decaying.
    \label{fig:NA64diagram}}
\end{figure*}

The fixed-target experiment NA64 searches for dark sector particles at the CERN SPS, employing a $100~\mathrm{GeV}$ electron beam.
The main search strategy relies on the fact that invisible dark photons can carry away a large amount of missing energy in hard electron-nucleus bremsstrahlung events~\cite{Gninenko:2013rka,Izaguirre:2014bca,NA64:2016oww,NA64:2017vtt,NA64:2019imj},
\begin{equation}
    e^- \mathcal{N} \to e^- \mathcal{N} A^\prime, \, A^\prime \to~\mathrm{invisible} \, ,
\end{equation}
where $\mathcal{N}$ is the target nucleus in the fixed target.
Like monophoton searches, the bremsstrahlung signal is proportional to $\epsilon^2$. 

The main parts of the detector relevant to our analysis are:
\begin{itemize}
    \item the electromagnetic calorimeter (ECAL), which also works as the active beam dump, made of Pb+Sc layers, with an average photon conversion length of $X_0 = 1.175~\mathrm{cm}$;
    \item a large high-efficiency veto counter downstream of the ECAL;
    \item a hadronic calorimeter (HCAL), made of three different modules.
\end{itemize}

The search for invisible decays of $A^\prime$ \cite{NA64:2019imj} was conducted on the total sample of electrons on target (EOT) collected during the period 2016--2018, $n_\textup{EOT} = 2.84 \times 10^{11}$.

NA64 also searched for semi-visible dark photon decays assuming the iDM model~\cite{NA64:2021acr}.
In this search, they considered the same data collected in the period 2016--2018, performing a recast-based analysis resembling that of their search for axion-like particles~\cite{NA64:2020qwq}, targeting visible final states coming from the dark photon decay chain and putting a model-dependent constraint on the kinetic mixing parameter.

\myparagraph{Event generation}
We simulated the production of dark photons and the detection of $e^+ e^-$ pairs in the NA64 detector with a fast MC generator.
From a comparison of the sensitivity curves with the limits obtained in Ref.~\cite{na64semivisible}, we find that our projections are comparable to the limits obtained using a proper GEANT4 detector simulation, with minor discrepancies that can be attributed to the complete description of the detector geometry, shower development, and energy collection within the different detectors.
The complete GEANT4 detector simulation, along with a discussion of the latest NA64 constraints on several semi-visible dark photon models, can be found in the accompanying~\cite{na64semivisible}.

We simulate bremsstrahlung events, producing an $A^\prime$ with an electron beam at energy $E_\textup{beam} = 100~\mathrm{GeV}$.
We consider the electron beam energy to be unaffected by any energy losses happening when entering the ECAL, so that $E_\textup{beam}$ can be considered constant.
The beam is considered to impact the ECAL at coordinate $x=y=z=0$.
The energy of the $A^\prime$ is distributed according to the following formula, obtained by applying the improved Weizsaker–Williams (IWW) approximation~\cite{Gninenko:2017yus}:
\begin{equation}
    \frac{\mathrm{d}\sigma}{\mathrm{d}x} \propto \biggl(m_{A^\prime}^2 \frac{1-x}{x} + m_{e}^2 x\biggr)^{-1} \biggl(1 - x + \frac{x^2}{3}\biggr) \,,
\end{equation}
where $x = E_{A^\prime} / E_\textup{beam}$, and the $A^\prime$ is in the $z$ direction.
After radiating an $A^\prime$, the beam electron (or main electron) will have energy $E_{e} = E_\textup{beam} - E_{A^\prime}$, and will shower inside the ECAL completely.
The $A^\prime$ decays promptly into a pair of HNFs, which are boosted and rotated to the lab frame according to the $A^\prime$ energy.
Each HNF will then decay according to its decay modes.
The simulation automatically handles the decay of the secondary HNFs in the same way.
The lightest HNF from each model is considered stable with respect to the size of the detector.
We assume a simplified shower development for the $e^+ e^-$ pairs produced inside the ECAL.
The energy loss can be computed assuming an exponential law:
\begin{equation}
\begin{split}
    \frac{\mathrm{d} E}{\mathrm{d} z} = -\frac{E}{X_0} \Rightarrow \Delta E(z) &= E(z) - E_0 \\
    {} &= E_0 \biggl[ 1 - \exp\biggl(- \frac{z - z_0}{X_0}\biggr) \biggr] \, ,
\end{split}
\end{equation}
where $z_0$ is the production point of the pair.
For pairs detected inside the HCAL, we assumed each pair is able to shower completely inside it.
Given the high energies of the $A^\prime$, the $e^+ e^-$ pairs are highly boosted and collimated.
We may then treat each pair as a single particle with energy $E(z_0) = E_{e^+}(z_0) + E_{e^-}(z_0)$.
The kinematical considerations on the $e^+ e^-$ pairs are corroborated by the distributions shown in \cref{app:distributions}.

Each event is made of invisible final states (the stable HNFs) and visible energy (the $e^+ e^-$ pairs):
\begin{gather*}
    e^- \mathcal{N} \to e^- \mathcal{N} A^\prime\\
    A^\prime \to (\psi_i \to (\psi_k \to \dots) \,e^+ e^-)(\psi_j \to (\psi_l \to \dots) \, e^+ e^- )
\end{gather*}
We check the visible energy collected in each event against the veto criteria applied to the different regions of the detector to see if the event is recorded as signal for the experiment.

\myparagraph{Semi-visible selection (S1)}
\label{sec:na64_semi_visible}
This signature corresponds to the one employed by NA64 to study semi-visible dark photon decays in the framework of the iDM model~\cite{NA64:2021acr}.
In this work, we recast their limits to the more complex models presented in \cref{sec:darkphoton}.

The following selection criteria were applied, according to the expected signal yield coming from this model, relying on the decay of the heaviest HNF $\psi_i = \psi_2$ or $\psi_3$ into a single $e^+ e^-$ pair:
\begin{enumerate}
    \item $\psi_{i}$ is expected to decay inside the HCAL: in particular, the analysis targets a fiducial HCAL volume composed by the last two modules of the HCAL detector in which a consistent amount of energy coming from a single
        $e^+e^-$ pair is expected to be detected.
    \item events with any other activity happening before the fiducial HCAL volume are vetoed;
    \item events with more than one visible decay vertex in the fiducial HCAL volume are vetoed.
\end{enumerate}

Additionally, in case $\psi_i$ decays beyond the HCAL, no other significant energy deposits are expected in the full detector, and the event resembles an invisible dark photon event.

This proposed analysis is tailored to the iDM model, focusing in particular in the parameter space where the $\psi_2$ lifetime is comparable to the size of the detector.
It rapidly loses sensitivity in the high mass region where $\psi_2$ decays promptly.
Other models involving large mass splittings between HNFs are also characterized by a shorter lifetime, due to $c\tau \propto \Delta^{-5}$, so this search will also be less sensitive to the high mass region.

\myparagraph{Invisible selection (S2)}
\label{sec:na64_invisible}
The analysis focused on detecting invisible decays of $A^\prime$ by constraining the amount of visible energy collected.
The set of energy cuts relies on the search strategy applied by the NA64 experiment to detect dark photon events containing missing energy, while limiting the possible background~\cite{NA64:2019imj}:
\begin{enumerate}
    \item
    {\bf pre-shower ECAL.} The total energy collected in the first layers of the ECAL should be compatible with the deposit expected for a primary electron.
    \item
    {\bf ECAL.} The total energy collected, $E_\textup{ECAL}$, including both the main electron and the $e^+ e^-$ pairs, is compared against the missing energy threshold chosen by the experiment:
    \begin{equation}
        E_\textup{ECAL} > 50~\mathrm{GeV} \, .
    \end{equation}
    \item
    {\bf Veto counter.} We do not expect any activity in the veto, in order to record an invisible event.
    In the case of semi-visible dark photon models, a particle can reach the veto in two cases:
    \begin{itemize}
        \item \label{fg-2:na64:veto_position_1}if they are produced inside the ECAL, they can shower until its end, releasing the remaining energy in the veto;
        \item \label{fg-2:na64:veto_position_2}if they are produced between the end of the ECAL and the veto, they will release their energy inside the veto.
    \end{itemize}
    We further assume that if a pair happens to reach the veto, it is able to release all of its energy inside it.
    This can be justified by the fact that the imposed veto threshold is sufficiently low that even the softest $e^+ e^-$ pairs we expect to produce would be able to trigger an event.
    Moreover, the thickness of the veto counter is large enough to guarantee a consistent energy deposit by the $e^+ e^-$ pairs.
    \item {\bf HCAL}: For an invisible event, no activity is expected inside the HCAL.
    Particles created between the veto and the HCAL or in the empty space between the three HCAL modules will be intercepted by the HCAL, eventually.
    The HCAL detector of NA64 is sufficiently long so that we can assume that any pair created inside it has enough space to shower completely, depositing its entire energy inside the HCAL.
    This approximation may not apply only to the small number of particles created at the very end of the HCAL.
\end{enumerate}
Using our simulation, we present a few distributions of the kinematics of the $e^+e^-$ pairs in \cref{app:distributions}.
It should be noted that in performing the recast of the invisible selection, we have assumed that NA64 is not able to distinguish the $e^+e^-$ showers due to short-lived HNF decays in the ECAL from the main electron beam.
We assume that the total energy deposition in the ECAL for these cases is the aggregate of the $e^+e^-$ energy and beam electron energy. 
Given the NA64 sensitivity to these prompt decays has not been previously studied, we take this recast constraint as a projection of NA64's potential sensitivity to decays of this kind.
The latest constraints from NA64, Ref.~\cite{na64semivisible}, tackle this region and show good agreement with our projections.

\myparagraph{Recasting the bound}
The derivation of the recast bound is done in a similar fashion to the BaBar simulation.
We start by considering the bound obtained by the invisible dark photon search performed in \cite{NA64:2019imj}, which we call $\varepsilon^\textup{NA64}$.
In addition, we extrapolated the bound above $1$~GeV, through a 2-degree polynomial fit.
Our toy Monte-Carlo yields the probability of obtaining an invisible event assuming the semi-visible dark photon decay $P^\textup{inv}$.
The recast bound can be found by solving the following equation, which matches \cref{eq:recast_eqn} discussed for the BaBar recast, with $\bm{x}$ being the set of model parameters:
\begin{equation}
    \epsilon^\textup{NA64} =  \epsilon \, g_D \sqrt{\text{P}^{\text{inv.}}\left(\bm{x}\right)}.
\end{equation}

\subsection{Other limits from beam dumps}

We also recast beam dump limits on displaced decays of dark sector particles.
In particular, the NuCal and E137 limits, originally recasted on the parameter space of iDM models~\cite{Izaguirre:2017bqb,Mohlabeng:2019vrz,Tsai:2019buq}.
We do not simulate the experimental setup but rescale the upper and lower bounds under a few approximations.
For the upper bound (small coupling regime), the rescaling is done by matching our models' production and detection rates with the iDM model.
In this region, the decay-in-flight rate is proportional to the number of particles produced times the decay rate into the respective signal channels, in our case, $\psi_i \to \psi_j \ell^+\ell^-$.
In some cases, more production channels are available than in the original study, such as when recasting iDM bounds onto the parameter space of mixed-iDM and i2DM, where $\psi_2$ can be produced in pairs in meson decays or bremsstrahlung.
We multiply the signal rate by the respective $A^\prime$ vertices and the corresponding long-lived particle decay rate to account for that.
For instance, to rescale a limit on $\epsilon$ from an iDM model to a mixed-iDM or i2DM model, we take
\begin{equation}\label{eq:recastBD}
    \frac{\varepsilon^4_{\rm iDM }}{\varepsilon^4_{\rm new}} =  \frac{\alpha_D^{\rm new}}{\alpha_D^{\rm iDM}} \sum_{i \geq j}|V_{ij}|^2  \frac{\Gamma_{\psi_i \to \psi_{i-1}e^+e^-}^{\rm new} + \Gamma_{\psi_j \to \psi_{j-1} e^+e^-}^{\rm new}}{\Gamma_{\psi_2 \to \psi_1 e^+e^-}^{\rm iDM}},
\end{equation}
where $\hat{\Gamma}$ are the decay rates deprived of all coupling pre-factors.
The $\alpha_D |V_{ij}|^2$ pre-factor takes care of rescaling the production of different dark states at the source or target, and the decay rate ratios account for the different mass splitting and mediator masses.
Note that we neglect decay cascades and that no cases require taking $\psi_i \to \psi_{i - 2} e^+e^-$ decays into account.
We also neglect the effect of the mass splitting on the kinematics, which can lead to a different acceptance for the new models.
For the lower-bound (upper line), we match the total lifetimes of the lightest unstable particles of the original bound to the one in our models.
That is, the new limit is found by setting $\Gamma_{\psi_2}^{\rm new} = \Gamma_{\psi_2}^{\rm iDM}$.
This is also an approximate procedure but relies on the fact that in this lower-limit regime, the number of particles produced is large, but the probability of the unstable state decaying within the detector is going to zero exponentially.
Our procedure also does not increase the mass reach of the constraints and therefore provides only an estimate of the excluded parameter space.
For E137 (scatter), the recast is similar to \cref{eq:recastBD}, but we rescale the scattering rate of dark particles instead, neglecting differences in mass splitting.

\section{Results}\label{sec:results}

We have studied a series of constraints on the theoretical models presented in Sec.~\ref{sec:darkphoton}, recasting the limits from NA64 and BaBar. 
As the parameter space is very large, we have fixed representative values for the gauge coupling $g_D$ and the masses of the HNFs,
parameterized in terms of $m_1 / m_{A^\prime}$ (known as $r$ in the literature) and $\Delta_{ij} \equiv (m_i - m_j)/m_j$. 
We have taken $\alpha_D \equiv g_D^2/(4 \pi)$ to be in the range $0.1-0.5$ as we are interested in the fast decays of the HNFs while maintaining perturbativity.
Except for the 3 HNF models, we fix the mass ratio $r = m_1 / m_{A^\prime}$ to be $1/3$ for most of the benchmarks, a standard value in the literature.
The values $\Delta_{ij}$ have been set to minimize the lifetime of the HNFs and maximize the amount of visible energy deposited by their decay products in the BaBar and NA64 detectors.
The constraints from NA64 are labeled according to the dark photon signature.
\begin{itemize}
    \item \emph{NA64 (S1)} (solid line) --- described in \cite{NA64:2021acr}, a model-dependent search for iDM was performed for the semi-visible dark photon signature of the model.
    \item \emph{NA64 proj. (S2)} (dash-dotted line) --- described in \cite{NA64:2019imj}, this constraint is on the invisible dark photon. Our recast expresses the potential future sensitivity of the experiment towards a dedicated semi-visible dark photon search, showing the capability to constrain the parameter space in a model-independent way.
\end{itemize}

In addition, for benchmarks BP1, BP2, and BP3, the lightest HNF can be a dark matter candidate. In this case, $\Delta_{ij}$ cannot be too large, to minimize the Boltzmann suppression in coannihilations.

We show the model-dependent bounds from NuCal~\cite{Blumlein:2011mv,Blumlein:2013cua,Tsai:2019buq}, E137 (scatter)~\cite{Izaguirre:2017bqb,Mohlabeng:2019vrz,Tsai:2019buq}, and the model-independent limits discussed in \cref{sec:darkphoton}.

\begin{figure*}[t]
    \centering
    \includegraphics[width=\columnwidth]{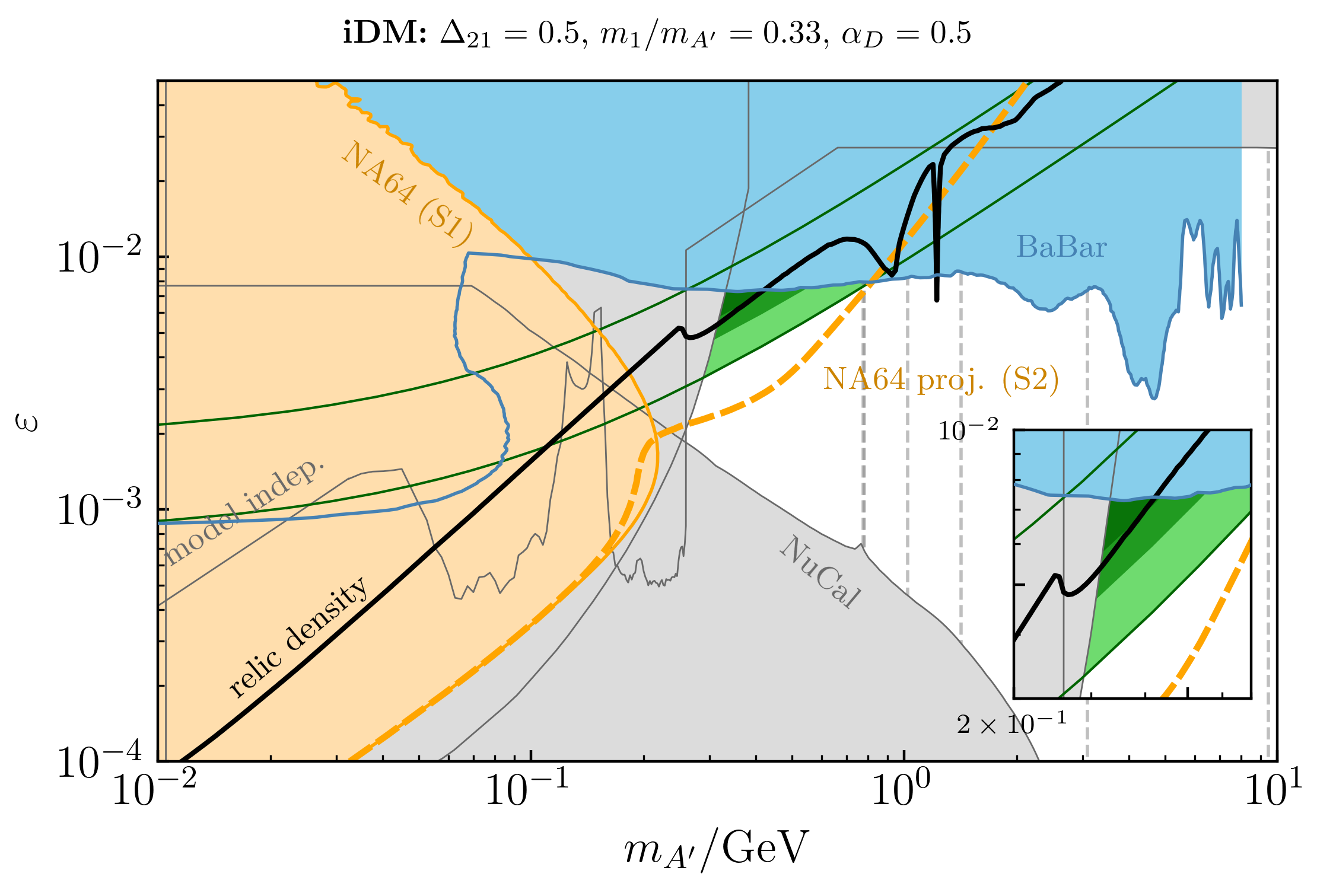}
    \includegraphics[width=\columnwidth]{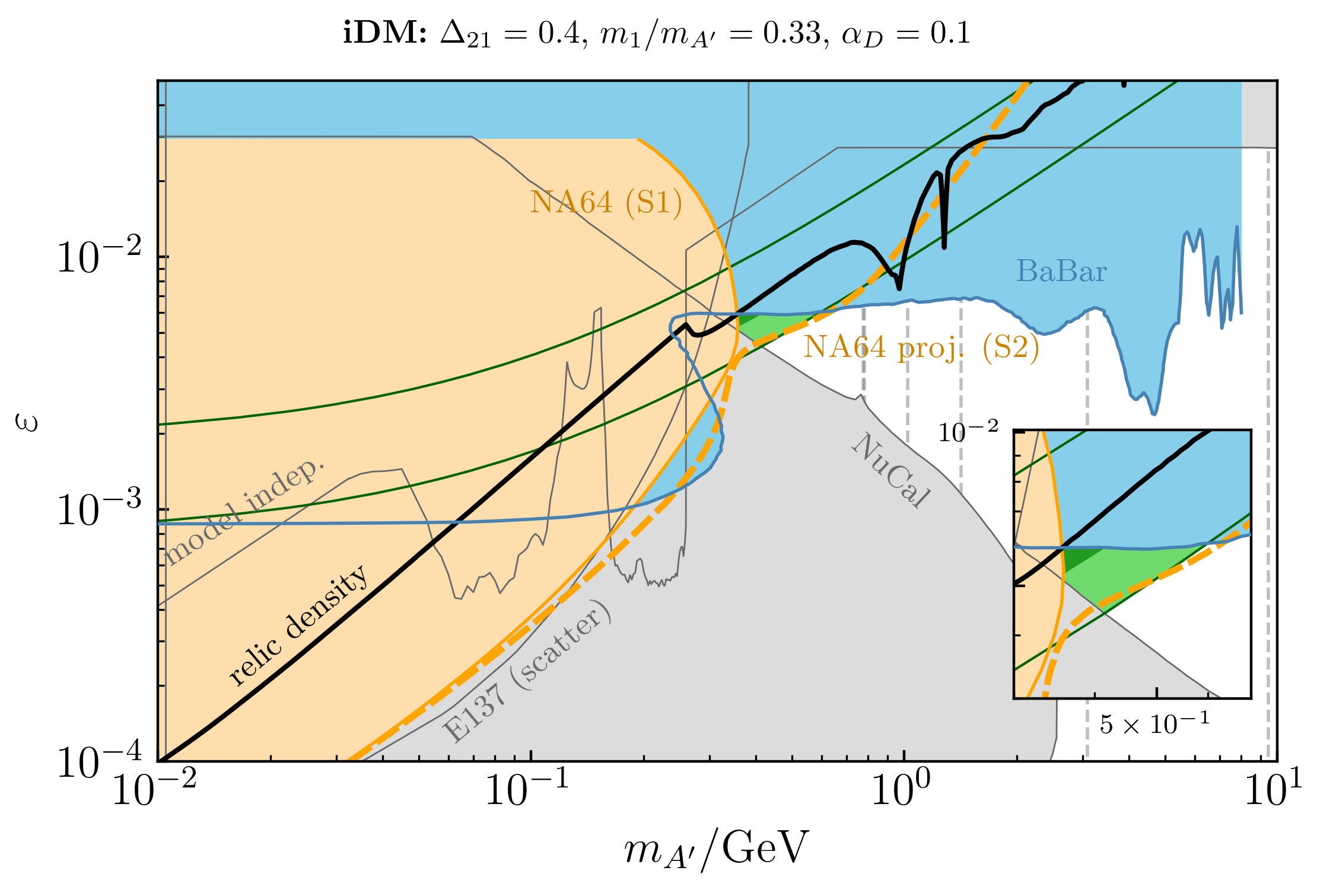}
    \caption{
    The kinetic mixing $\epsilon$ as a function of the dark photon mass $m_{A^\prime}$ for BP1a (left) and BP1b (right) in the inelastic dark matter (iDM) model. 
    We show the $\deltaamu$-preferred $1$, $2$, and $3\sigma$ regions in shades of green. 
    The recast constraints from BaBar and NA64 are shown in blue and orange, respectively.
    The NA64 curves show the recast constraints using the dedicated \emph{semi-visible} search, corresponding to those derived using displaced decays (S1 in \cref{fig:NA64diagram}), and the projected sensitivity of a search for invisible and promptly-decaying particles (S2 in \cref{fig:NA64diagram}).
    Assuming the lightest HNF to be dark matter, the relic density line is shown in black. 
    Other constraints from $(g-2)_e$, EWPO, DIS, and NA62 are shown with thin gray lines and light gray regions and are referred to as model independent constraints (see \cref{sec:model_independent}).
    The constraint imposed by NuCal and E137 (scatter) are shown with the same style. 
    The masses of vector meson resonances are shown as vertical grey dashed lines.
    \label{fig:recast_1}}
\end{figure*}

\myparagraph{Inelastic dark matter (BP1a/b)}
The results for the iDM benchmark are shown in Fig.~\ref{fig:recast_1}, expressed in the $\varepsilon$/$m_{A^\prime}$-plane.
In BP1a, we see a significant relaxation of the NA64 and BaBar bounds, with a sizeable $\deltaamu$ preference region now open. 
Due to a large DS coupling, $\alpha_D=0.5$, the decay rates of the HNFs are enhanced, allowing for more semi-visible events in the detectors.
Benchmark BP1b corresponds to the choice of parameters used in Ref.~\cite{Mohlabeng:2019vrz}.
Assuming the lightest HNL, $\psi_1$, is a dark matter candidate, we find that the correct dark matter abundance can be achieved in both benchmarks BP1a and BP1b.

In BP1b we find a much less significant relaxation of the bounds compared to BP1a, leaving very little open parameter space for a $\deltaamu$ explanation.
In particular, we find the BaBar bound to be much more constraining than in Ref.~\cite{Mohlabeng:2019vrz}.
This is predominantly due to the difference in selection criteria used in the two analyses.
In Ref.~\cite{Mohlabeng:2019vrz}, an energy cut of $60$~MeV is applied to charged particles produced in semi-visible $A^\prime$ decays.
In this work, we take the larger value of $100$~MeV, corresponding to the energy threshold used in the analyses to veto additional tracks in the BaBar drift chamber~\cite{Li:2022amw,BaBar:2017tiz}.
The impact of the higher energy threshold is to veto a greater proportion of the semi-visible decays, leading to a stronger constraint from BaBar.
In addition, we cut on the polar angle of the charged lepton pairs, which allows us to exclude events in which the leptons are produced in the direction of the beam pipeline. We find the fraction of these ``pipeline'' events to be small and that the relative strength of the BaBar bound is predominantly a consequence of the energy threshold.
Contrary to the more conservative analysis of Ref.~\cite{Duerr:2019dmv}, in which a threshold energy of $150$~MeV is taken, a very small region of preference for $\deltaamu$ remains open at the $2\sigma$ level for $m_{A^\prime}\sim 300 - 500$~MeV.

In both BP1a and BP1b, we find that a search for promptly decaying HNFs at NA64, with signatures of the type S2, can cover the newly open $\deltaamu$ parameter space (see also the companion paper in Ref.~\cite{na64semivisible}).
This region of parameter space is also accessible to other lower-energy $e^+e^-$ colliders, including KLOE/KLOE-2~\cite{Bossi:2008aa,Amelino-Camelia:2010cem} and BES-III~\cite{BESIII:2020nme}.

\begin{figure*}[t]
    \centering
    \includegraphics[width=\columnwidth]{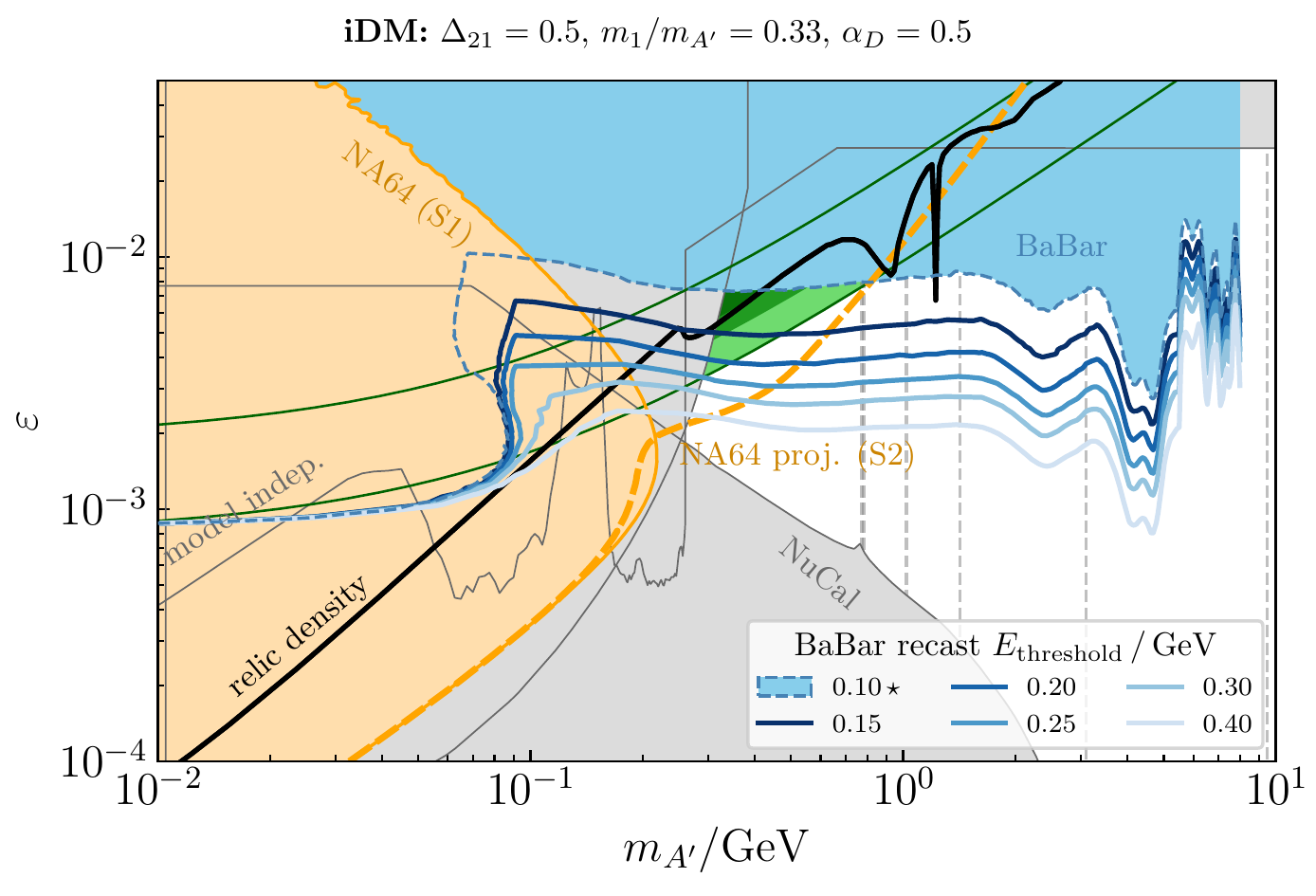}
    \includegraphics[width=\columnwidth]{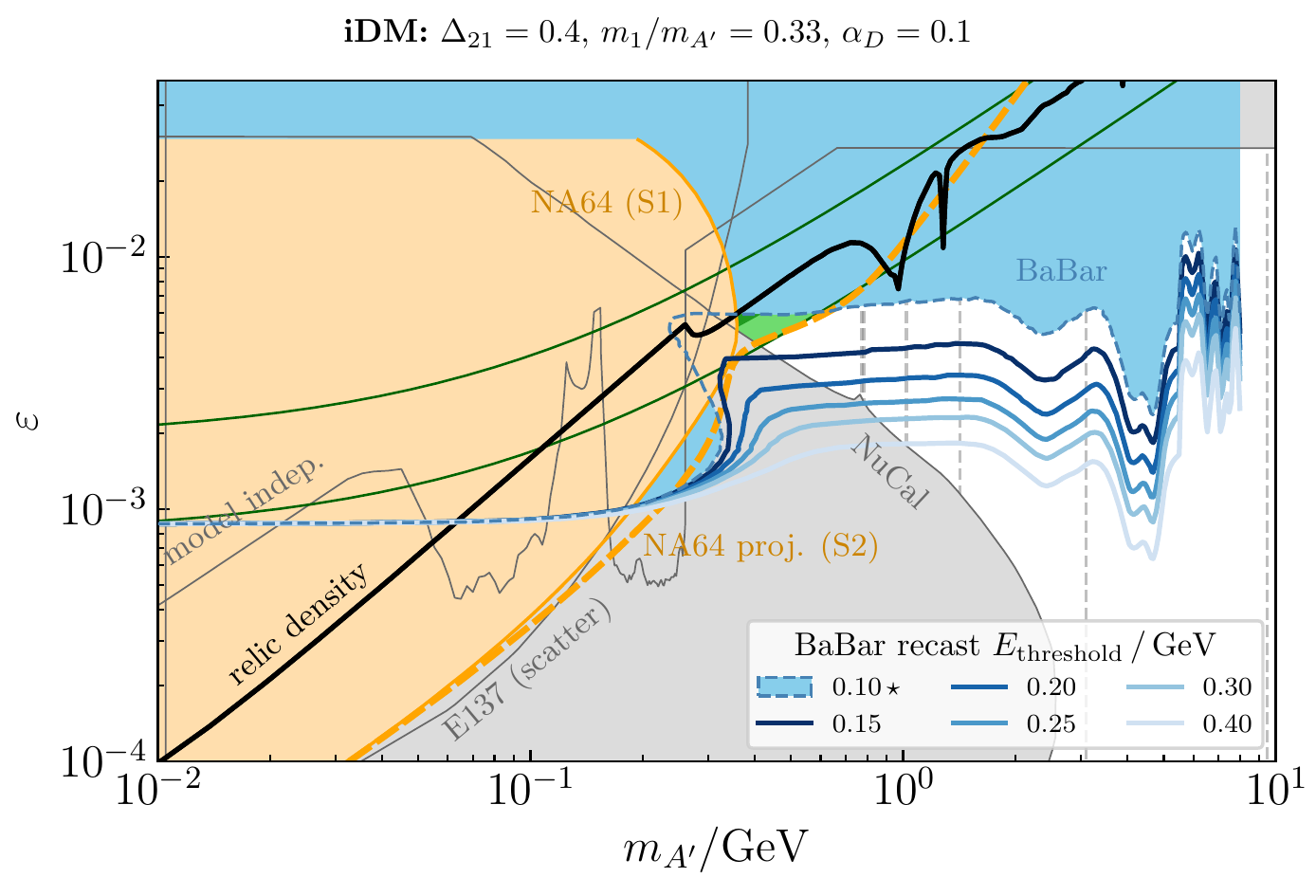}
    \caption{Same as Fig.~\ref{fig:recast_1}, with the addition of multiple BaBar recasts obtained under different assumptions for the energy threshold for the $e^+/e^-$ tracks. 
    The different energy thresholds are represented by shades of blue with the adopted values shown in the legend. 
    The threshold chosen for the analysis performed in this work is marked with a $\star$ symbol.\label{fig:recast_bp1_babar_thresholds}}
\end{figure*}

In the BaBar signal selection, events with charged tracks above $100~\mathrm{MeV}$ were vetoed, justifying our choice for the energy threshold~\cite{BaBar:2017tiz}.
However, the probability of detection of such low-energy tracks may not be exactly unity, setting an effective energy threshold.
Therefore, our hard requirement on the energy of leptons is an important source of uncertainty.
In Fig.~\ref{fig:recast_bp1_babar_thresholds}, we show the effects of changing this value on the BaBar limits for the iDM model in BP1a and BP1b.
Increasing the energy threshold makes it more difficult to veto events, leading to a stronger bound.
For a threshold larger than $250~\mathrm{MeV}$, BaBar would cover the entire parameter space allowed by $\deltaamu$ in the case of BP1a, while a threshold of $150~\mathrm{MeV}$ is already sufficient to rule it out completely in BP1b.
We note, however, that thresholds exceeding $200~\mathrm{MeV}$ are likely unrealistic as the probability of missing such tracks within the inner BaBar detectors is very small.

\begin{figure*}[t]
    \centering
    \includegraphics[width=\columnwidth]{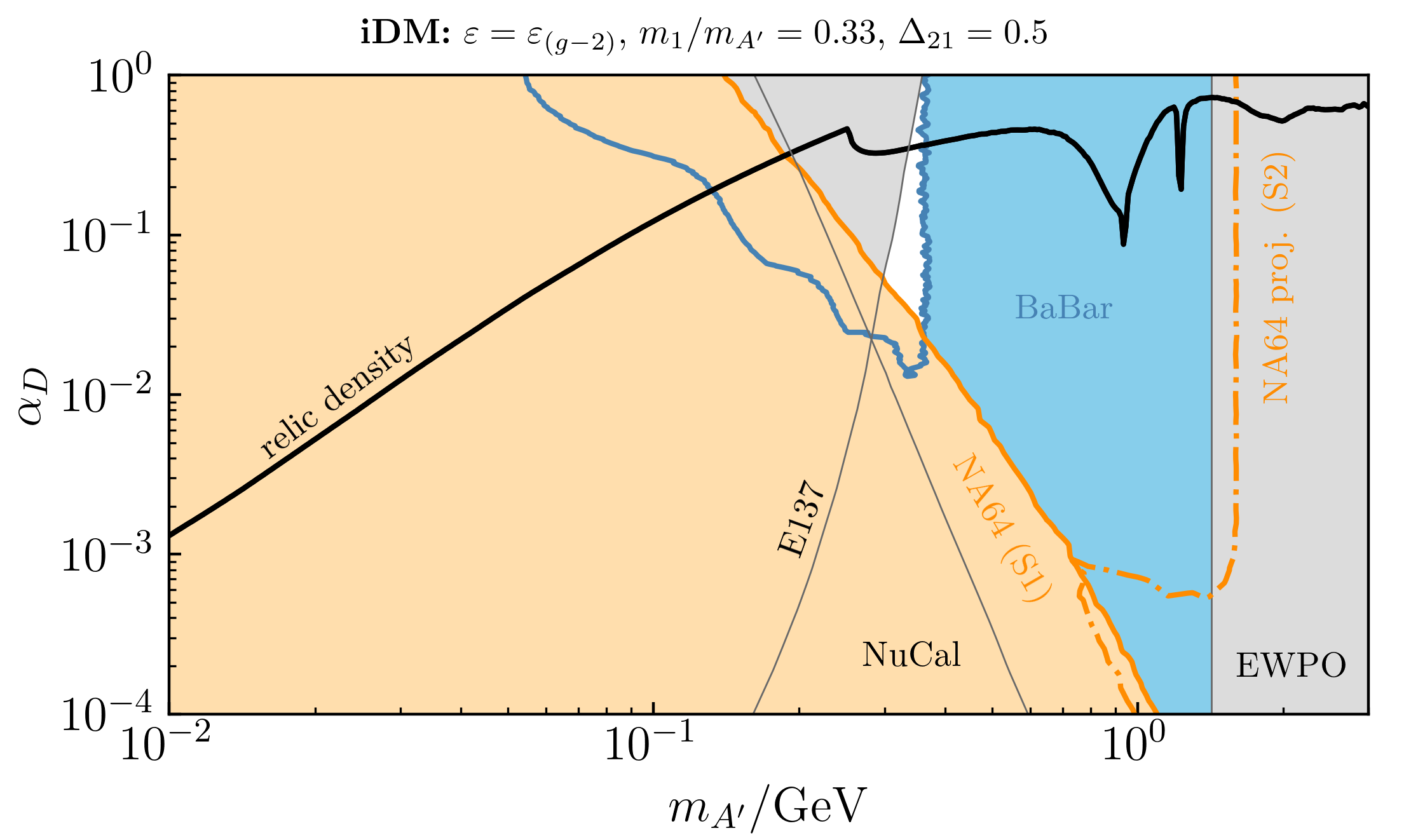}
    \includegraphics[width=\columnwidth]{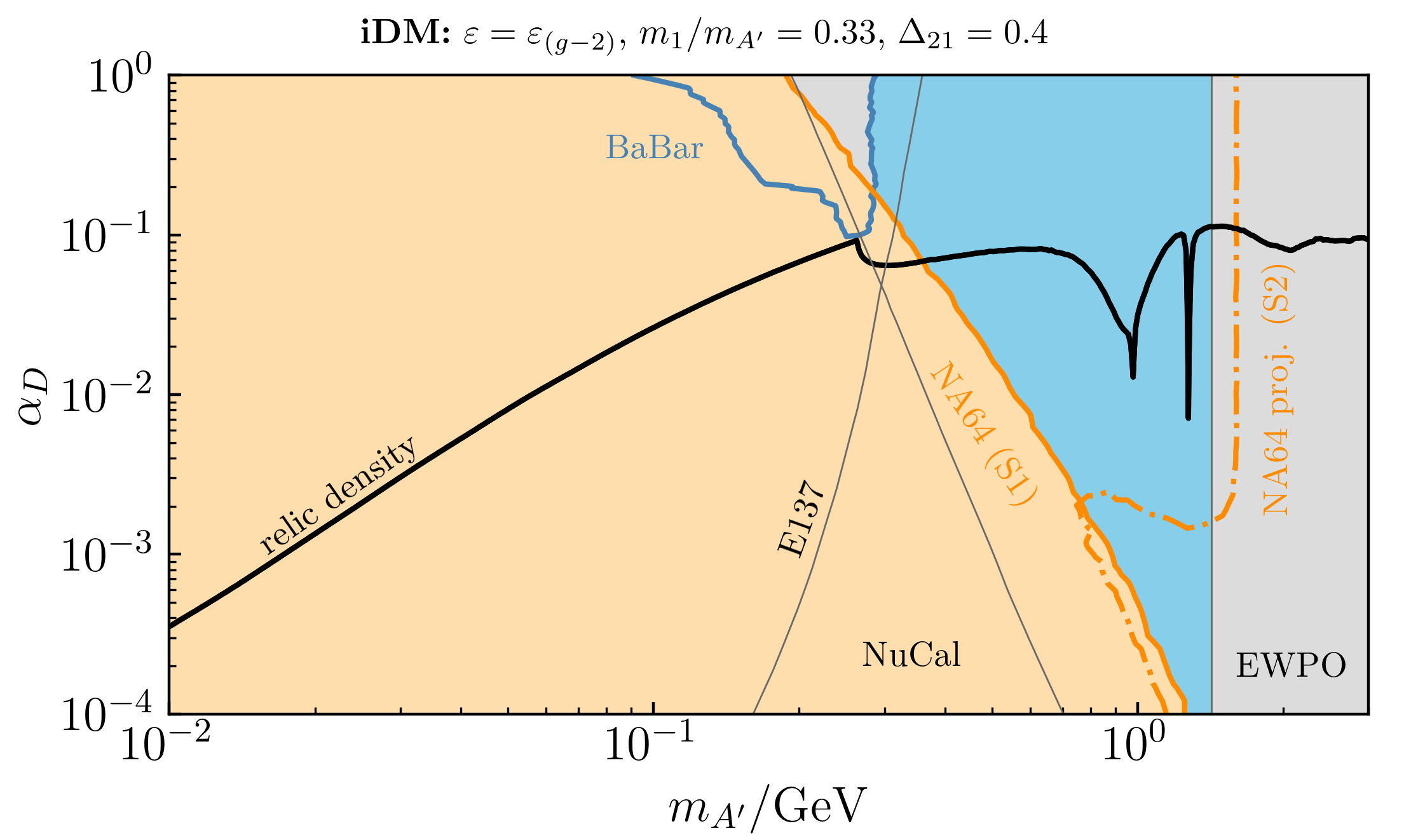}
    \caption{
    Parameter space of $\alpha_D$ versus $m_{A^\prime}$ for the iDM model, fixing to the value needed to explain the $\deltaamu$, $\epsilon = \epsilon_{g-2}$.
    The mass splitting is fixed at $\Delta_{21}=0.5$ (left) and $\Delta_{21}=0.4$ (right).
    The constraints are shown with the same style used in \cref{fig:recast_1}.
    \label{fig:alphaD_v_maprime}}
\end{figure*}

\begin{figure*}[t]
    \centering
    \includegraphics[width=\columnwidth]{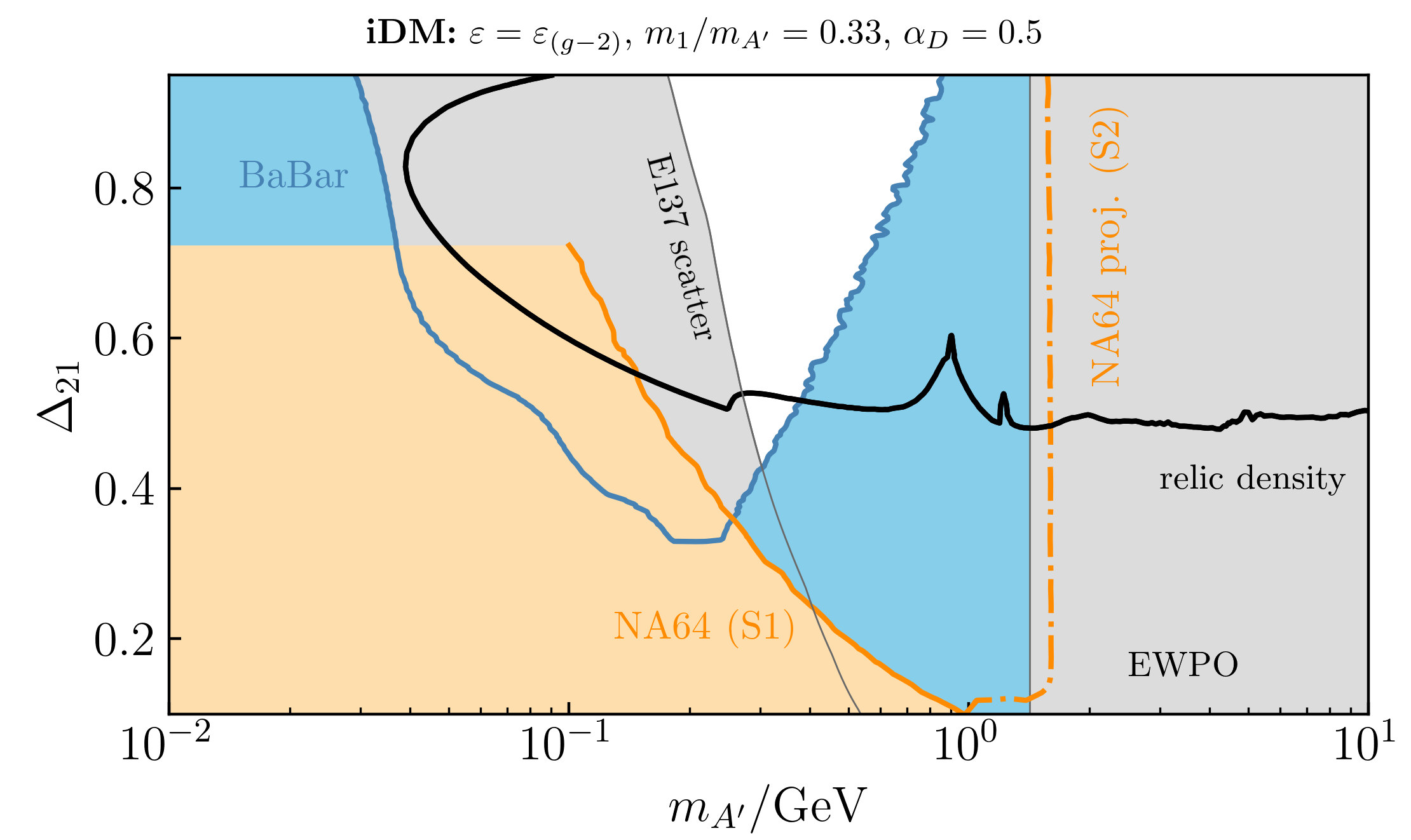}
    \includegraphics[width=\columnwidth]{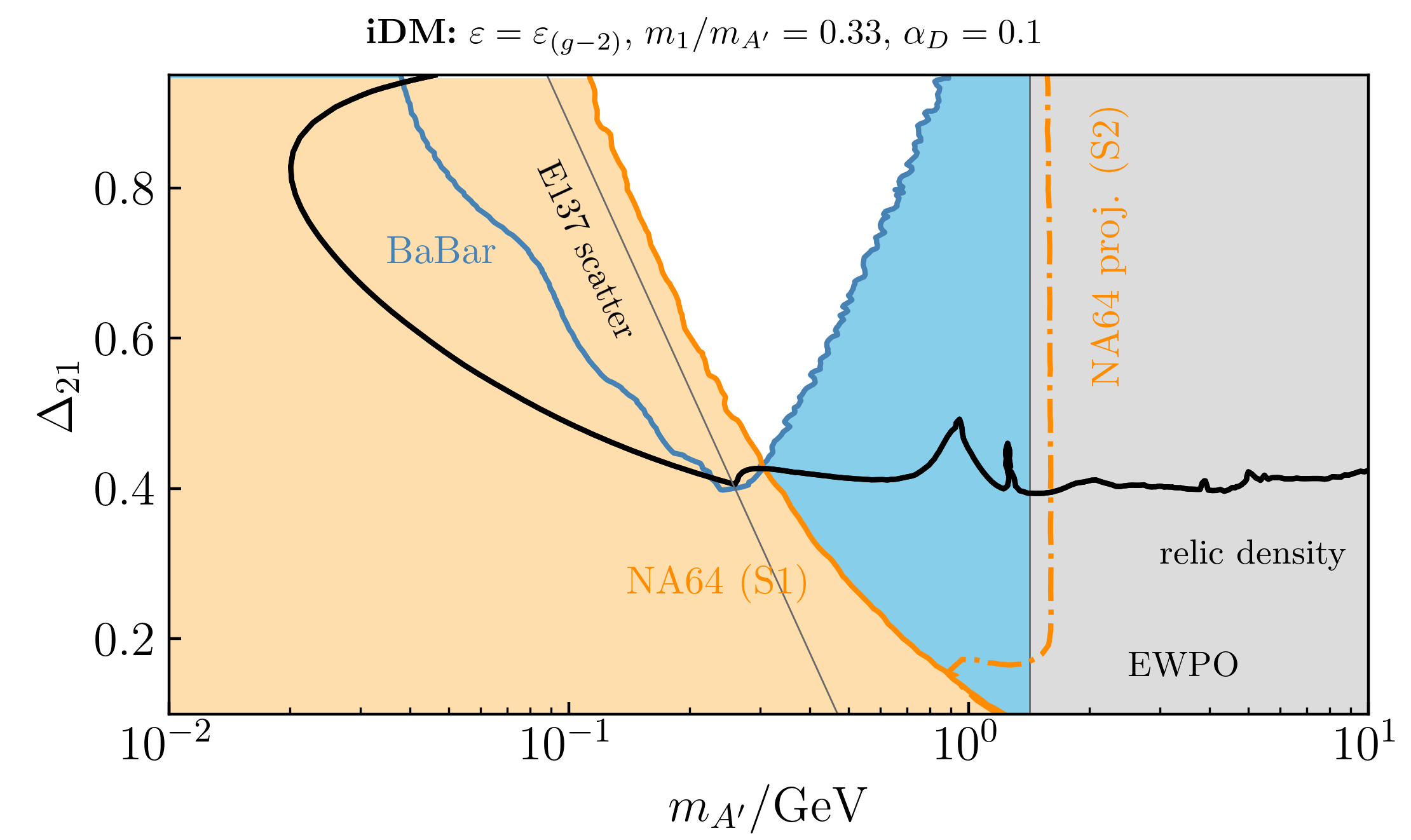}
    \caption{
    Parameter space of $\Delta_{21}$ versus $m_{A^\prime}$ for the iDM model, fixing to the value needed to explain the $\deltaamu$, $\epsilon = \epsilon_{g-2}$.
    The dark coupling is fixed at $\alpha=0.5$ (left) and $\alpha=0.1$ (right).
    The constraints are shown with the same style used in \cref{fig:recast_1}.
    \label{fig:delta_v_maprime}}
\end{figure*}

To understand whether it is possible to accommodate the different constraints and the relic density, we can inspect the parameter space in $\alpha_D$/$m_{A^\prime}$ and $\Delta_{21}$/$m_{A^\prime}$ plots, shown respectively in \cref{fig:alphaD_v_maprime} and in \cref{fig:delta_v_maprime}.
In \cref{fig:alphaD_v_maprime}, we fix the value of $\epsilon$ to that required to explain $\deltaamu$ and we vary $\alpha_D$ and $m_{A^\prime}$.
Increasing $\alpha_D$ affects the lifetime of $\psi_2$, making them short-lived and allowing their decays to happen inside the detector, increasing the amount of semi-visible events that can be identified.
Both scenarios are strongly constrained by BaBar, NA64, E137, and NuCal, with only a small part of the parameter space allowed in the left panel.
That region is also able to accommodate the DM relic density.
The two scenarios differ only by the value of the mass splitting $\Delta_{21}$: the right panel is characterized by a lower value.
Decreasing $\Delta_{21}$ means that the two HNFs become more degenerate and the lifetime of $\psi_2$ increases, becoming larger than the size of the detector.
This effect reduces the amount of $\psi_2$ decays happening inside the detector and makes the bound more constraining.
The right panel is indeed entirely constrained.
%

In \cref{fig:delta_v_maprime} we fix the value of $\epsilon$ to that required to explain $\deltaamu$ and we vary $\Delta_{21}$ and $m_{A^\prime}$.
In this case, the E137 constraint shown with a thin gray line has been extrapolated at large $\Delta_{21}$ value.
As discussed, previously, decreasing $\Delta_{21}$ makes most of the $\psi_2$ decays to happen outside of the detector.
No semi-visible event would be detected in this case, and the bound would resemble the original invisible $A^\prime$ bound, covering the region in which $\epsilon$ can explain $(g-2)_\mu$.
In this case, the NA64 (S1) constraint follows a similar trend as for the BaBar constraint, becoming weaker for larger mass splitting.
In the case of NA64 (S2), at large $m_{A^\prime}$, the experiment loses sensitivity, and the bound becomes naturally weaker, independently of the value of $\Delta_{21}$.
This corresponds to the loss of sensitivity in the original invisible bound posed by NA64, caused by a lack of event rate.

Nevertheless, increasing the mass splitting between the two HNFs to accommodate the constraints affects the dark matter relic abundance of $\psi_1$.
A larger mass splitting increases the Boltzmann suppression of $\psi_2$ number density in the early universe, depleting it faster and suppressing the coannihilation contribution to the cross section.
This results in an overabundant scenario, which can be controlled only assuming secluded annihilations within the dark sector, and it is expressed by the parameter space above the relic density line in \cref{fig:delta_v_maprime}.
The smaller $\alpha_D$ value in BP1b translates into a shift of the relic density line towards lower $\Delta_{21}$, because of its effect in decreasing the annihilation cross section.
A smaller $\Delta_{21}$ ensures a smaller Boltzmann suppression of $\psi_2$ number density and a larger coannihilation cross section.

It is interesting to notice that the projections shown by the NA64 (S2) line could constrain the free parameter space.
The search for promptly decaying HNFs in the detector can address whether this minimal model can simultaneously explain $\deltaamu$ and dark matter.

\begin{figure*}[t]
    \centering
    \includegraphics[width=\columnwidth]{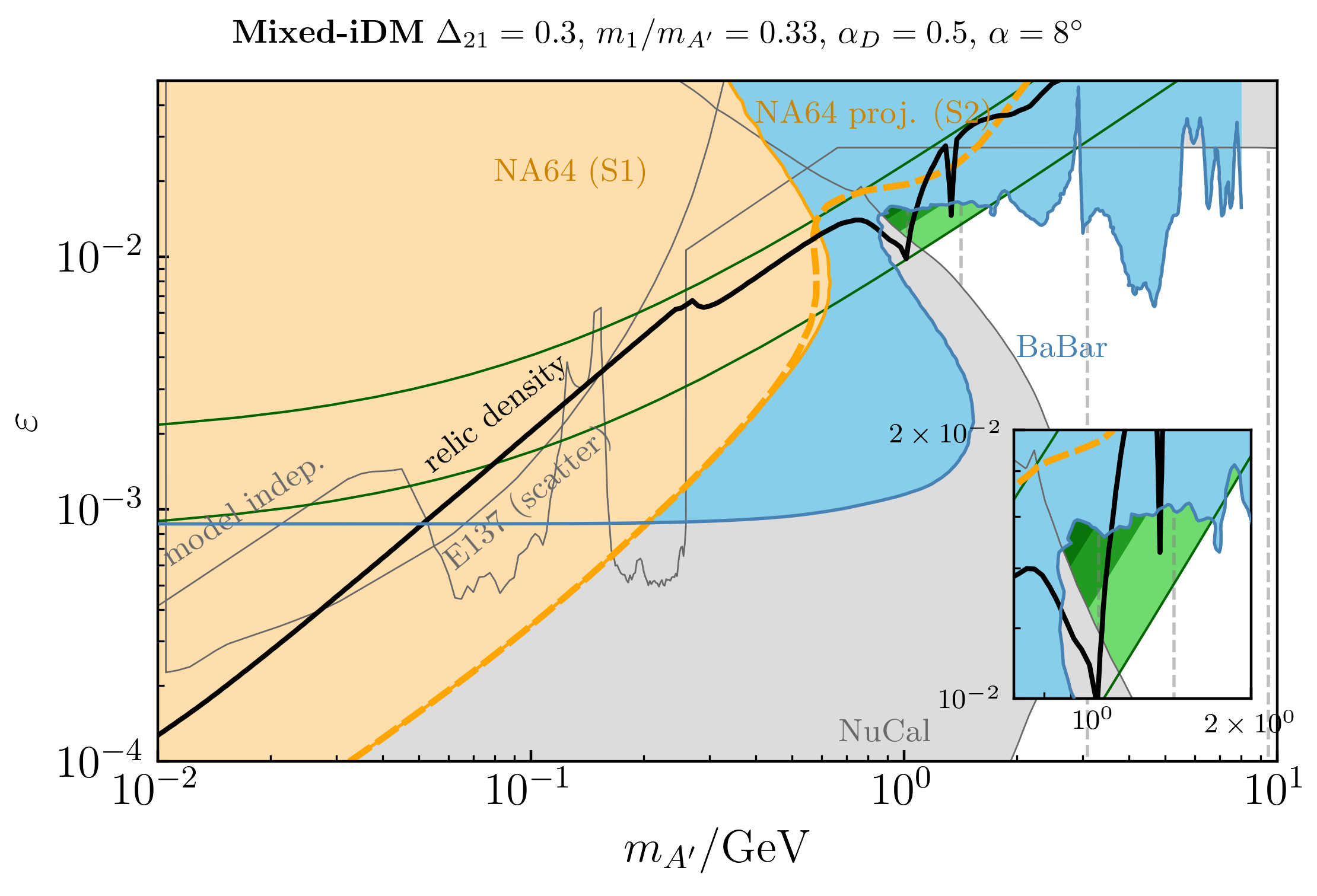}
    \includegraphics[width=\columnwidth]{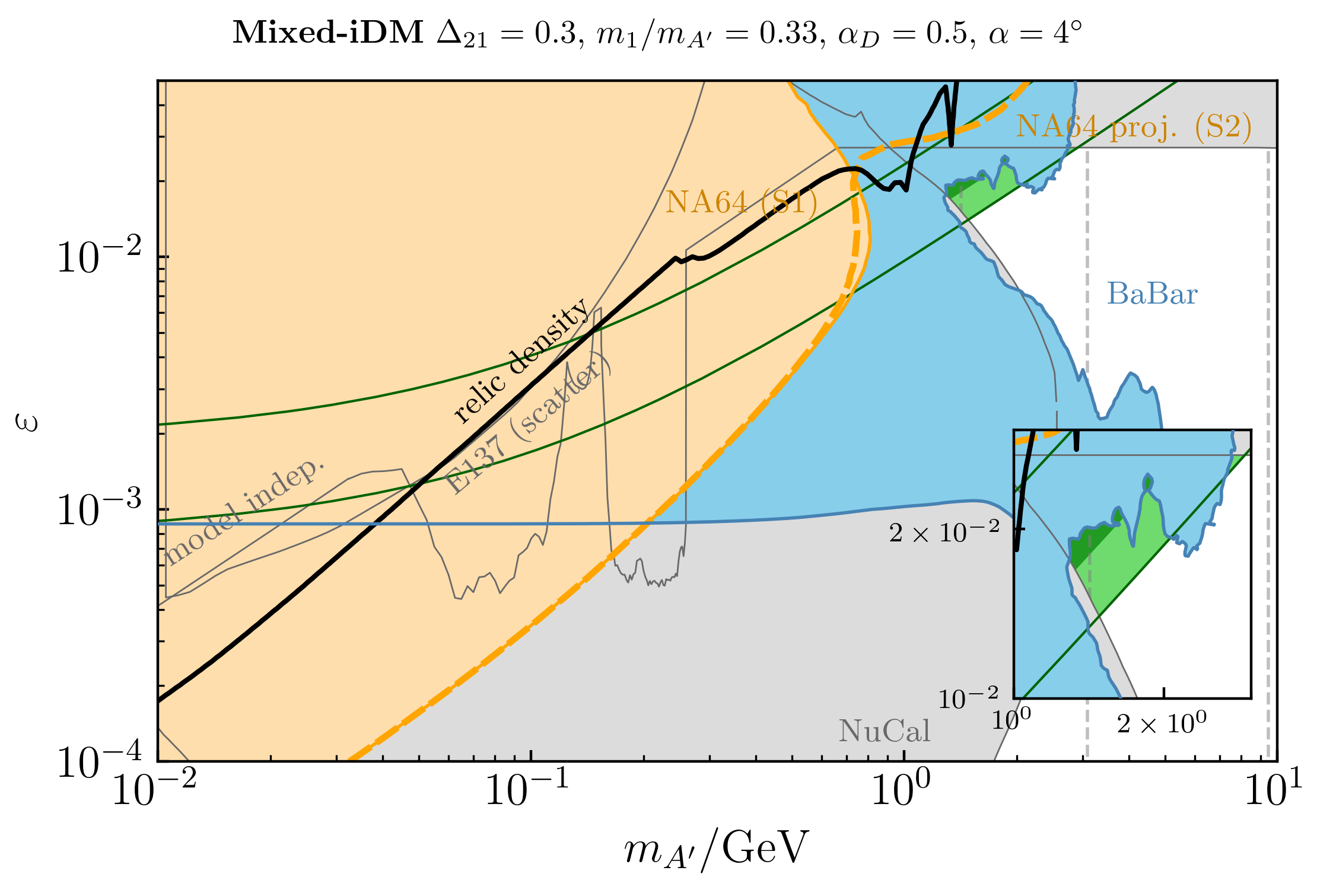}
    \caption{
    Same as \cref{fig:recast_1} but for BP2a (left) and BP2b (right), corresponding to the mixed inelastic Dark Matter (mixed-iDM) model.  
    The two panels represent two different realizations of the model, obtained by varying the $\alpha$ mixing angle. 
    \label{fig:recast_2ab}
    }
\end{figure*}

\myparagraph{Mixed Inelastic Dark Matter (BP2a/b)}
We show the constraints for the mixed-iDM model in \cref{fig:recast_2ab}, expressed as $\varepsilon$/$m_{A^\prime}$ constraints.
This model main feature, which is also expected in BP3, is an enhanced $A^\prime$ decay BR into $\psi_2\psi_2$.
The branching ratio to $\psi_2\psi_1$ is suppressed by a factor of the mixing angle $\alpha$,
 and the one into two $\psi_1\psi_1$ is forbidden by the $C$ symmetry.
Because most dark photon events come accompanied by two unstable particles, the additional energy deposition is missed even less often, relaxing the bounds further.

The relic density of $\psi_1$ for this model depends strongly on the efficiency of coannihilations and co-scattering processes.
In the realization shown in \cref{fig:recast_2ab}, we find that a simultaneous explanation of $\deltaamu$ and dark matter relic density, along with the constraints discussed, can be achieved in the region $0.9~\mathrm{GeV} \lesssim m_{A^\prime} \lesssim 1.2~\mathrm{GeV}$ for BP2a.
In the case of BP2b, the coannihilation processes are inefficient due to the smaller $\alpha$, so that $\psi_1$ is overabundant in the region of parameter space that can explain $\deltaamu$.

\begin{figure*}[th]
    \centering
    \includegraphics[width=0.9\textwidth]{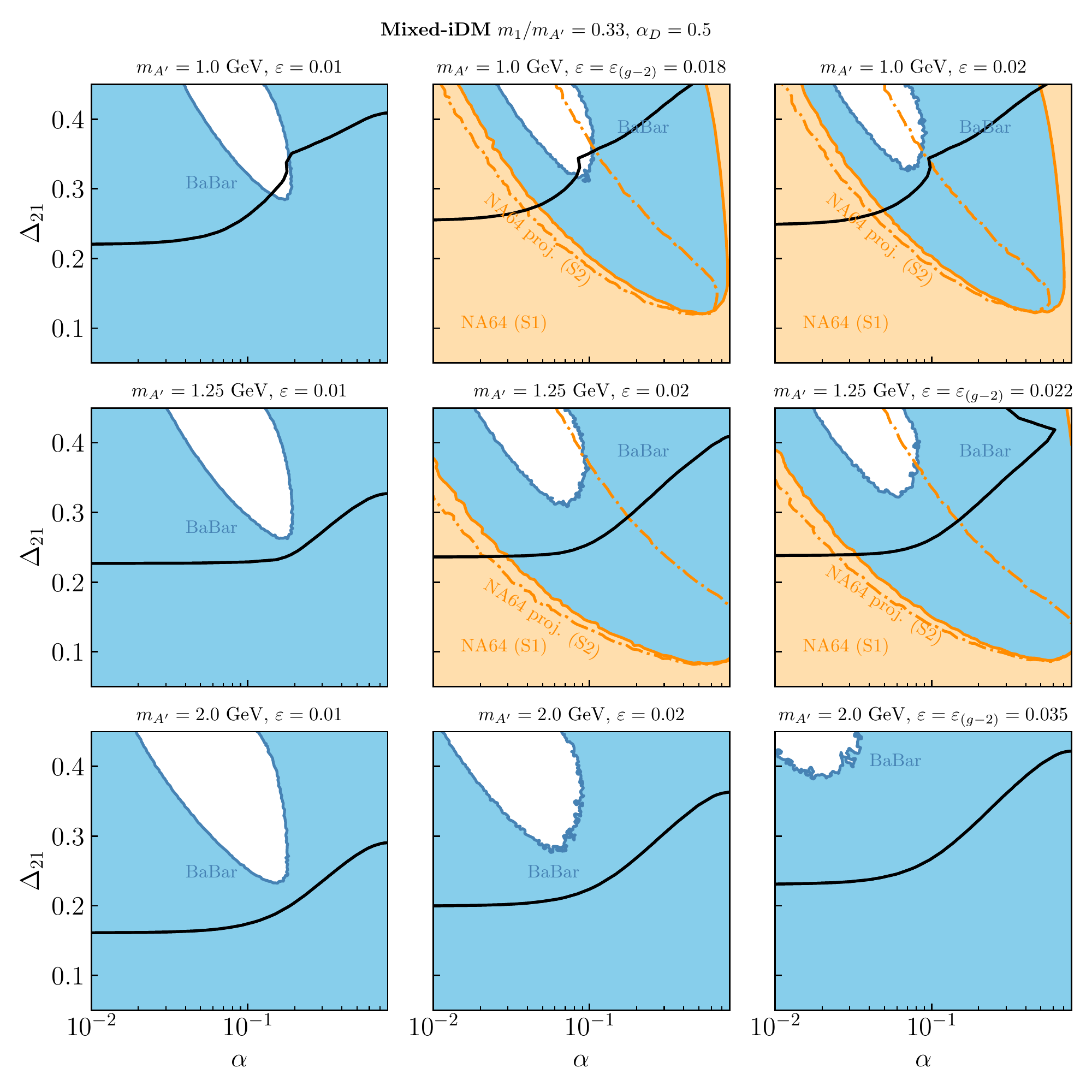}
    \caption{Parameter space of the mixed-iDM in the plane of $\Delta_{21}$ and the mixing angle $\alpha$.
    The parameter $m_{A^\prime}$ has been fixed to $1.0$, $1.25$ and $2.0~\mathrm{GeV}$ (column-wise), while the kinetic mixing $\varepsilon$ has been chosen among $0.01$, $0.02$, and value $\varepsilon_{g-2}$ (row-wise), corresponding to the central value of the $\deltaamu$ explanation.
    BaBar and NA64 recast bounds are shown with the same style used in \cref{fig:recast_1}.}
    \label{fig:delta_vs_theta_2}
\end{figure*}

In addition, we report an analysis on the $\Delta_{21}$ and $\alpha$ parameters, showing the constraints in a $\Delta_{21}/\alpha$ in \cref{fig:delta_vs_theta_2}, fixing the mass of the dark photon to $m_{A^\prime} = 1,\, 1.25,\, 2~\mathrm{GeV}$ and $\epsilon = 0.01,\, 0.02,\, \epsilon_{(g-2)}$.
For some combinations of the parameters, the NA64 constraints are not present because they are too weak, given the efficient relaxation that this model can provide.
The dependence on the angle $\alpha$ is expressed by the branching ratio: a larger value favors a larger branching ratio to the $\psi_2\psi_1$ channel, with respect to $\psi_2\psi_2$; it has the effect of decreasing the amount of visible energy, and ultimately the possibility to detect a semi-visible event.
On the contrary, a smaller $\alpha$ affects the decay rate of $\psi_2$, suppressing its decay, and recovering the original invisible bound.
The behavior of the constraints is similar to BP1 for what concert $\Delta_{21}$ dependence: a larger $\Delta_{21}$ means a shorter $\psi_2$ lifetime, and the energy of its decay is released inside the detector.
On the other hand, a lower $\Delta_{21}$ resembles the case of fully invisible dark photon decays, as the lifetime becomes larger than the size of the detector.

\begin{figure*}[t]
    \centering
    \includegraphics[width=\columnwidth]{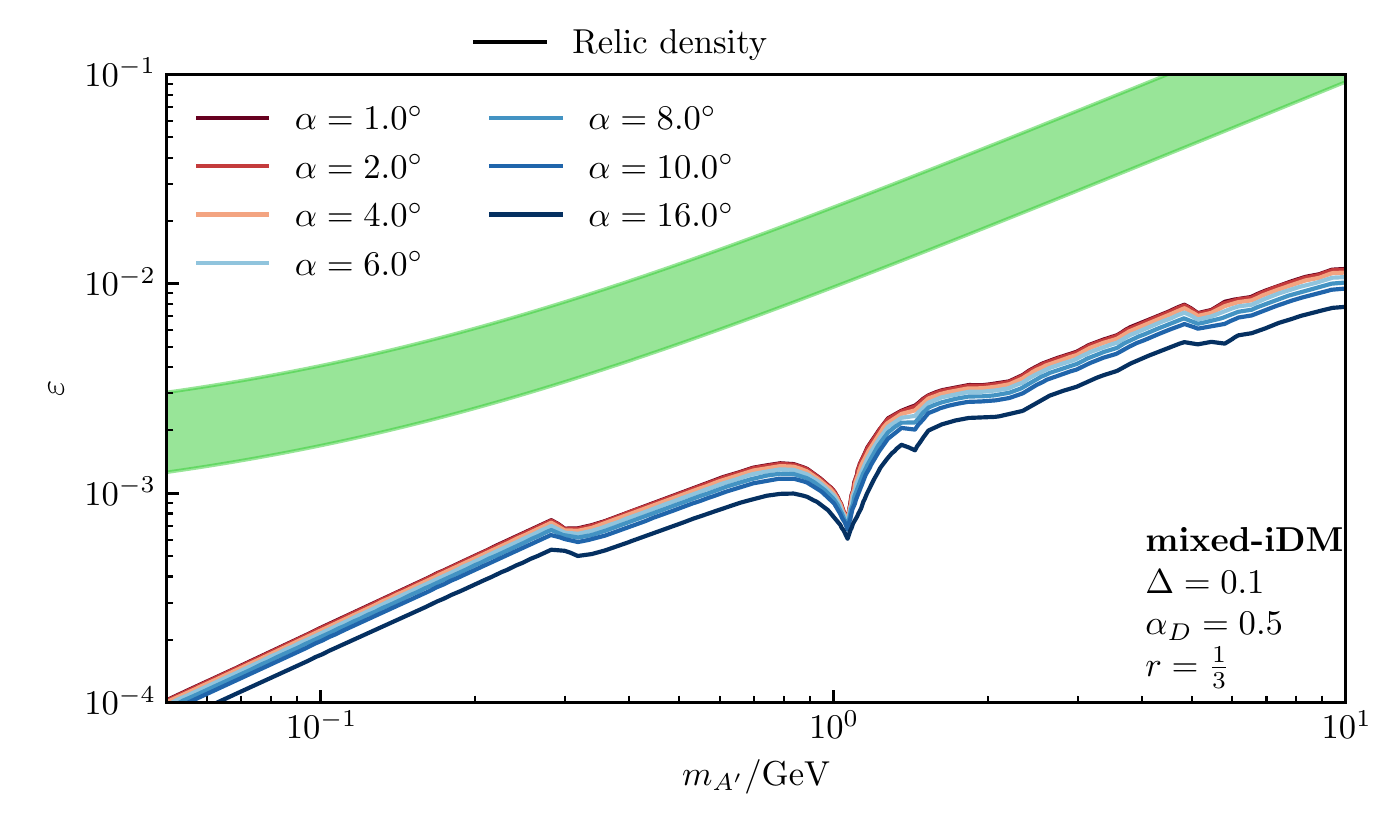}
    \includegraphics[width=\columnwidth]{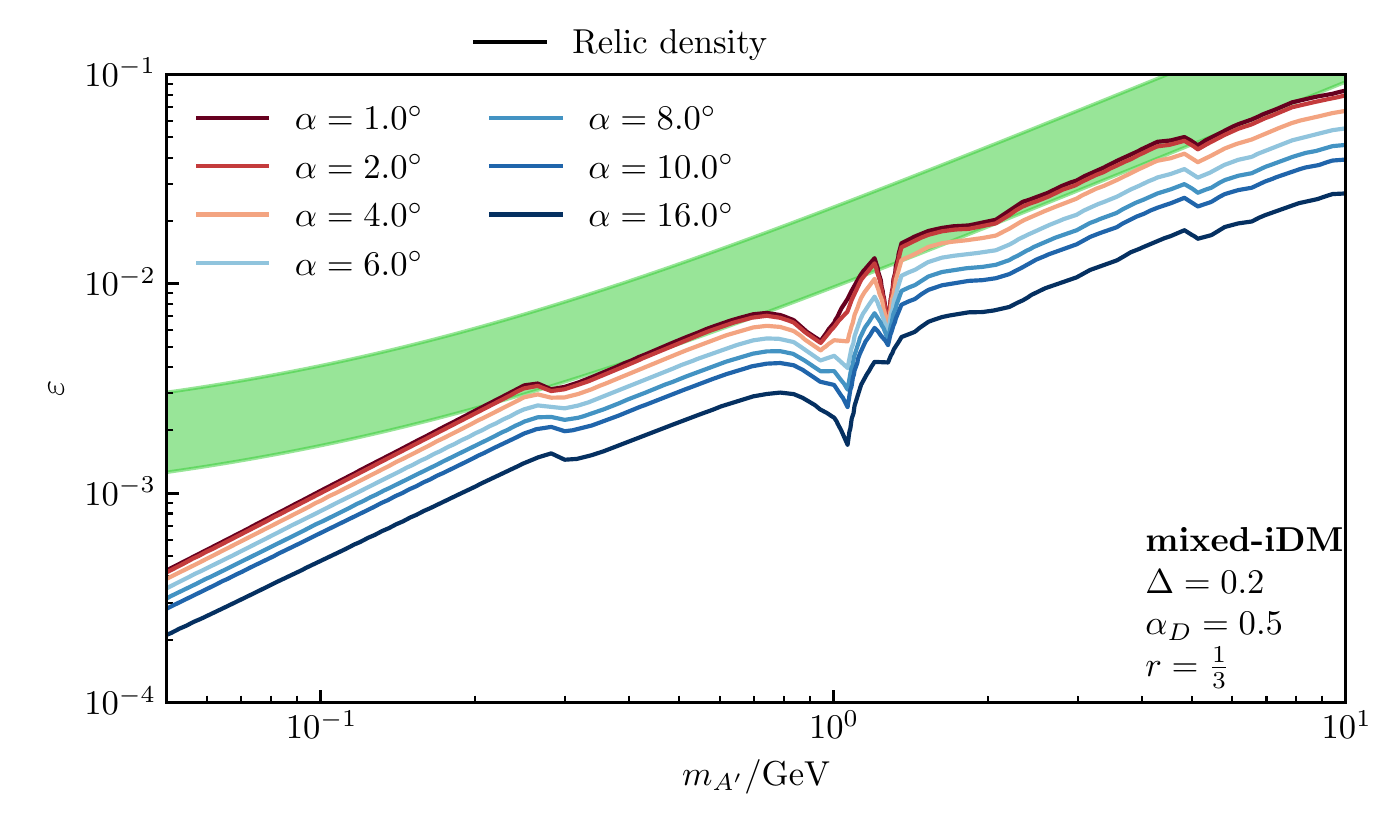}
    \includegraphics[width=\columnwidth]{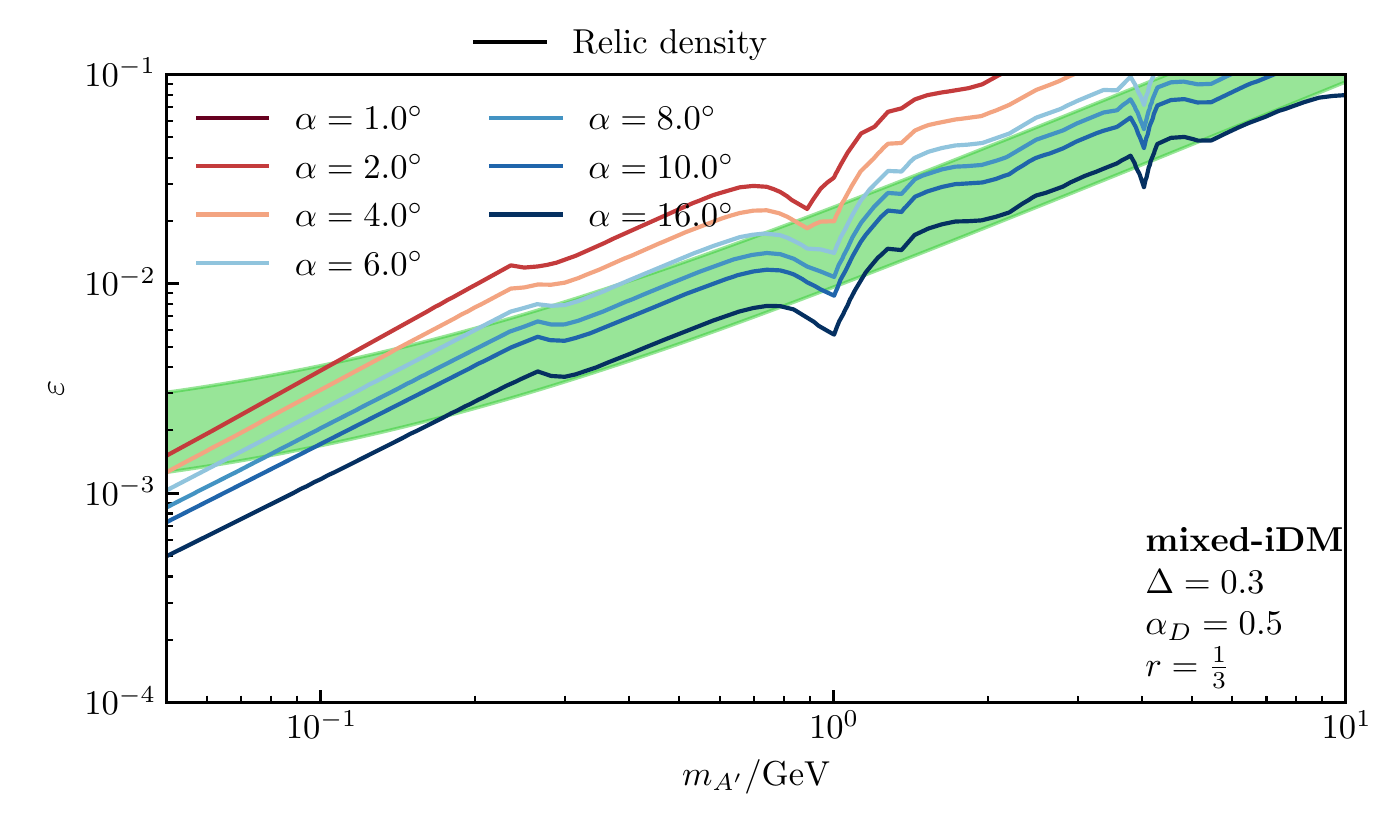}
    \includegraphics[width=\columnwidth]{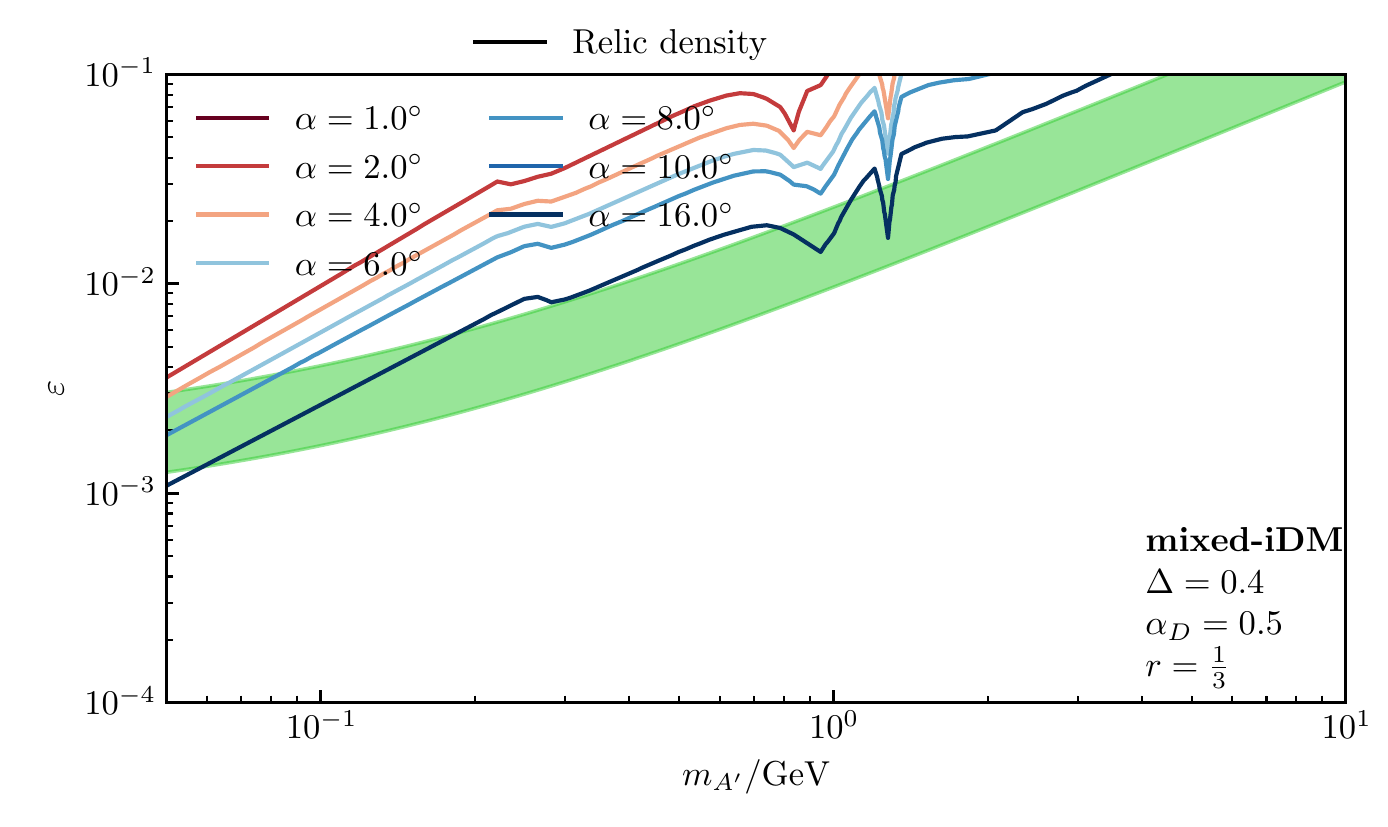}
    \caption{
    The relic density (solid) in the parameter space of mixed-inelastic DM model in the plane of kinetic mixing $\epsilon$ versus dark photon mass $m_{A^\prime}$ for fixed values of $\alpha = 0.5$, $r = m_1/m_{A^\prime} = 1/3$,
    and $\Delta_{21}$.
    CMB limits are not applicable as the dark matter self-annihilation $\psi_1 \psi_1 \to f^+f^-$ is forbidden in the $C$ symmetric limit.
    \label{fig:relic_mixediDM}}
\end{figure*}

The relic density lines in \cref{fig:delta_vs_theta_2} identify the overabundant region, corresponding to high $\Delta_{21}$ and low $\alpha$.
In that case, coannihilations and coscattering processes become too inefficient.
For small $\Delta_{21}$, the choice of mixing angle does not have a strong impact, as coscattering remains efficient for longer, and the self-annihilation of $\psi_2$ sets the relic abundance of $\psi_1$.
For large $\alpha$ the dependence on $\Delta_{21}$ on the relic density is relaxed: the enhanced coscattering ratio obtained with a larger $\alpha$ allows to afford a larger $\Delta_{21}$ value before ending up in an underabundant scenario.
The kink present in the dark matter relic density for $m_{A^\prime} = 1.0~\mathrm{GeV}$ is due to the presence of a resonance region, that can be observed also in \cref{fig:relic_mixediDM}.
In that figure we show the trend of the relic density line for different choice of the parameters of the model, along with the $3\sigma$ region accounting for the $\deltaamu$ explanation.

\begin{figure*}[th]
    \centering
    \includegraphics[width=\columnwidth]{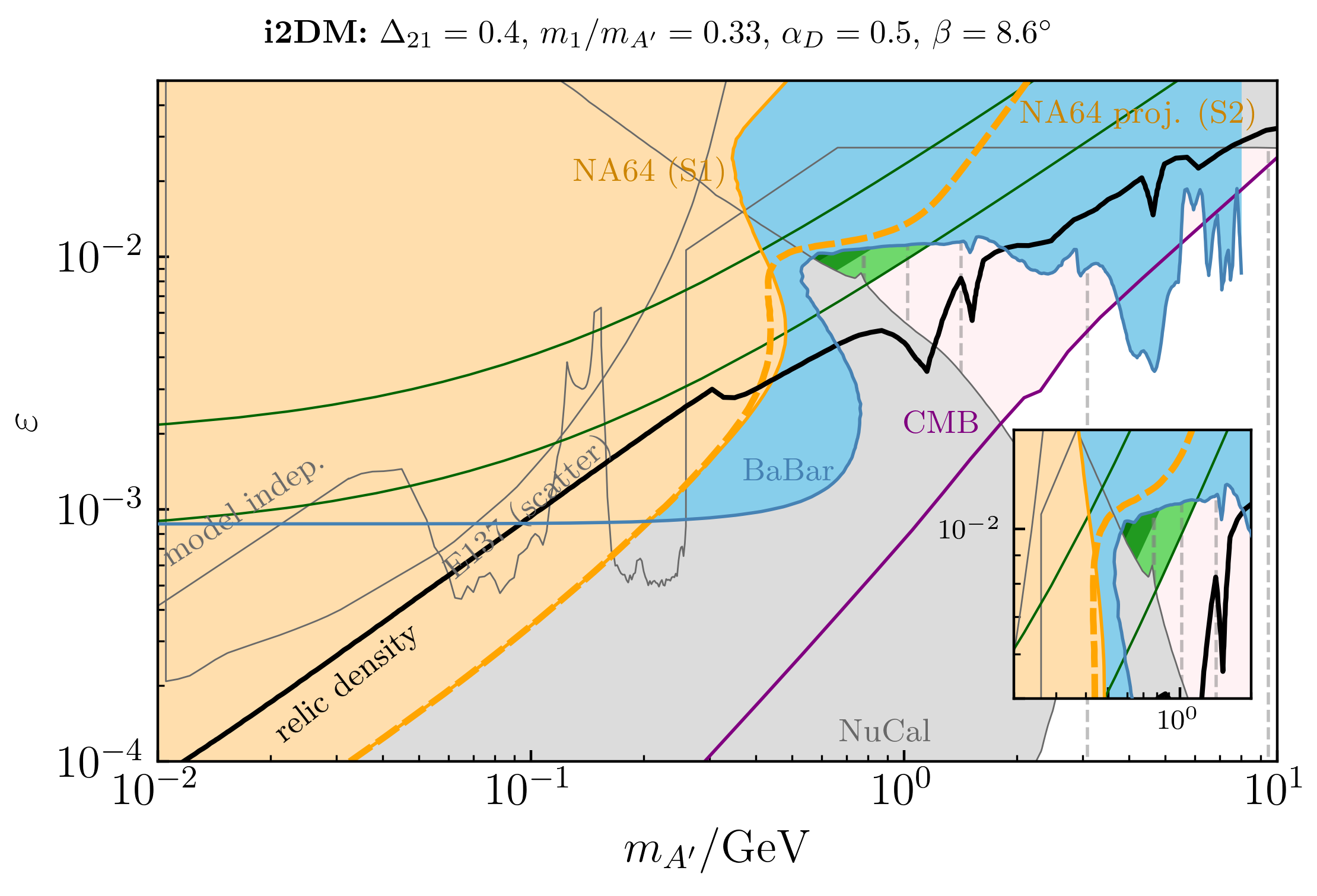}
    \includegraphics[width=\columnwidth]{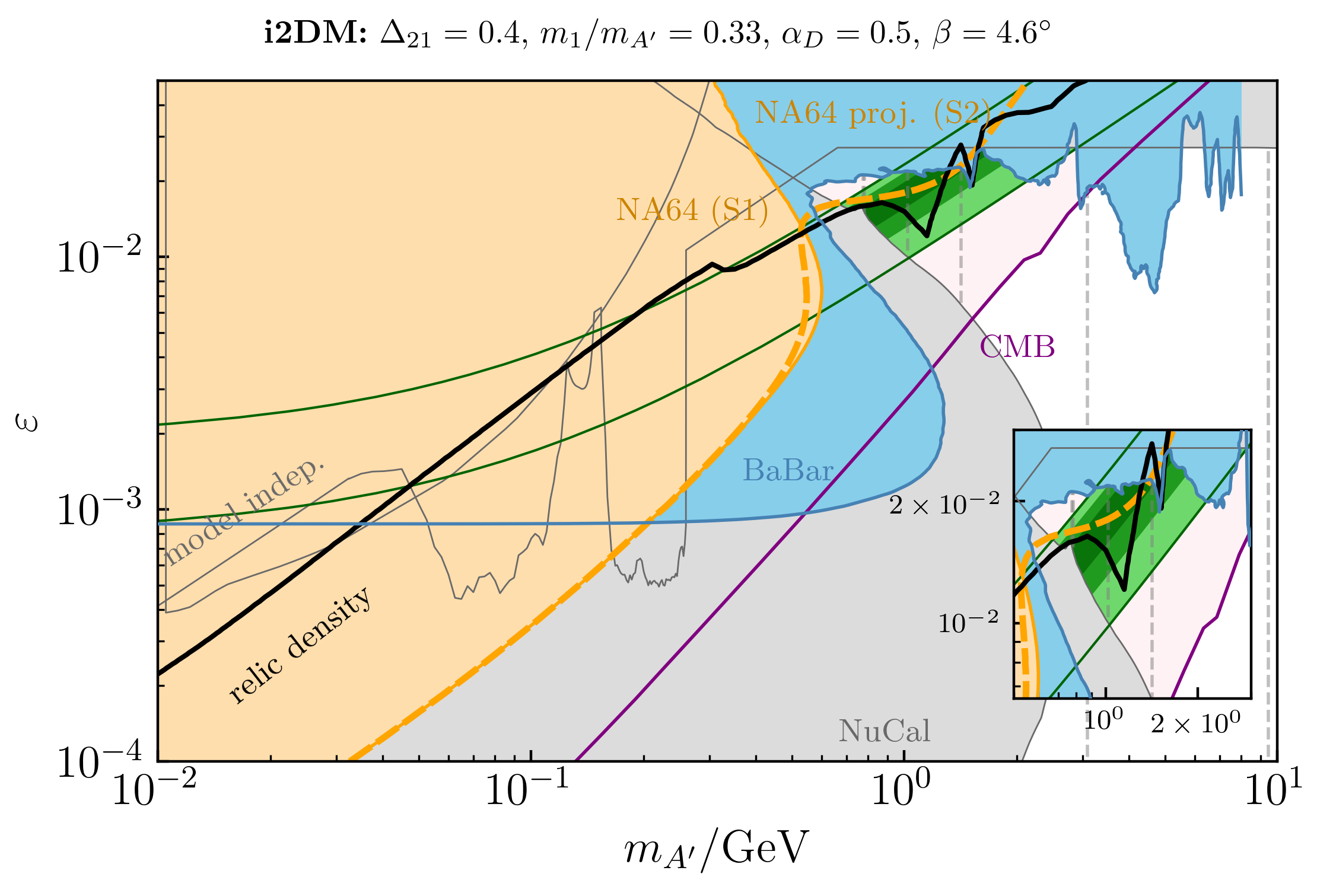}
    \includegraphics[width=\columnwidth]{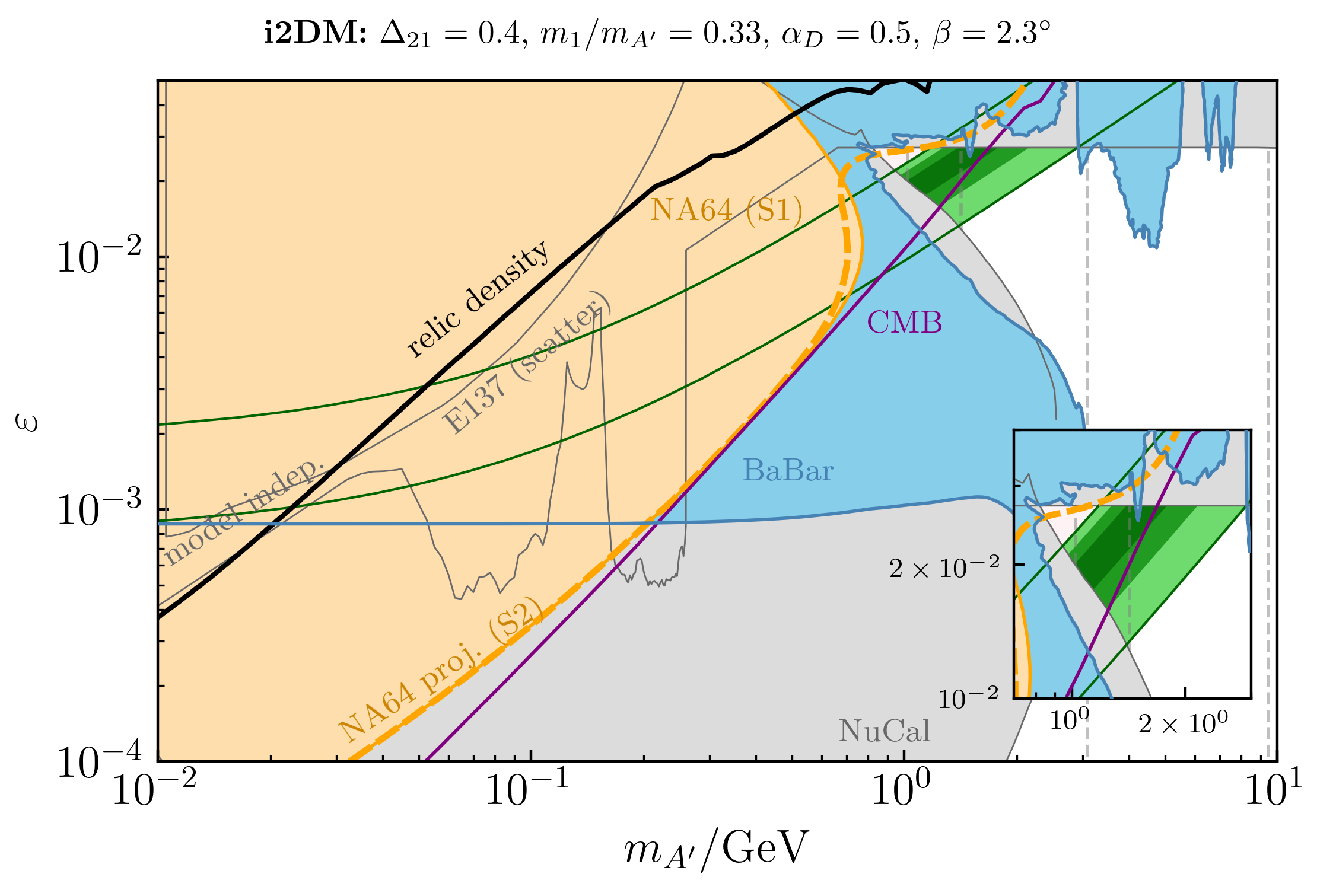}
    \includegraphics[width=\columnwidth]{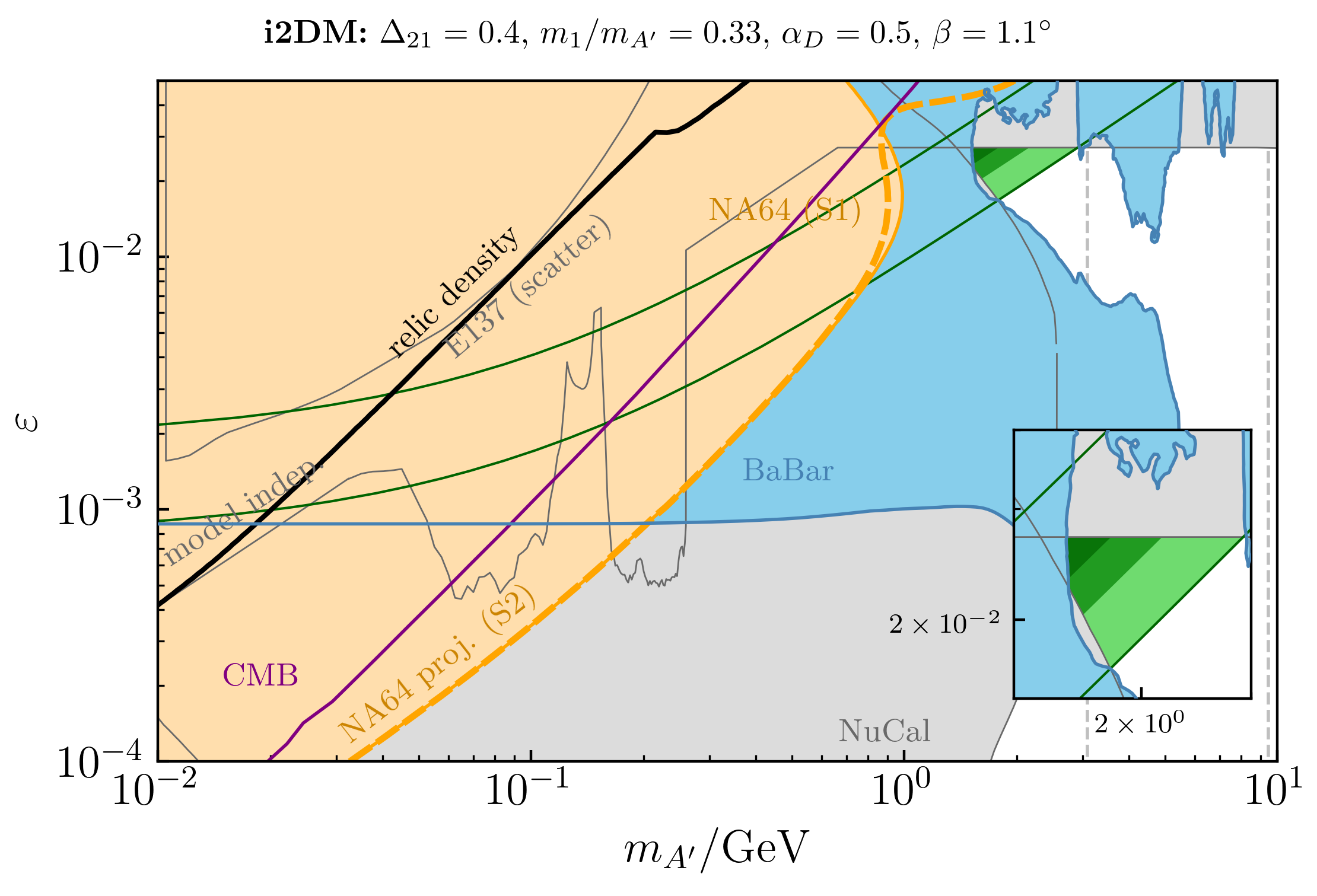}
    \caption{
    Same as \cref{fig:recast_1} but for BP3a (top left) and BP3b (top right), BP3c  (bottom left), and BP3d (bottom right), corresponding to the inelastic Dirac dark matter (i2DM) model.  
    The four panels represent the same choice of parameters, varying solely the mixing angle, $\beta$. 
    \label{fig:recast_3}
    }
\end{figure*}

\myparagraph{Inelastic Dirac Dark Matter (BP3a--d)}
We show the constraints for the mixed-iDM model in \cref{fig:recast_3}, expressed as $\varepsilon$/$m_{A^\prime}$ constraints.

Similarly to the previous model, the constraints are relaxed, and a new region of the parameter space opens up.
This model is characterized by an enhanced dark photon decay rate to the channel $\psi_2\psi_2$ as it happens in the case of the mixed-iDM model.
The rate into the channels $\psi_2\psi_1$ and $\psi_1\psi_1$ is suppressed by respectively a factor $\beta$ and $\beta^2$.
Differently from the mixed-iDM case, this model allows for dark photon decays to $\psi_1\psi_1$ channel.
A larger branching ratio to the heaviest HNF increases the possibility to detect a semi-visible event in the detector, given the larger abundance of those particles releasing $e^+e^-$ pairs after their decay.

The relic density of $\psi_1$ for this model depends on the efficiency of coannihilations and coscattering processes.
The main difference with respect to the previous model is that it is not possible to evade the CMB bounds, because of the possibility for the dark matter candidate $\psi_1$ to annihilate through the vertex $\psi_1\psi_1$.
Even though this vertex is suppressed, it can have a sizable contribution to late time annihilations, injecting additional energy into the CMB.
In the realizations shown in \cref{fig:recast_3}, we find that, despite the relaxation of the main constraints, the CMB bounds are unavoidable and can exclude large parts of the parameter space, with the only exception being the choice of a small $\beta$ parameter, as represented by benchmarks BP3c and BP3d.
This choice corresponds to suppress further the channels $\psi_2\psi_1$ and $\psi_1\psi_1$.
However, it has an impact on the relic abundance of $\psi_1$, because it suppresses the contributions of coscattering and annihilations, resulting in an overabundant scenario, which can be set under control only assuming secluded annihilations within the DS.

\begin{figure*}[th]
    \centering
    \includegraphics[width=0.9\textwidth]{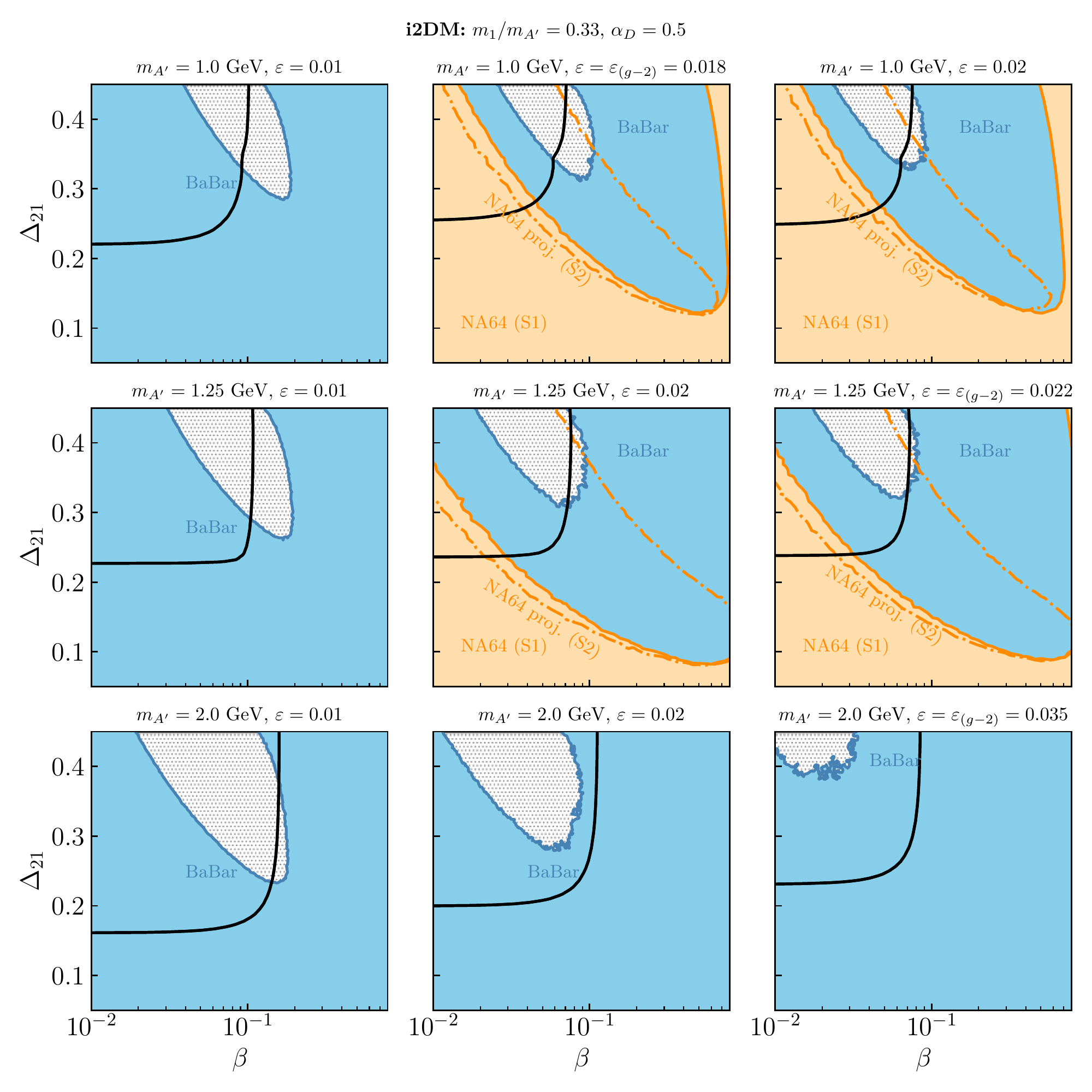}
    \caption{
    Same as \cref{fig:recast_1} but for BP3, corresponding to the inelastic Dirac dark matter (i2DM) model.
    The parameter space is shown as a function of $\Delta_{21}$ and the mixing angle $\beta$.
    The region dotted in black is excluded by CMB limits, providing a full exclusion of these slices of parameter space.
    \label{fig:delta_vs_theta_3}}
\end{figure*}

\begin{figure*}[t]
    \centering
    \includegraphics[width=\columnwidth]{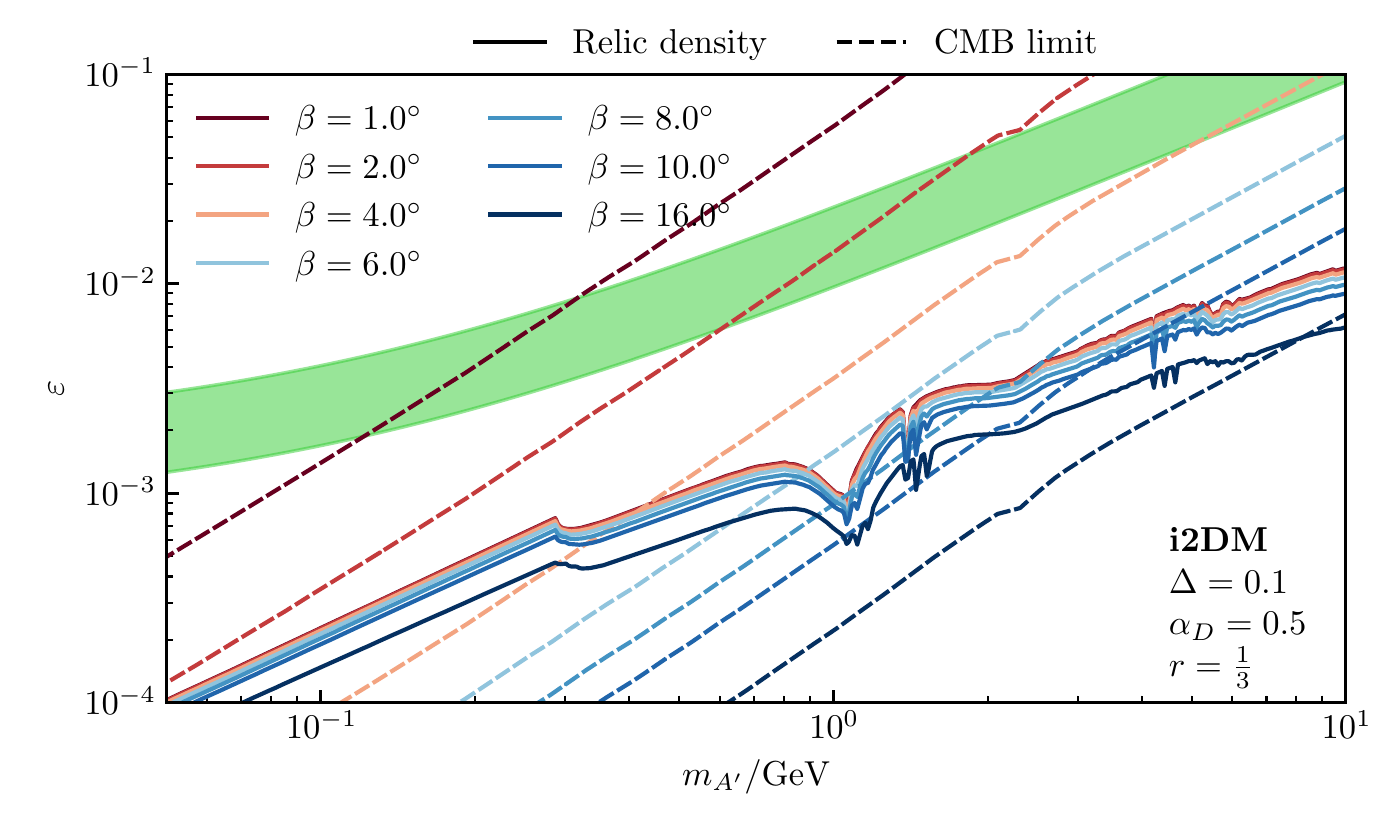}
    \includegraphics[width=\columnwidth]{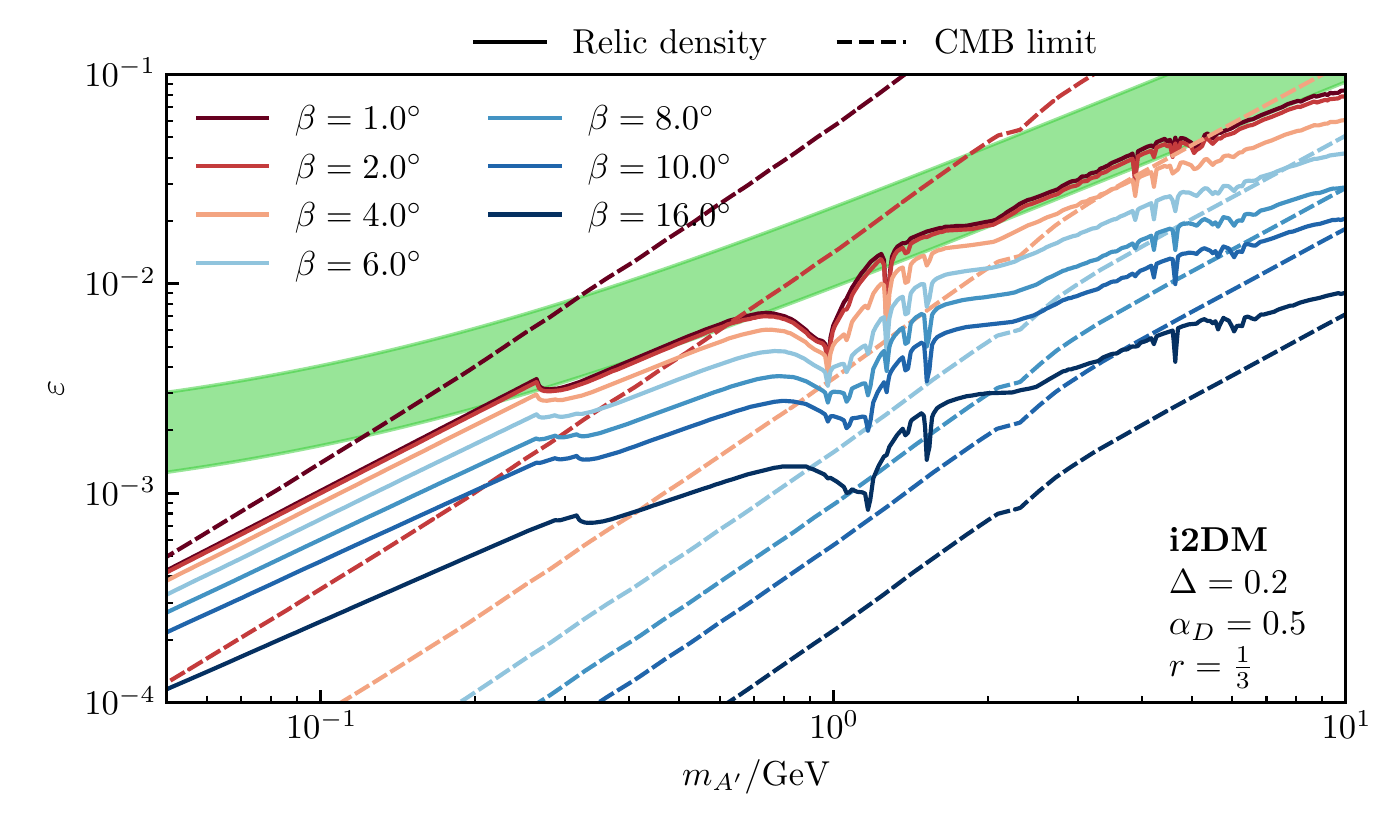}
    \includegraphics[width=\columnwidth]{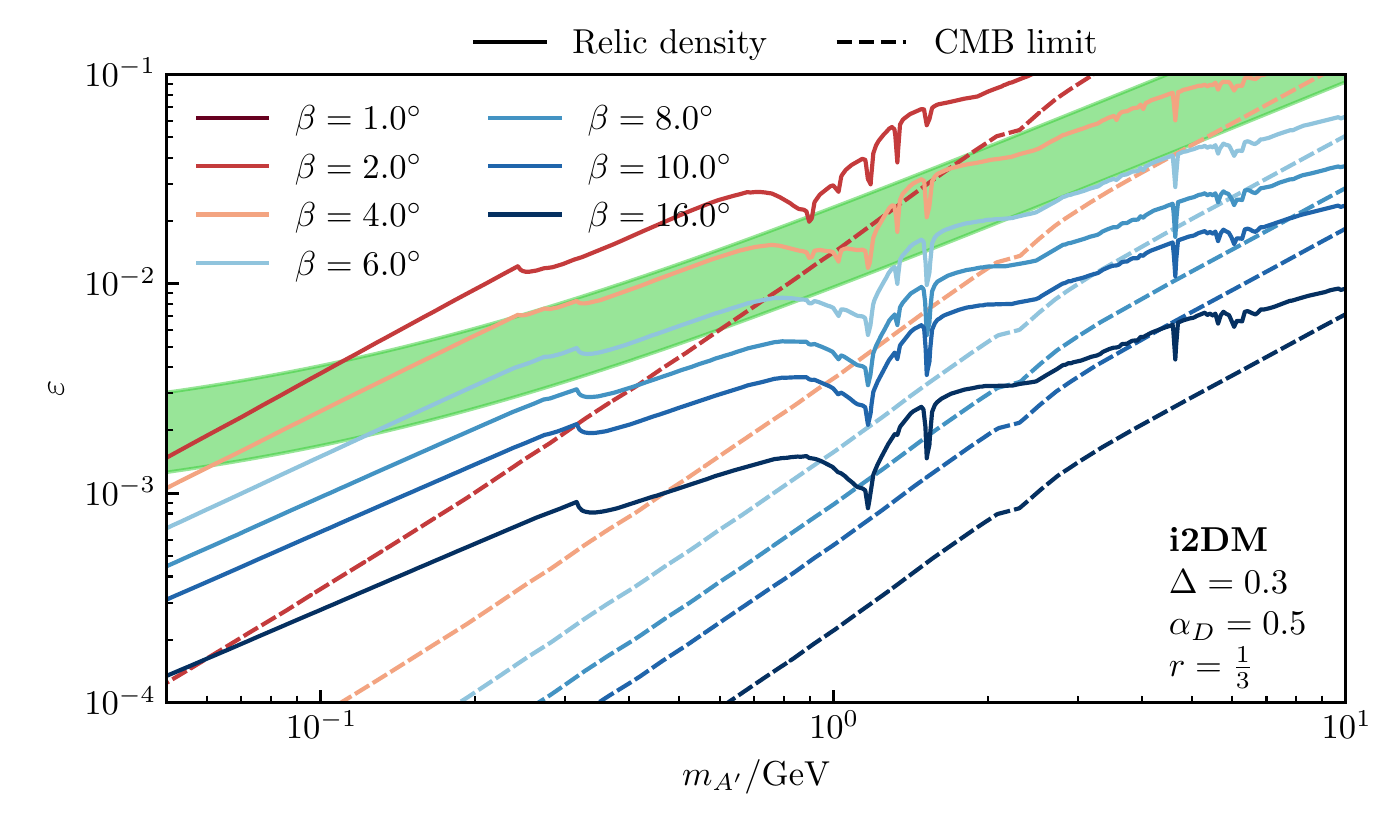}
    \includegraphics[width=\columnwidth]{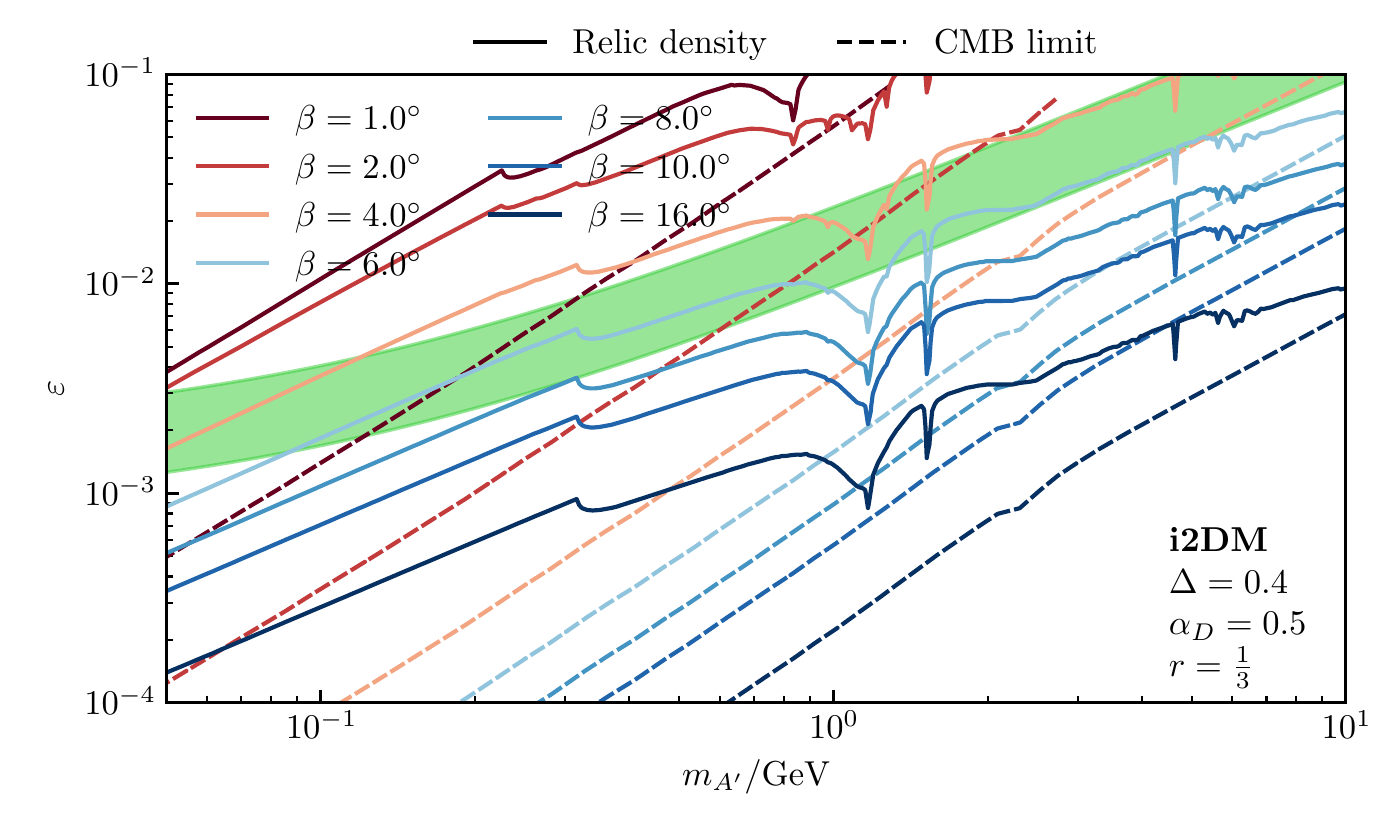}
    \caption{
    The relic density (solid) and CMB limits (dashed) in the parameter space of inelastic Dirac Dark Matter (i2DM) in the plane of kinetic mixing $\epsilon$ versus dark photon mass $m_{A^\prime}$ for fixed values of $\alpha_D = 0.5$, 
    $r = m_1/m_{A^\prime} = 1/3$, and $\Delta_{21}$.
    Each curve corresponds to a different value of the mixing angle $\beta$.
    The CMB limits exclude large values of $\epsilon$. The $\deltaamu$ $3\sigma$ preferred region is shown as a green band.
    \label{fig:cmb_relic_i2DM}}
\end{figure*}

Additionally, the relic density depends also on $\Delta_{21}$, because it affects the Boltzmann suppression of the coannihilation $\psi_2$.
It is interesting to understand the interplay of $\Delta_{21}$ and $\beta$ in the different constraints.
In \cref{fig:delta_vs_theta_3}, the bounds are shown in a $\Delta_{21}/\beta$, fixing the mass of the dark photon to $m_{A^\prime} = 1,\, 1.25,\, 2~\mathrm{GeV}$ and $\epsilon = 0.01,\, 0.02,\, \epsilon_{(g-2)}$.
We can draw similar conclusions as the ones discussed for the Mixed-iDM model in \cref{fig:delta_vs_theta_2}, with the difference that the $x$-axis now represents the parameter $\beta$.
In addition, the CMB constraints are shown: their exclusion region corresponds to large $\beta$ values, because of the enhancement of $\psi_1\psi_1$ annihilation rate.
Regarding the relic density, we can draw a similar conclusion as for benchmarks BP2a and b based on \cref{fig:delta_vs_theta_2} and \cref{fig:cmb_relic_i2DM}. 
The overabundant region corresponds to large $\Delta_{21}$ and low $\beta$, due to both inefficient coannihilations and suppressed $\psi_1$ self-annihilations.
However, the dependence on $\beta$ is stronger due to the presence of $\psi_1$ self-annihilations, which can dominate over coannihilations in depleting the DM density.
For this reason, the relic density is underabundant for large $\beta$ independently on the value of $\Delta_{21}$: the coannihilator does not play a crucial role anymore in the determination of the dark matter relic density.
Nevertheless, the small regions in which dark matter density is underabundant and compatible with other constraints are excluded by the CMB bounds.

\begin{figure*}[th]
    \centering
    \includegraphics[width=\columnwidth]{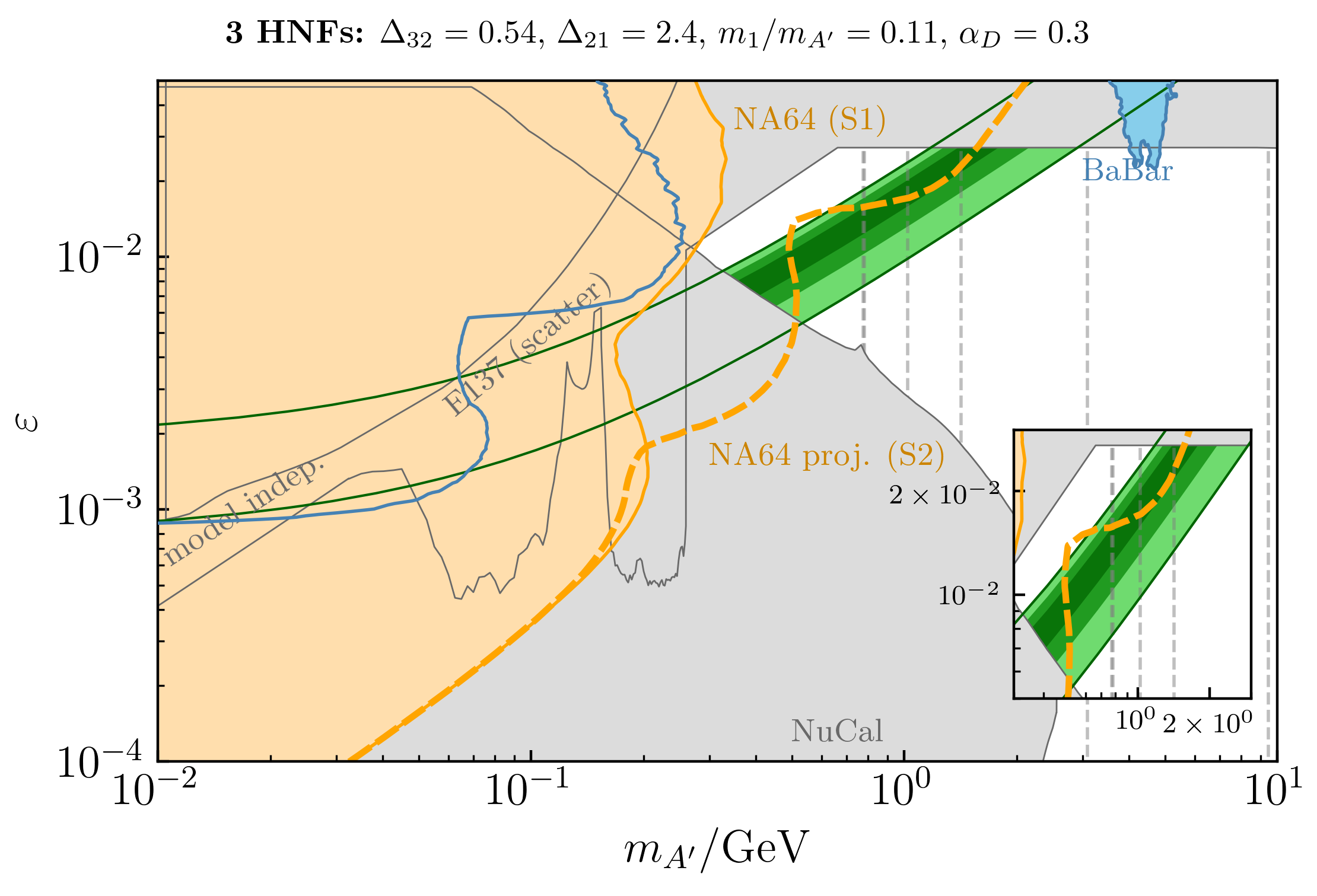}
    \includegraphics[width=\columnwidth]{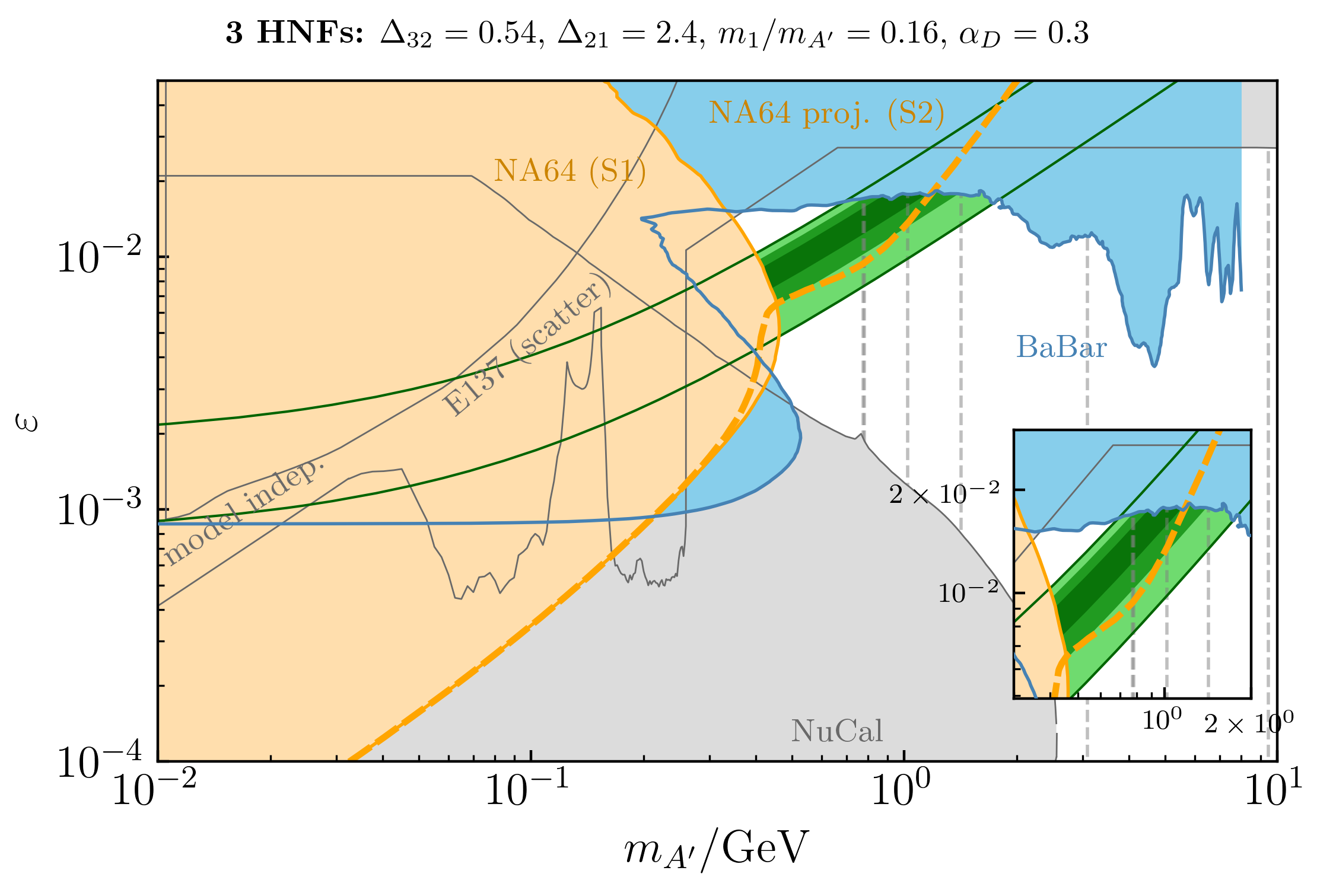}
    \includegraphics[width=\columnwidth]{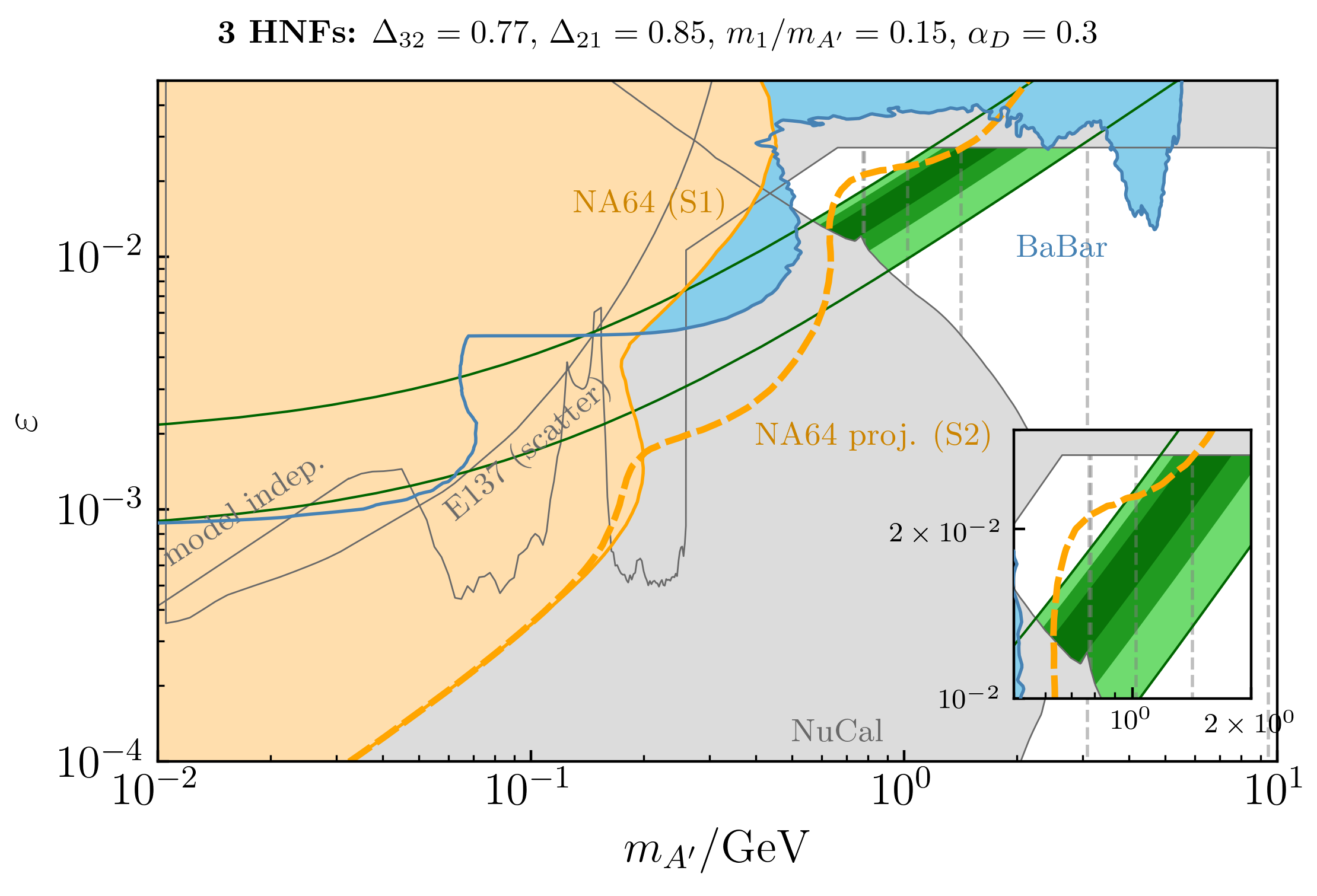}
    \includegraphics[width=\columnwidth]{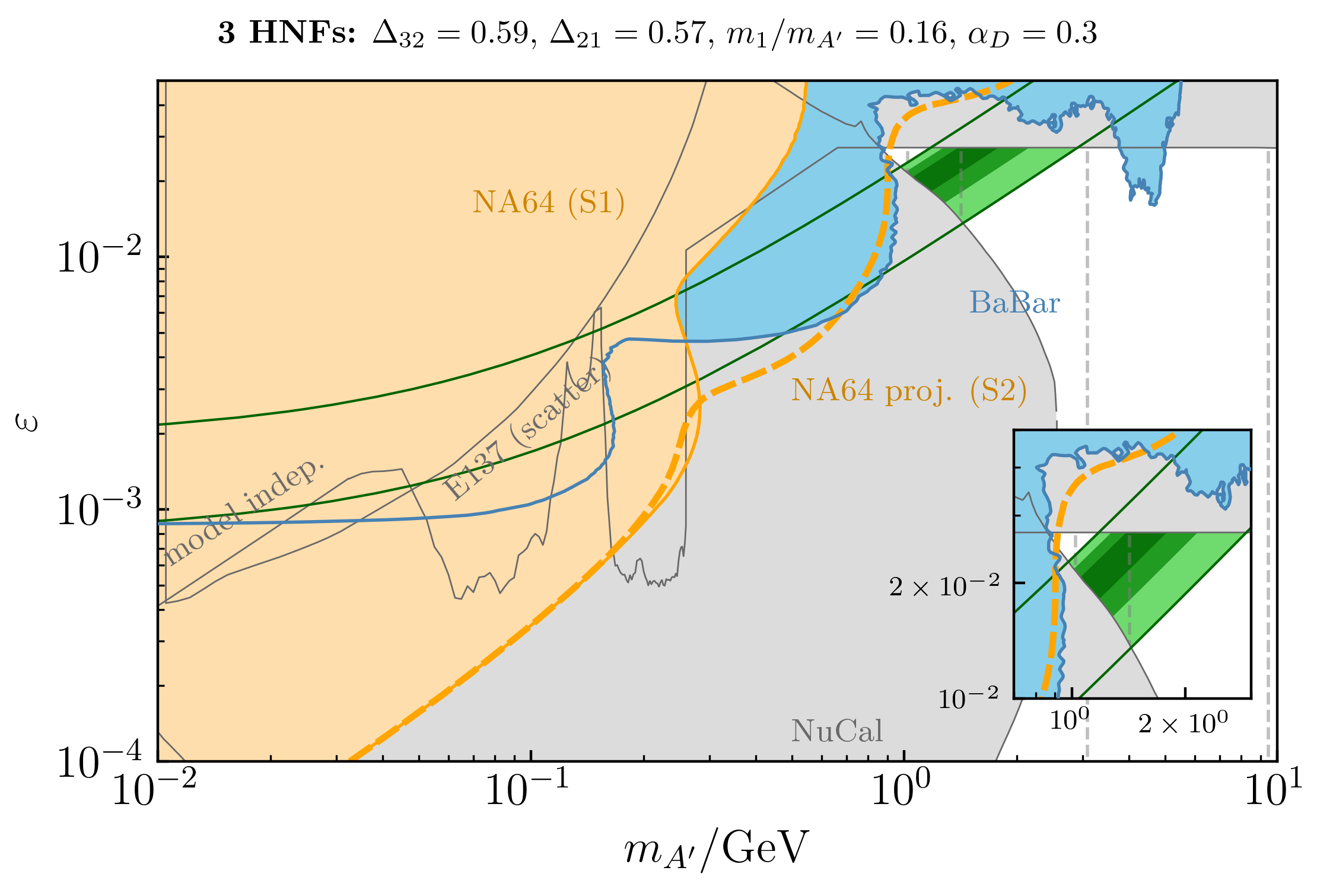}
    \caption{Same as \cref{fig:recast_1} but for BP4a (top left), BP4b (top right), BP4c (bottom left), BP5 (bottom right), corresponding to the models with 3 heavy neutral fermions.
    The dark photon branching ratios are dominated by $A^\prime \to N_2 N_3$ decays.
    \label{fig:recast_3HNFs}
    }
\end{figure*}

\myparagraph{Three Heavy Neutral Fermions (BP4a--c, BP5)} 
We show the constraints for the Three Heavy Neutral Fermions models in \cref{fig:recast_3HNFs}, expressed as $\varepsilon$/$m_{A^\prime}$ constraints.

Similarly to the previous cases, the constraints coming from both BaBar and NA64 are relaxed, and a new region of the parameter space opens up.
All benchmarks are characterized by having three HNFs, and by a sizable $V_{32}$ coupling, which enhances the dark photon decay rate to $N_3 N_2$ final states.
Furthermore, the produced HNFs can promptly decay, releasing $e^+e^-$ pairs in the detectors.
The presence of a new fermion, and the enhanced annihilation rate to the heaviest HNFs make it possible to have a larger number of $e^+e^-$ pairs, and, consequently, a larger probability of detecting a semi-visible event.
The main difference between the models is that BP4a--c are characterized by only off-diagonal couplings, with the only possible decay chains being:
\begin{align*}
    A^\prime &\to (N_2 \to N_1 e^+ e^-) \, (N_3 \to (N_2 \to N_1 e^+ e^-) e^+ e^-),
    \\
    A^\prime &\to (N_2 \to N_1 e^+ e^-) N_1.
\end{align*}
Differently, BP5 allows for any possible coupling among the HNFs.

The downward shift of the BaBar bound happening between BP4a and BP4b benchmarks is due to the different values assumed by the parameter $r = m_1/m_{A^\prime}$.
This parameter affects both the HNF lifetime and the $A^\prime$ branching ratios.
The HNF lifetime depends on $r$ according to $c\tau \sim m_{A^\prime}^{-1} r^{-5}$, so a larger value would imply a shorter lifetime, translating into a more relaxed bound, because the potential larger fraction of events releasing energy inside the instrumented regions of the detector.
However, as it can be observed, the bound becomes stronger from BP4a to BP4b, even with a larger $r$.
The new value modifies the branching ratio of the dark photon decay and forbids the channel $A^\prime \to N_3 N_2$, because the kinematics requires:
\begin{equation}
    m_{A^\prime} > m_2 + m_3 \Rightarrow r < \frac{1}{(\Delta_{32} + 2) (\Delta_{21} + 1)} \approx 0.116,
\end{equation}
which is satisfied for BP4a, but not for BP4b.
Being the decay to the two heavy HNFs forbidden means that the potential production of $e^+e^-$ pairs is suppressed, because $A^\prime$ can decay only to $N_2 N_1$, and only $N_2$ can decay further.

The constraints show that the NA64 (S2) projected bound has the capability of excluding new regions of the parameter space, demonstrating the capability of the experiment to be sensitive to promptly decaying HNFs.

\section{Prospects and constraints on HNL and DM interpretations}
\label{sec:discussion}

\subsection{HNFs as dark matter candidates}

The U$(1)_D$ symmetry can be responsible for the stability of the lightest HNF, and, therefore, provide a dark matter candidate.
Light dark matter models with self-annihilating Dirac fermions are excluded by CMB data due to the s-wave, velocity-unsuppressed annihilation.
Majorana fermions or scalar particles have p-wave, velocity-suppressed, and self-annihilations; however, in this case, the required values of self-interactions render the dark photon fully invisible, and, therefore, excluded as an explanation of $\deltaamu$.
 Self-annihilation near the $A^\prime$ resonance, $r = m_1/m_{A^\prime} \lesssim 1/2$, can significantly enhance cross sections, but such mass spectrum would leave no room for semi-visible, promptly-decaying fermions. 

Coannihilations, $\psi_1 \psi_{2,3}\to (A^\prime)^* \to f^+f^-$, are therefore the most natural possibility to achieve freeze-out. 
The coannihilation cross section of opposite $C$ states, $\psi_i$ and $\psi_j$, is given by
\begin{widetext}
\begin{equation}\label{eq:sigmav}
    \sigma v (\psi_i \psi_j \to f^+f^-) = 8\pi \alpha_D \alpha \epsilon^2 |V_{ij}|^2 \frac{(2 m_{ij}^2+m_f^2)}{(4m_{ij}^2-m_{A^\prime}^2)^2+m_{A^\prime}^2\Gamma_{A^\prime}^2}\sqrt{1 - \frac{m_f^2}{m_{ij}^2}} + \mathcal{O}(v^2),
\end{equation}
\end{widetext}
where $m_{ij} = (m_i+m_j)/2$. 
Just like the self-annihilation of Dirac fermions, the cross section is velocity unsuppressed.
Nevertheless, the annihilation is exponentially suppressed at late times due to the mass splitting between the two states and the subsequent decays of the co-annihilator.

To calculate the DM relic density of $\psi_1$ in our benchmarks, we assume that all dark sector fermions are in chemical equilibrium at the time of freeze-out, and employ the formalism of Ref.~\cite{Griest:1990kh}.
We find good agreement with the literature on iDM~\cite{Duerr:2019dmv,Berlin:2020uwy} and i2DM~\cite{Filimonova:2022pkj}.
We find a $50\%$ disagreement with the relic curves of Ref.~\cite{Izaguirre:2015zva} for $m_{1} \gtrsim 100$~MeV.

\myparagraph{Direct detection} 
Direct detection of a dark matter particle of mass $m_\chi \sim \mathcal{O}(100)$~MeV, with large kinetic mixing, would provide strong evidence for the DM nature of the HNFs.
For the parameter space we consider in iDM and mixed-iDM models, low-energy direct detection can only probe the loop-induced elastic scattering of DM.
The tree-level upscattering rates are exponentially suppressed as only the largest DM velocities can overcome the kinematical threshold of the large mass splittings.
Direct detection prospects are instead dominated by the loop-induced, elastic DM-quark coupling.

In the case of kinetic mixing, the scalar-current dominates, $c_5^q (\overline{\chi}\chi)(\overline{q}q)$~\cite{Bell:2018zra}.
In terms of coupling to nucleons,
\begin{equation}
    \sigma_{\chi N}^{1{\rm -loop}} = \frac{(c_1^N \mu_{N})^2}{\pi}
\end{equation}
where $\mu_{N} = m_\chi m_N/ (m_{\chi}+m_N)$ is the reduced mass of the DM and the nucleon $N$, and $c_1^N$ is the loop-induced, DM-nucleon coupling.
The matching to nucleon currents gives~\cite{Bishara:2017nnn}
\begin{equation}
    c_{1}^N = 4 \frac{\alpha_D \alpha_{\rm QED} \varepsilon^2}{m_{A^\prime}^4} m_{\chi} F_{3}(r^2) \sum_q F_{S}^{q/N} Q_q^2 \sum_{i}|V_{1i}|^2,
\end{equation}
where $Q_q$ is the quark electric charge, $r = m_\chi/m_{A^\prime}$, $F_{3}(x)$ is the loop function from~Ref.~\cite{Bell:2018zra}, and $F_{S}^{q/N}(Q^2)$ the nucleon scalar form factors.
Approximating the form factors to their $Q^2=0$ value and $F_{S}^{q/p}(0) \simeq F_{S}^{q/n}(0) \simeq (15, 35, 40)$~MeV for $q = (u,d,s)$, we estimate the elastic DM-nucleon cross sections.
At a typical point of parameter space,
\begin{equation}
    \sigma_{\chi N}^{1-{\rm loop}} \simeq 1.4 \times 10^{-3} \text{ pb} \times \left(\frac{\alpha_D}{0.3}\right)^2\left(\frac{\epsilon}{10^{-2}}\right)^4
\end{equation}
where we assumed $m_{A^\prime} = 3 m_\chi = 1$~GeV and $ \sum_i |V_{1i}|^2=1$.
This value is not currently probed by any low-energy direct detection experiments.
In this mass region, CREST-III (2019)~\cite{CRESST:2019jnq} provides the leading limits on elastic DM-nucleus scattering using nuclear recoils.
Those limits are over two orders of magnitude above our estimate.
The next-generation SuperCDMS detectors at SNOLAB may be able to probe part of the parameter space of interest using nuclear recoils~\cite{SuperCDMS:2016wui,SuperCDMS:2022kse}.
A more promising avenue in sub-GeV DM direct detection, however, is the use of nuclear-inelastic processes.
In this case, DM can impart all of its kinetic energy into excitating a target nucleus, which  subsequently de-excited emitting an electron through the Migdal effect, or a photon.
The electron recoil, in this case, can significantly improve the prospects for sub-GeV DM direct detection~\cite{Essig:2011nj,Kouvaris:2016afs,Ibe:2017yqa,Berghaus:2022pbu}.
This method has been used by the LUX~\cite{LUX:2018akb}, SENSEI~\cite{SENSEI:2020dpa}, XENON1T~\cite{XENON:2019zpr}, and DarkSide-50~\cite{DarkSide:2022dhx} collaborations to set limits in our mass region of interest.
The best constraints come from DarkSide, where $\sigma_{\chi N} \lesssim 0.1$~pb at $m_\chi = 300$~MeV at 90\% C.L.

Unfortunately, the loop-induced scattering on electrons is very suppressed in iDM and mixed-iDM models due to the small electron mass.

In the i2DM model, direct detection is sensitive to the mixing-angle-suppressed elastic scattering of $\chi\equiv\psi_1$ on electrons.
For a heavy dark photon, the total cross section is
\begin{align}
    \overline{\sigma}_{e} &= \frac{16 \pi \sin^4{\beta} \alpha_D \alpha_{\rm QED}\epsilon^2 \mu_e^2}{m_{A^\prime}^4}
    \\\nonumber &\simeq 
    4\times 10^{-7} \text{\,pb} \times \left(\frac{\sin{\beta}}{0.14}\right)^{\!\!4} \!\!\left(\frac{\alpha_D}{0.3}\right)\!\left(\frac{\epsilon^2}{10^{-2}}\right)\!\left(\frac{1\text{ GeV}}{m_{A^\prime}}\right)^{\!\!4}\!\!.
\end{align}
The leading limits in this parameter space are from XENON1T~\cite{XENON:2019gfn,XENONCollaborationSS:2021sgk}, PandaX-II~\cite{PandaX-II:2021nsg}, and SENSEI~\cite{SENSEI:2020dpa}.
At $m_\chi \sim 100$~MeV, XENON1T constrains $\overline{\sigma}_{e}\lesssim 10^{-4}$~pb, still orders of magnitude above our estimates for $\beta = 8^\circ$.
The prospects are more interesting for scattering on protons, where the bounds discussed just above apply as well.
In this case, DarkSide-50 already probes the largest values of kinetic mixing and $\beta$ for $m_{A^\prime} \gtrsim 1.5$~GeV.
However, these are already excluded by BaBar and CMB constraints.

Another possibility for direct detection is to search for a boosted DM population~\cite{Bringmann:2018cvk,Ema:2018bih,Ema:2020ulo,Yin:2018yjn}.
Cosmic rays can interact with the DM background, upscatter $\chi \to \psi_{2,3, \dots}$, which subsequently decay to fast DM particles.
This cosmic-ray-boosted DM population can then be searched for in direct detection and neutrino experiments.
Refs.~\cite{Bell:2021xff,Feng:2021hyz} derive limits on similar models using XENON1T data, from where we can conclude that current limits are still too weak to constrain our parameter space, in all models of interest. Large neutrino detectors can further enhance the sensitivity thanks to their large mass and excellent detector performance~\cite{Ema:2018bih}.
A more detailed study is needed to assess the flux of boosted DM particles in our models and their testability via this strategy.

\myparagraph{Cosmic Microwave Background}
Precision measurements of the Cosmic Microwave Background (CMB) also provide significant limits on the models we consider when the HNFs are dark matter.
If the dark matter fermions significantly annihilate or decay to charged particles at the time of recombination, they can inject additional energy into the SM plasma, re-ionize Hydrogen, and delay the formation of the CMB~\cite{Slatyer:2009yq,Madhavacheril:2013cna,Slatyer:2015jla,Slatyer:2015kla,Liu:2016cnk}.
The latest constraints from Planck~\cite{Planck:2018vyg} rule-out light and thermal dark matter candidates with s-wave annihilations for $m_{\chi}\simeq $.
This constraint is much weaker and, therefore, not significantly constraining for models with co-annihilating dark matter candidates, like iDM and mixed-iDM, and in models with p-wave annihilations, like the case of Majorana dark matter fermions.

Out of the models and mass splittings we consider, only the i2DM model is subject to such constraints. 
This is because the light Dirac dark matter fermion can undergo velocity-unsuppressed self-annihilations, even if suppressed by the fourth power of a small mixing angle $\beta$.
The self-annihilation cross section to charged leptons in the i2DM model is given by \cref{eq:sigmav}, with $m_{ij} \to m_1$ and $V_{ij} \to \beta^2$.
The curves where the correct relic density of DM can be achieved are compared with the limits from CMB in \Cref{fig:recast_2ab}.
For the typical lifetimes and mass splittings considered in this work, the late-time annihilations involving $\psi_2, \psi_3, \dots$ can be safely neglected.

\subsection{HNFs as heavy neutral leptons}
\label{sec:HNL_pheno}

Having discussed the theoretical aspects of a HNL interpretation in \cref{sec:HNL_theory}, we now turn to the phenomenological consequences. Searches of HNLs require mixing with active neutrinos, which emerges from the Yukawa coupling between the sterile neutrinos and the leptonic doublet after symmetry breaking.
While a successful $\deltaamu$ explanation does not lead to any constraint on this mixing, the latter will be constrained from below by BBN, such that $\tau_{N_4} \lesssim 0.1$~s, and from above by laboratory searches. 
As highlighted in Refs.~\cite{Ballett:2019pyw,Abdullahi:2020nyr}, the phenomenology of the models under consideration can be very different from the minimal case.  

For most of the parameter space of interest in this paper, the heavier HNLs will decay very fast into lighter HNLs and dark photons, into 3 lighter HNLs, and into HNLs and dilepton pairs, depending on which channels are kinematically allowed. We focus on a hierarchy of HNF and dark photon masses such that the latter decay dominates, allowing to evade BaBar bounds as discussed in the previous section. 
The lightest HNL decays primarily into a dilepton pair and missing energy.

Thanks to the presence of a light dark photon both HNL scattering and decays are enhanced compared to the standard case in which HNL interact with the SM via mixing with the neutrinos.

HNLs are tested experimentally mainly via peak searches and via visible decays.
In the former case, the emission of a HNL is a pion or kaon decay leads to a small peak in the charged lepton spectrum at a lower energy. These bounds are very robust as they rely uniquely on the kinematics of the meson decay and pose some of the strongest constraints in the sub-GeV HNL mass region~\cite{NA62:2020mcv}. 
Similarly, for HNLs coupled exclusively to the tau flavor, peak searches in $\tau$ and $D$ decays, such the recent BaBar analysis~\cite{BaBar:2022cqj}, provide strong limits.
In the models we propose, even peak searches can be affected as, for very fast decays, these events would be vetoed by the requirement of no additional charged particles. A weakening of the bounds can be expected in certain ranges of parameter space. 
A more detailed discussed is provided in Ref.~\cite{Abdullahi:2020nyr}.

GeV-scale HNLs can be produced via mixing in meson decays and in neutrino scatterings, typically in beam dump and neutrino experiments. In the first setup, high energy protons impinge on a target producing copious amounts of pions, kaons and, for sufficiently high energies, heavier mesons, which subsequently decay producing HNLs. These travel some distance before decaying into missing energy and visible particles that can be revealed in dedicated or multi-purpose neutrino detectors. 
Due to the kinetic mixing and light dark photon mass,$\Gamma^{A^\prime}_{N_4 \to \nu e^+e^-} \gg \Gamma^{\rm W,\,Z}_{N_4\to {SM}}$, where $\Gamma^\textup{Md}$ is the decay rate mediated by the particle $\textup{Md}$.
There are two cases: in the long-lived regime, $c \tau_4^{\rm LAB} > L$ with $L$ the baseline of the experiment, the event rate of $N_4$ decays in DIF searches can be enhanced as it scales as $\Gamma^{A^\prime}_{N_4 \to \nu e^+e^-}/L$ and the bounds get stronger. Alternatively, if $N_4$ is too short-lived, $c \tau_4^{\rm LAB} \ll d$, with $d$ being the distance between the source and the detector, the limits do not apply at all as the HNLs do not even reach the detector.

The strongest limits on $U_{e4}$ and $U_{\mu 4}$ are set by T2K~\cite{T2K:2019jwa} and MicroBooNE~\cite{Kelly:2021xbv} (see also PS191 limits~\cite{Bernardi:1987ek} and the discussion in Refs.~\cite{Arguelles:2021dqn,Gorbunov:2021wua}), while $U_{\tau 4}$ is only constrained by higher-energy experiments like CHARM~\cite{CHARM:1985nku} (see also~\cite{Orloff:2002de,Boiarska:2021yho,Barouki:2022bkt}), NOMAD~\cite{NOMAD:2001eyx}, and LEP~\cite{DELPHI:1996qcc,L3:1999ymc,L3:2001zfe}. 
As discussed, these bounds need to be revisited in the light of the considerations above and depend critically on the choice of parameters. A more in depth discussion is available in Ref.~\cite{Abdullahi:2020nyr}.

In the second type of setups, HNLs are produced by a neutrino beam via upscatterings in the detector itself and subsequently decay leading to visible signatures. In this case, the upscattering cross-section can be very significantly higher than the standard HNL one. This, combined with much shorter decay lengths, can lead to striking signatures in neutrino experiments with short baselines, such as at the SBN programme at Fermilab, and at near detectors of long baseline accelerator neutrino experiments, including T2K, NoVA, DUNE. 

A particularly interesting signature is the electron-positron pairs from HNLs decays produced by neutrino upscattering in the MiniBooNE experiment. It has been shown that for suitable values of the parameters, this signature can explain the anomalous excess events at MiniBooNE~\cite{Abdullahi:2020nyr}, as well as the $(g-2)_\mu$ anomaly. The particle $\psi_2$ could be efficiently produced and decay into a dilepton pair which can mimic an electron neutrino scattering, as either the two leptons are very collimated or one of them is not reconstructed being too soft~\cite{Ballett:2018ynz}. This explanation critically relies on the large values of kinetic mixing that are allowed in our models. Specifically, in order to explain the MiniBooNE anomalous excess, it is necessary for the HNLs to decay within the detector, that has a typical size of few meters. Decay lengths $\tau_{HNL}\ll 1$~m cannot be obtained in the standard HNL scenario and require light dark photons and large kinetic mixing.

\subsection{Prospects for detection} \label{sec:future_prospects}

In this subsection, we discuss future prospects for the detection of semi-visible dark photons and HNFs.

\myparagraph{Low-energy $e^+e^-$ colliders}
Direct tests of the parameter space present in our work can be achieved with a dedicated semi-visible search at BaBar, KLOE, BES-III, and Belle-II $e^+e^-$ colliders.
These searches can be divided into two categories.
On-shell production of dark photon through initial state radiation, $e^+e^- \to A^\prime \gamma$, or the production of HNFs through off-shell dark photons, $e^+e^- \to (A^{\prime})^* \to \psi_i \psi_j$.
In particular, low-energy machines like KLOE and BES-III ran with center-of-mass energies close to the dark photon mass, significantly enhancing their prospects for direct production of HNFs.

\emph{Initial State Radiation ---} One advantage of keeping the ISR topology is the kinematic imbalance in the photon-dark-photon center-of-mass system.
Since multiple invisible particles are emitted in the decay cascade, it is not possible to reconstruct the dark photon mass with visible energy.
However, by isolating the photon, a resonance on $M_X^2 = s - 2E_\gamma \sqrt{s}$ would still be visible.
A detailed sensitivity study of the Belle-II reach to iDM through this channel was carried out in Refs.~\cite{Duerr:2019dmv,Duerr:2020muu}.
We expect the sensitivity to be even better in models with two or more HNF decays and leave a detailed study for mixed-iDM and HNL models to future literature.

\emph{S-channel Production ---} Unlike the ISR channel, the s-channel production cross section is proportional to the dark coupling $\alpha_D$, and so can be large for models where $\alpha_D \gg \alpha$~\cite{Abdullahi:2020nyr,Kang:2021oes}.
In terms of the cosine of the angle of $\psi_i$ with respect to the beam in the COM, $c_\theta$, the differential cross section is given by,
\begin{widetext}
\begin{equation}
    \frac{\dd \sigma}{\dd c_\theta} = |V_{ij}|^2 \frac{\pi \alpha_D \alpha_{\rm QED} \epsilon^2}{(E_{\rm CM}^2 - m_{A^\prime}^2)^2 + \Gamma_{A^\prime}^2 m_{A^\prime}^2 } \frac{E_{\rm CM}^2}{2}  
    \left[
          1 + \left( c_\theta - (1 - c_\theta) \frac{m_i^2\Delta_{ij}(2 + \Delta_{ij})}{E_{\rm CM}^2} \right)^2
    \right],
 \end{equation}
 \end{widetext}
 where $\Delta_{ij} = (m_i - m_j)/m_j$, $E_{\rm CM} = \sqrt{s}$ is the center-of-mass collision energy, and $\Gamma_{A^\prime}$ is the total decay width of the dark photon.
 Unlike ISR events, where the displaced HNF vertices would depend on the direction of travel of $A^\prime$, s-channel production would provide a source of back-to-back displaced vertices.
 The dileptons would not point allow pointing back to the collision point due to the missing energy.
 In addition, for secondary decays, like $\psi_3 \to \psi_2 \to \psi_1$, a third, lower-energy dilepton pair could be visible, keeping the two primary decay vertices and the collision point on the same line. 
 In Ref.~\cite{Kang:2021oes}, the authors have explored these events in iDM models, finding that Belle-II can cover open regions of parameter space.
 The sensitivity reaches values of $\epsilon <10^{-3}$ for dark photon precisely in the region of interest for $\deltaamu$, $m_{A^\prime} \gtrsim 500$~MeV.
The case of mixed-iDM and i2DM have not been studied, but the additional semi-visible vertices can provide additional discrimination from backgrounds, and extend its reach into parameter space.
A detailed study of these events is left to future literature.

In the presence of a signal at Belle-II, both of the channels above would shed light on the mass splittings and masses of the HNFs.
Firstly, the ISR channel could reconstruct the dark photon mass via $m_{A^\prime} \overset{!}{=} M_X^2 =\equiv s - 2E_\gamma \sqrt{s}$.
Then, in both ISR and s-channel events, the dilepton invariant mass of  $\psi_i \to \psi_j \ell^+\ell^-$ decays would constrain the mass HNF splittings through the inequality $m_{\ell \ell} < \Delta_{ij} m_j$.
In this case, displaced vertices would help isolate the different HNF decay cascades.
Finally, with displaced vertices in both ISR and s-channel, the experiment would be able to extract more information on the boosts of the HNFs.
The boosts will be larger for s-channel HNFs than in ISR.

We now comment on a short-term possibility that can be pursued.
Current published datasets from searches for visible $A^\prime$ at KLOE/KLOE-2~\cite{Bossi:2008aa,Amelino-Camelia:2010cem} and BES-III~\cite{BESIII:2020nme} can shed light on semi-visible $A^\prime$ by looking for a broad invariant mass spectrum of dileptons on top of their smooth backgrounds.
While visible resonances are better constrained due to the smaller backgrounds in a bump hunt, the smooth but often narrow distributions of dilepton invariant masses $m_{\ell \ell}$ can still be searched for. 
We identify the following channels as promising datasets for a semi-visible analysis: 
\begin{itemize}
    \item $e^+e^- \to \gamma (X\to\ell^+\ell^-)$~\cite{Bossi:2008aa,BaBar:2014zli,KLOE-2:2014qxg,KLOE-2:2016ydq},
    \item $\phi \to \eta (X\to e^+e^-)$~\cite{KLOE-2:2012lii,KLOE-2:2014hfk}, and
    \item $e^+e^- \to (X\to \ell^+\ell^-) X_{\rm invisible}$~\cite{KLOE-2:2015nli}.
    \item $e^+e^- \to (X\to \ell^+\ell^-) (Y\to (X\to \ell^+\ell^-) (X\to \ell^+\ell^-))$~\cite{BaBar:2012bkw},
\end{itemize}
where $X$ and $Y$ are some fully visible resonances.
We leave the evaluation of these constraints to future literature.

\myparagraph{Couplings to the $Z$ boson}
In addition to the $A^\prime$ coupling to the EM current, one can also explore the SM $Z$ boson coupling to the \emph{dark} current, shown in \cref{eq:couplings_to_currents}.
For the large values of kinetic mixing explored here, $Z$ decays can produce HNFs with branching ratios of 
\begin{align}
    \mathcal{B} (Z\to \psi_i \psi_j)  &\simeq \tau_{Z} \frac{ |V_{ij}|^2 G_F m_Z^3}{12 \sqrt{2}\pi}  \left(\frac{2 g_X s_W \epsilon}{g}\right)^2
    \\\nonumber
    & = |V_{ij}|^2 \times 10^{-7} \times \left( \frac{\alpha_D}{0.1}\right) \left( \frac{\epsilon}{10^{-2}}\right)^2,
\end{align}
where we neglected the small mass of the HNFs.
While this BR is too small to be constrained by invisible $Z$ decays, it can be used to look for lepton jets, as done at LEP by the DELPHI~\cite{DELPHI:1996qcc} and L3~\cite{L3:1999ymc,L3:2001zfe} collaborations.
The signature considered at LEP was a single HNL decaying to leptons or quarks, produced alongside a neutrino, $Z\to \nu N$. 
This search can, in principle, also be used to constrain semi-visible dark photons, using the channels $Z\to \psi_1 \psi_{2}$, where $\psi_2$ decays either promptly or displaced inside the detector and $\psi_1$ would constitute missing energy.
The limits will be modified due to the small splitting between parent and daughter HNFs, which decreases the dilepton energy.
Channels like $Z\to \psi_{2,3,\dots} \psi_{2,3,\dots}$ could be much more common than in the HNL scenario, where they are doubly suppressed by neutrino mixing.
We leave a detailed study of this interesting probe for future literature.

\myparagraph{Neutrino experiments} 
Neutrino experiments can test semi-visible dark photon models in two ways.
Firstly, in a DM interpretation, the HNFs can be produced in neutral meson decays and bremsstrahlung at the target, and the DM could travel to the detector, where it can coherently interact with nuclei $\mathcal{N}$ to produce its coannihilation partners~\cite{Izaguirre:2017bqb,Tsai:2019buq,Batell:2021ooj},
\begin{equation}
    \psi_1 \, \mathcal{N} \to (\psi_{2,3, \dots} \to \psi_1 \ell^+ \ell^-) \, \mathcal{N}.
\end{equation}
The decays of the co-annihilators can then produce displaced dileptons. 
This displacement is especially interesting for multi-component detectors like MINERvA and the near detector of T2K, ND280~\cite{Arguelles:2022lzs,Kamp:2022bpt}, and can be explored at future experiments like DUNE and Hyper-Kamiokande~\cite{Schwetz:2020xra,Atkinson:2021rnp}.
At high energies, experiments like IceCube and KM3NET can search for the upscattering signature by using the atmospheric production of DM particles. 
The DM particle can then upscatter via deep-inelastic scattering and the subsequent decay of $\psi_{2,3, \dots}$ would be sufficiently displaced from the vertex to form a double-bang signature.
This has been explored in the context of HNLs in Refs.~\cite{Coloma:2017ppo,Coloma:2019qqj}, but can be adapted to DM particles.
Co-annihilators can also be produced at the target alongside $\psi_1$.
For small mass splittings, they would be long-lived and can be constrained using high-energy beam dumps, like CHARM and NuCAL~\cite{Tsai:2019buq}, or searched for at forward or surface detectors at the LHC~\cite{Bertuzzo:2022ozu}
Their prompt decays, however, would contribute to the flux of $\psi_1$.

In Ref.~\cite{Batell:2021ooj}, the authors study the sensitivity of the SBN program at Fermilab to the production of iDM.
The three Liquid Argon detectors placed along the Booster Neutrino Beam (BNB) are sensitive to the production of DM in the BNB, as well as those produced in the NuMI beam, which is located at an off-axis location.
In addition, the BNB has the ability to run in off-target mode, directing the proton straight into the beam dump.
In this way, the flux of neutrinos going through the detector is minimized due to the absence of focus from the magnetic horns.
The sensitivity of SBN to iDM models can improve on CHARM and NuCAL constraints~\cite{Tsai:2019buq}, especially in the mass range of interest for our work.

If instead an HNL interpretation is assumed, the decays in flight of the lightest HNF, $N_4 = \psi_1$ can be searched for.
The HNL $N_4$ can only decay via the small mixing with SM neutrinos, through CC, NC, or dark photon interactions.
Typically, for the masses and large values of $\alpha_D \epsilon^2$ considered here, the decays of $N_4$ will proceed predominantly through the dark photon, and will make the branching ratios into $\psi_i \to \nu_{1,2,3} \ell^+\ell^-$ dominate.
The leading limits on this type of decay come from the T2K experiment~\cite{T2K:2019jwa}.
The dark photon contribution increases the decay rate of $N_4$, and enhances the decay-in-flight event rate, opening new parameter space between cosmological and laboratory-based limits on the mixing of $N_4$ with active neutrinos~\cite{Arguelles:2021dqn}.

Also in a HNL interpretation, there is a second possibility to produce the HNLs.
As discussed in ~\cref{sec:HNL_pheno}, the scattering of neutrinos in the beam with the dirt or in the detector can produce HNLs, which subsequently decay.
Through the exchange of a dark photon, active neutrinos coherently upscatter on nuclei $\mathcal{N}$,
\begin{equation}
    \nu_\alpha \, \mathcal{N} \to (N_i \to N_j \ell^+\ell^-) \,\mathcal{N}.
\end{equation}
The event rate is proportional to the mixing between the HNLs and active neutrinos.
The mixing with the muon flavor is the most relevant in this case as most of the flux at accelerator and atmospheric experiments is composed of $\nu_\mu$ and $\overline{\nu}_\mu$.

 While both DIF and upscattering signatures are proportional to parameters that have no impact on the semi-visible signatures at collider and fixed-target experiments, they cannot be uniquely determined in the parameter space shown in \cref{sec:results}.
Nevertheless, evidence for either of these would indicate that HNFs mix with active neutrinos, $\psi_i = N_i$, and confirm an HNL hypothesis.

\myparagraph{Kaon factories} 
Kaon decays can further constrain the parameter of semi-visible dark photons in two ways:
i) through the direct production of dark photons,
or ii) through the direct production of the HNFs.
The latter possibility is, in particular, a powerful probe mixing between neutrinos and the HNFs.
As discussed in \cref{sec:model_independent}, dark photons can be produced directly via kinetic mixing in $K \to \pi A^\prime$ as well as $K^+ \to \ell^+\nu_\ell A^\prime$.
The subsequent semi-visible decays of $A^\prime$ would then lead to multi-lepton final states~\cite{Hostert:2020xku}, albeit with at least two invisible particles.
Direct production of $A^\prime$ via kinetic mixing is, however, suppressed by $m_{A^\prime}^2/m_K^2$, and has limited reach (cf. \cref{fig:model_independent}).
A much more promising channel, however, is the direct production of HNLs through their mixing with the electron or muon flavor.
Just as the upscattering signatures discussed in the paragraph above, direct production of HNFs in kaon decays would provide direct evidence for their heavy neutral lepton interpretation, $\psi_i = N_i$.
As pointed out in Ref.~\cite{Ballett:2019pyw}, NA62 can use a three-track search to look for the production and the decay products of $N_i$,
\begin{equation}
    K^+ \to \ell^+_\alpha (N_i \to N_j \ell^+_\beta \ell^-_\beta ), \text{  where } \alpha, \beta \in \{ e, \mu\}.
\end{equation}
The striking feature of this signature is the reconstruction of the dark particle masses via the reconstructed quantities,
$m_{N_i}^2 \sim (p_K - p_{\ell_\alpha})^2$ and $m_{N_j}^2 = (p_K - p_{\ell_\alpha} - p_{\ell_\beta^+} - p_{\ell_\beta^-})^2$.
The event rate is proportional to $|U_{\alpha N_{i}}|^2$, and the subsequent primary as well as any secondary decays would provide the additional lepton tracks at no additional cost to the rate.
Displaced vertices in NA62 can be identified for proper lifetimes of the HNFs as small as $c\tau^0 \sim 10$~ps thanks to the $\mathcal{O}(10)$~cm vertex resolution of NA62.

\myparagraph{Future fixed target experiments (LDMX)} 
The next-generation fixed-target experiment LDMX~\cite{LDMX:2018cma,Berlin:2018bsc} provides a unique setup to search for semi-visible signatures.
The proposed design is focused on searches for bremsstrahlung-production of dark sector particles, $e^- \mathcal{N} \to e^- A^\prime \mathcal{N}$, through the missing-\emph{momentum} technique. 
Differently from NA64, LDMX aims to measure both the energy and transverse momentum of the recoil electron, having more access to the kinematics of $A^\prime$ production.
The proposal considers a primary beam of electron of $\sim 4 - 16$~GeV at SLAC impinging on a thin target inside a magnetic field~\cite{Akesson:2022vza}.
The beam would be tracked with low-mass trackers up and downstream of the target 
and then stopped by a large detector with ECAL and HCAL components, where the total energy of the recoil electron can be measured. 

Because the primary electrons will not shower until reaching the ECAL, the production of additional $e^+e^-$, $\mu^+\mu^-$, and $\pi^+\pi^-$ pairs in prompt semi-visible $A^\prime$ decays would provide a striking signature in the experiment.
In contrast to NA64, LDMX offers a lower-energy beam, enlarging its reach in HNF lifetime, and tracking capabilities, allowing the additional tracks to be seen in association with the recoiling electron.
While a detailed background study is needed, we note that QED processes like trident production, $e^- \mathcal{N} \to e^- e^+e^- \mathcal{N}$, and hard-photon conversions, $e^- \mathcal{N} \to e^- (\gamma_{\rm brem} \to e^+e^-) \mathcal{N}$, would not carry the missing momentum of the semi-visible signal.
In this regard, semi-visible events have an advantage over fully invisible ones because of the high multiplicity of tracks.
Finally, we note that the $\deltaamu$ region of interest overlaps with many vector meson $V$ resonances. 
The mass mixing of $V$ and $A^\prime$ can provide an additional and powerful production mechanism of semi-visible HNFs~\cite{Schuster:2021mlr}.

\section{Conclusions} \label{sec:conclusions}

Semi-visible dark photons typically arise in rich $U(1)_D$ dark sectors with a non-minimal particle content. 
If a symmetry distinguishes the DS fields from the SM fields, the lightest DS particle provides an ideal candidate for dark matter below the GeV-scale. 
If no such symmetry is present, the HNFs can mix with the SM light neutrinos and are identified with heavy neutral leptons. 
In this case, the lightest HNL decays to SM particles with long lifetimes.

In this paper, we have systematically studied a range of models with increasing fermionic content in the dark sector.
In particular, we have discussed the role of a charge-conjugation symmetry $C$ in the dark sector, which ensures that the $A^\prime i\overline{\psi}_i\gamma^\mu \psi_j$ interactions are predominantly off-diagonal in the $i$ and $j$ indices, generalizing the idea behind the popular iDM model.
This is necessary to suppress dark photon decays into two $\psi_1$ particles, as such decays contribute to the invisible branching ratio of the dark photon, which is severely constrained.
As an addition to iDM, we propose the three fermion mixed-iDM model, where a mostly-sterile Majorana DM $\psi_1$ particle co-annihilates with a mostly-dark and heavier Dirac fermion, $\Psi_2$.
Due to the $C$ symmetry, the DM can only couple to $\Psi_2$ through a small mixing $\alpha$, while $\Psi_2$ can have $\mathcal{O}(1)$ self-couplings.
This possibility has not been explored before in the context of DM.
Within the three fermion scenario, we also consider more general models of three Majorana particles, with and without enforcing $C$ symmetry.
These models typically favor more pronounced mass hierarchies and a HNL interpretation over that of DM, since coannihilations are strongly suppressed.

We follow this with a discussion of the exact Dirac limit of a four Weyl-fermion model, recovering the inelastic Dirac dark matter (i2DM) scenario~\cite{Filimonova:2022pkj}.
In this case, a $U(1)_D$-charged Dirac fermion mixes with a sterile Dirac particle.
In contrast to the mixed-iDM case, the Dirac DM particle now has self-interactions, albeit suppressed by a small mixing angle $\beta$.

If $A^\prime$ is heavier than the HNFs, its decay to pairs of HNFs can be followed by their subsequent decays, $\psi_i \to \psi_j \ell^+\ell^-$ or $\psi_i \to \psi_j \pi^+\pi^-$. 
As the decays of the $A^\prime$ do not lead to any visible resonances, resonance searches in the invariant mass spectra of dilepton pairs are not constraining.
For large couplings and kinetic mixing, these HNF decay lengths can be much smaller than the size of a typical particle detector.
The presence of visible particles and missing energy within the detector vetoes these semi-visible decays, modifying existing constraints on invisible dark photons.

In order to quantitatively assess these effects, we develop our own fast MC simulation of dark photon production and decay at BaBar and NA64, recasting the bounds from these experiments on the parameter space of semi-visible $A^\prime$ models.

We find a significant modification of the bounds on kinetic mixing in the region of $m_{A^\prime}\sim 0.3-1.3$~GeV.
This opens up new parameter space at large values of $\epsilon \sim \mathcal{O}(10^{-3} - 10^{-2})$, which has been fully excluded for visible and invisible $A^\prime$.
This region is of particular interest as dark photons with masses in this range can explain the discrepancy between dispersive calculations and the measurement of the anomalous magnetic moment of the muon, $\deltaamu^{\rm disp}$.
In both iDM (cf., \cref{fig:recast_1}) and mixed-iDM scenarios (cf.. \cref{fig:recast_2ab}), we find that the lightest HNF can also constitute a thermal DM candidate.
In i2DM, however, the mixing-suppressed self-annihilations of DM are still significant at late times, so CMB constraints exclude the entire $\deltaamu$ region (cf., \cref{fig:recast_3}).

We point out that in the newly-open parameter space, the bremsstrahlung production of $A^\prime$ in the fixed target of NA64 can still probe the semi-visible dark photon without the need for displaced decays, the method employed in Ref.~\cite{NA64:2021acr}.
By aggregating the additional energy of $e^+e^-$ pairs produced by short-lived HNFs to the energy of the primary electron beam in the electromagnetic calorimeter, NA64 would be sensitive to the missing energy carried away by stable or long-lived HNFs.
Our sensitivity curves show that this new signal definition could cover most of the open $\deltaamu^{\rm disp}$ parameter space in an iDM model.
A companion paper derives new NA64 limits and future sensitivity curves for iDM and i2DM models based on this method~\cite{na64semivisible}. Our projections are in good agreement with the experimental results given in Ref.~\cite{na64semivisible}, where a sophisticated detector simulation is used.

In addition, the newly-open regions also provide a realistic target for $e^+e^-$ colliders.
Re-analyses of existing BaBar, KLOE/KLOE-II and BESS-III data can target HNF production with multiple leptons associated, with or without initial state radiation.
This also includes LEP, where the $Z$ boson coupling to the dark current can be constrained using $Z\to \psi_1 \psi_{2,3,\dots}$ events.
In the near future, monophoton searches at Belle-II can improve on the BaBar limits on invisible $A^\prime$.
The veto on additional leptons will be even more important in this case due to the improved hermeticity of the detector.
Dedicated searches, such as those discussed here and in Refs.~\cite{Duerr:2019dmv,Duerr:2020muu,Kang:2021oes}, will be essential in constraining a semi-visible $A^\prime$.

Following a possible detection of HNFs in the semi-visible decays of $A^\prime$ in fixed-target or collider experiments, a key question would arise on whether they are a DM or HNL particles, revealing the presence or absence of additional symmetries in the theory.
Direct detection of non-relativistic DM particles would be challenging due to the large mass splittings and inelastic interactions.  Boosted DM, produced via the DM upscattering by cosmic rays, could provide an interesting detection avenue to be further investigated, in particular, the potential to exploit large neutrino detectors with low energy thresholds.
On the other hand, signatures associated with the HNF mixing with neutrinos would provide decisive evidence for the HNL interpretation and a possible connection to the origin of neutrino masses. 
The three most promising experimental strategies for this scenario are decay-in-flight searches for $\psi_1 \to \nu \ell^+ \ell^-$, neutrino upscattering to $\psi_{2,3,\dots}$, and direct production of $\psi_i$ particles in leptonic kaon decays. 

A semi-visible option for GeV-scale dark photons keeps the door open for large kinetic mixings that can be directly probed at low-energy experiments. 
This scenario provides a last chance for the kinetically-mixed dark photon interpretation of the muon $\deltaamu^{\rm disp}$, which has already been ruled out in the visible and invisible options. 
The class of semi-visible dark photon models is certainly within present experimental reach and may give us a clue as to whether nature prefers a rich and complex dark sector.

\acknowledgments
We want to thank Martina Mongillo, Paolo Crivelli, Laura Molina Bueno, and Benjamin Banto Oberhauser for in-depth discussions about the NA64 experiment.
We also thank Maxim Pospelov and Yue Zhang for illuminating discussions.
This project has received partial funding from the European Union’s Horizon 2020 research and innovation programme under the Marie Sklodowska-Curie grant agreement No. H2020-MSCA-ITN-2019/860881-HIDDeN (ITN HIDDeN) and from the European Union’s Horizon Europe research and innovation programme under
the Marie Sklodowska-Curie Staff Exchange grant agreement No. 101086085 - ASYMMETRY.
 The research at the Perimeter Institute is supported in part by the Government of Canada through NSERC and by the Province of Ontario through the Ministry of Economic Development, Job Creation and Trade, MEDT.
This work is also supported by the Fermi National Accelerator Laboratory, managed and operated by Fermi Research Alliance, LLC under Contract No. DE-AC02-07CH11359 with the U.S. Department of Energy.

\appendix
\onecolumngrid

\section{Simulations and event distributions}\label{app:distributions}

\begin{figure}[t!]
\centering
\includegraphics[width=\textwidth]{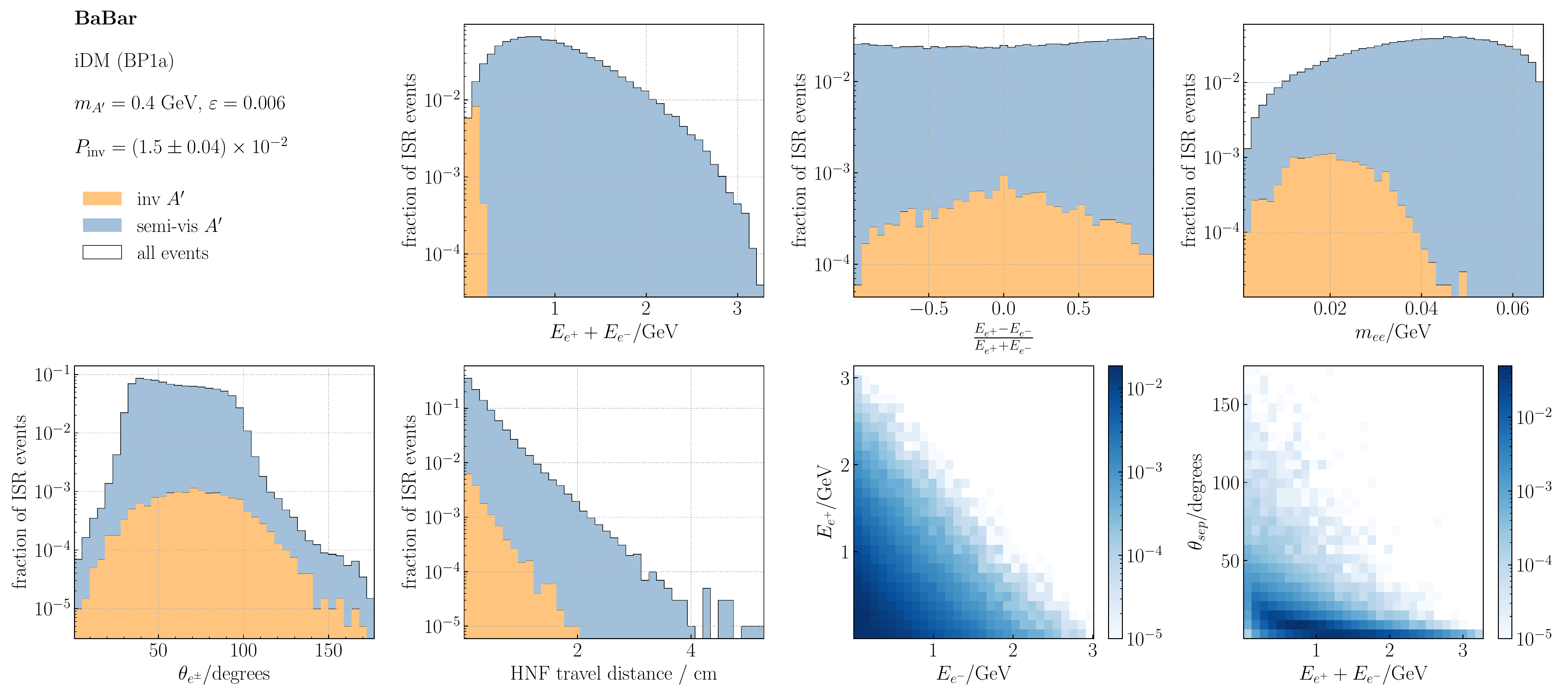}
\caption{
The kinematical distributions dark photon decay products for the initial-state-radiation signatures at BaBar.
In a clockwise order, we show the distribution of the total $e^\pm$ energies, their energy asymmetry, invariant mass $m_{ee}$, 
the 2D distribution of total energy and separation angle, the 2D distribution of each electron and positron energies, the distance traveled by the HNF between its production point and its decay position
the angle of $e^\pm$ with respect to the beam pipe, and the opening angle of $e^\pm$ pairs.
\label{fig:fraction_of_events1_babar}
}
\end{figure}

\begin{figure}[t!]
\centering
\includegraphics[width=\textwidth]{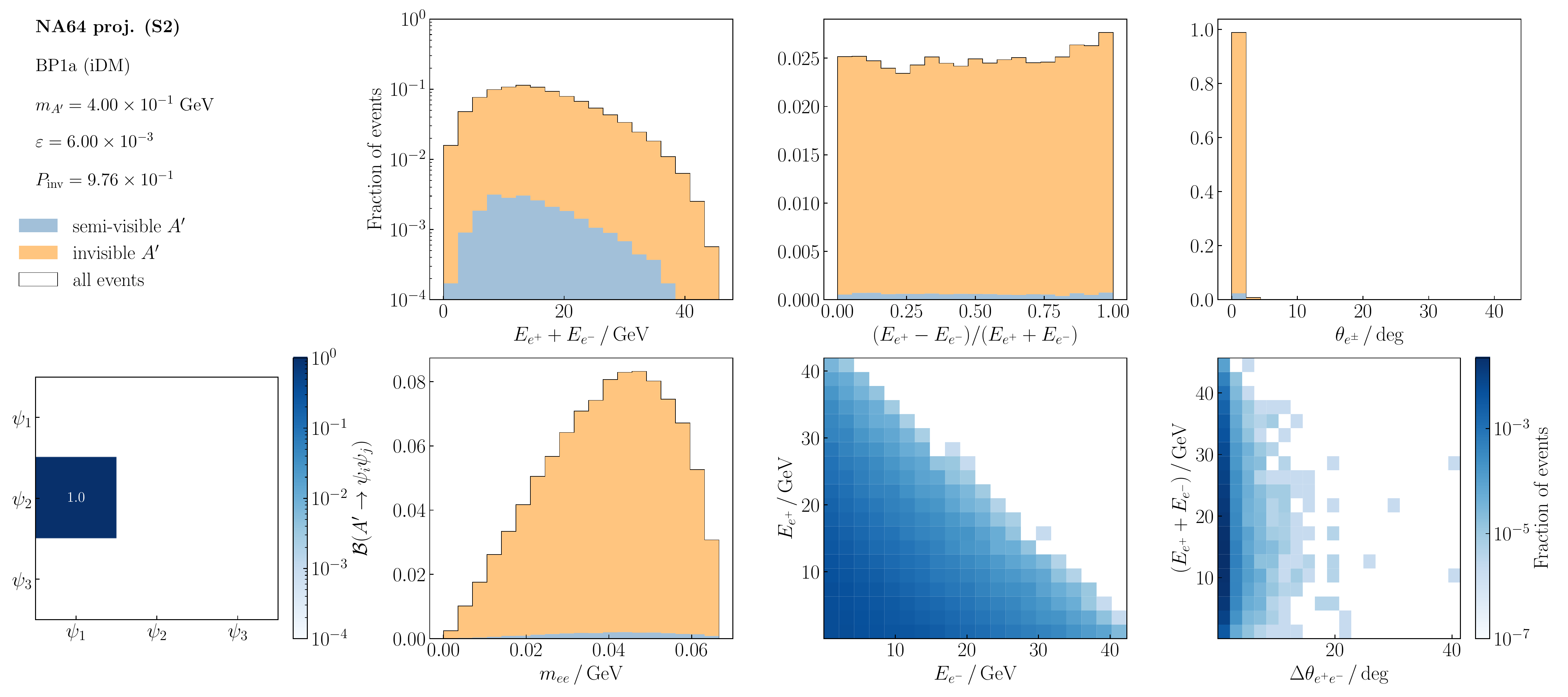}
\caption{
The kinematical distributions dark photon decay products for bremsstrahlung $A^\prime$ in NA64.
In a clockwise order, we show the distribution of the total $e^\pm$ energies, their energy asymmetry, the angle of $e^\pm$ with respect to the beam pipe, the 2D distribution of separation angle and total energy, the 2D distribution of each electron and positron energies, the invariant mass $m_{ee}$, and the branching ratios of the dark photon decays to the HNFs.
\label{fig:fraction_of_events1_na64}
}
\end{figure}

\begin{figure}[th]
\centering
\includegraphics[width=\textwidth]{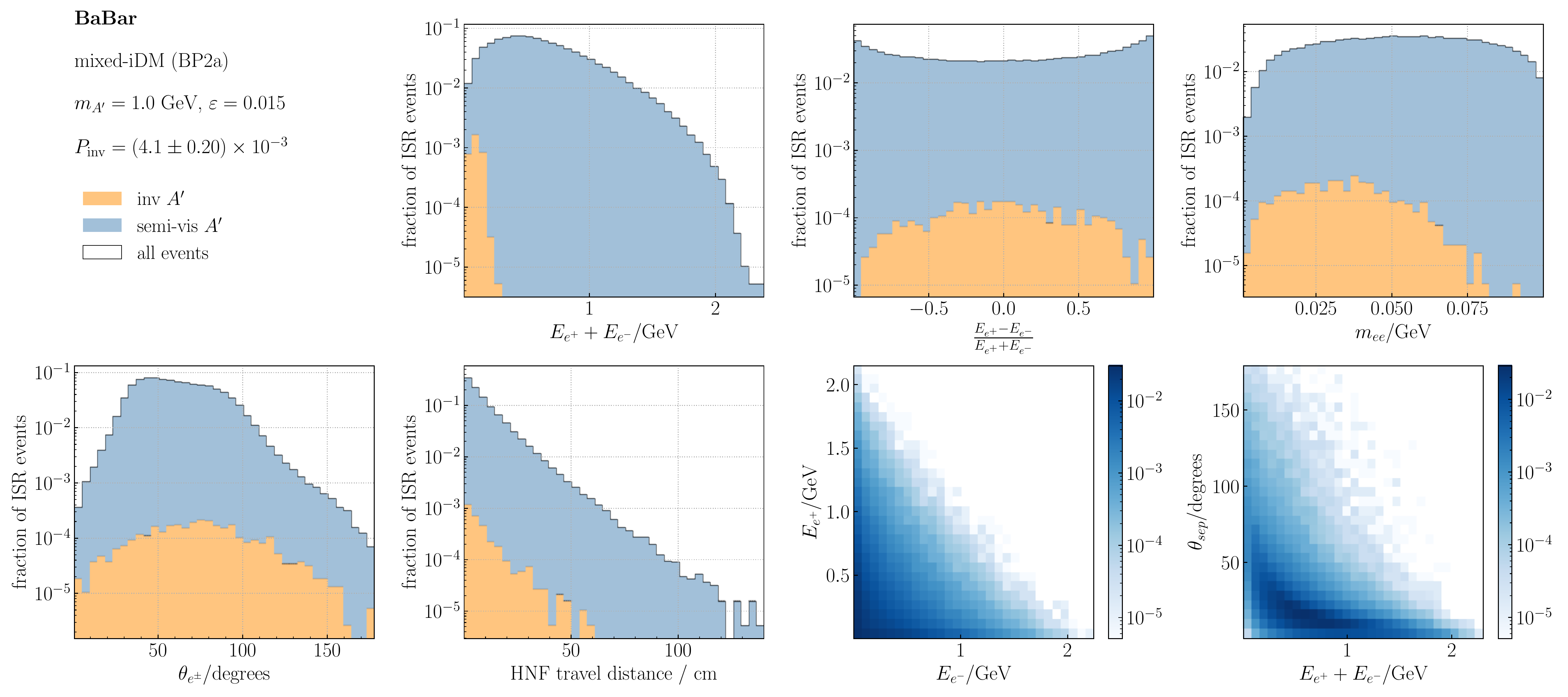}
\caption{Same as \cref{fig:fraction_of_events1_babar} but for BP2a.\label{fig:fraction_of_events2_babar}
}
\end{figure}

\begin{figure}[ht]
\centering
\includegraphics[width=\textwidth]{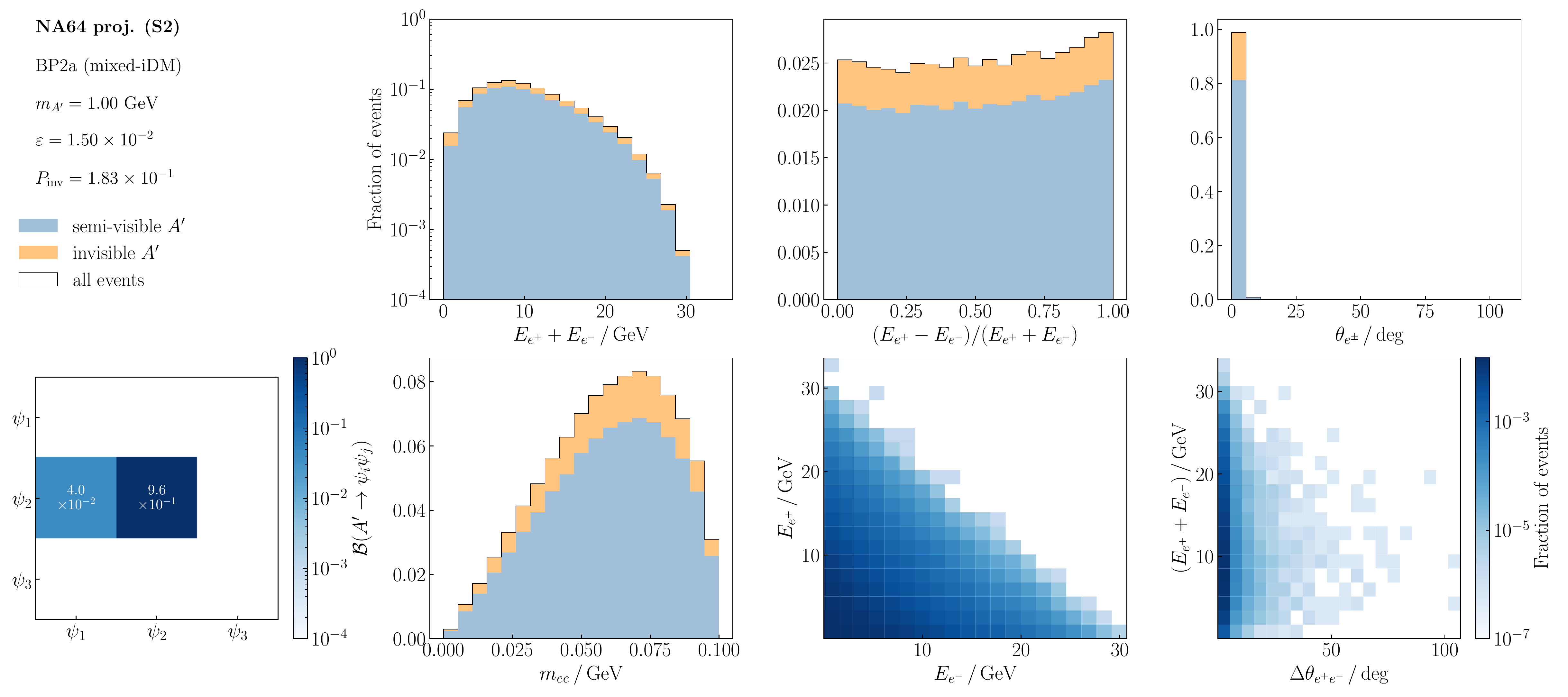}
\caption{Same as \cref{fig:fraction_of_events1_na64} but for BP2a.\label{fig:fraction_of_events2_na64}
}
\end{figure}

\begin{figure}[th]
\centering
\includegraphics[width=\textwidth]{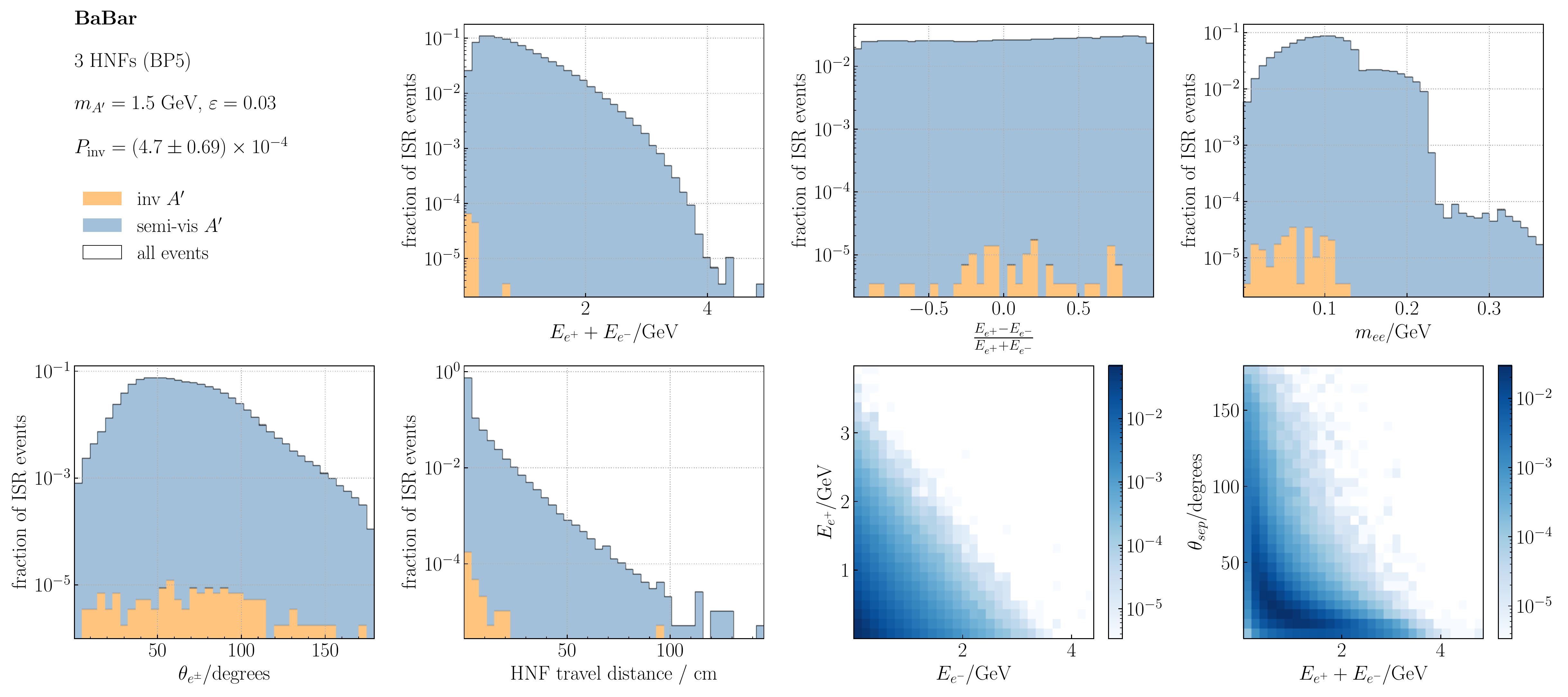}
\caption{Same as \cref{fig:fraction_of_events1_babar} but for BP5.\label{fig:fraction_of_events3}
}
\end{figure}

\begin{figure}[ht]
\centering
\includegraphics[width=\textwidth]{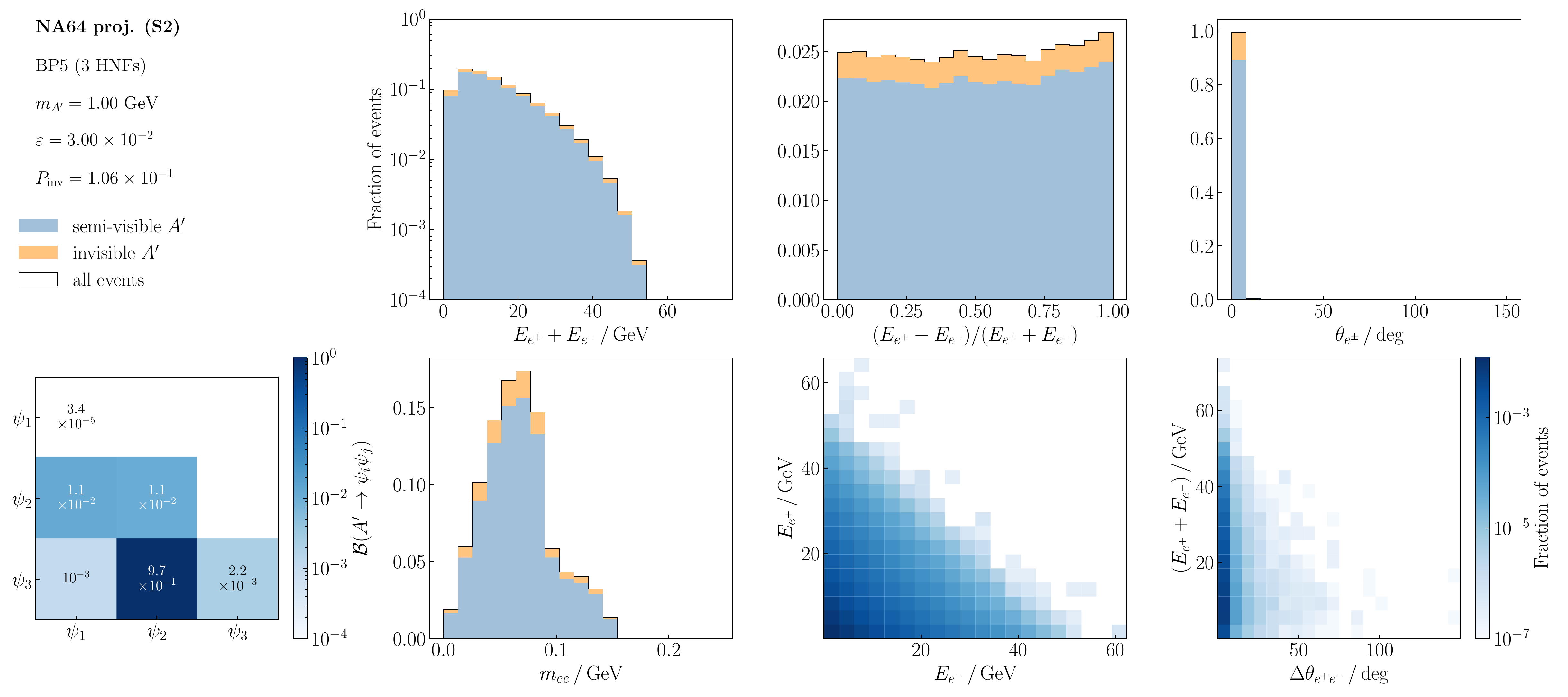}
\caption{Same as \cref{fig:fraction_of_events1_na64} but for BP5.\label{fig:fraction_of_events3_na64}
}
\end{figure}

In this appendix, we show some more details of the BaBar and NA64 simulations. 
The kinematical distributions of the final state leptons in semi-visible events at BaBar and NA64 are shown for representative benchmarks in \cref{fig:fraction_of_events1_babar} to \cref{fig:fraction_of_events3_na64}.
Each panel shows the truth-level distribution of events in total energy ($E_{e^+}+E_{e^-}$), energy asymmetry ($|E_{e^+}-E_{e^-}|/(E_{e^+}+E_{e^-})$), the lutput is actepton angle with respect to the beam pipeline ($\theta_{e^\pm}$), the lepton pair invariant mass ($m_{ee}$), as well as two-dimensional distributions.
We also show the total dark photon branching ratios in terms of the final state particles in a two-dimensional grid.

\section{New physics in the anomalous magnetic moment of the muon}
\label{app:amu_explanations}

Several BSM theories have been put forward to explain $\Delta a_\mu$. Among them are supersymmetric models~\cite{Athron:2021iuf}, leptoquarks~\cite{Chakraverty:2001yg,Cheung:2001ip}, including those capable of explaining other flavor anomalies~\cite{Bauer:2015knc,Bigaran:2020jil,Greljo:2021xmg}, and several other theories with new, heavy degrees of freedom. One interesting possibility, however, is that the new physics lies below the EW scale.

At the one-loop level, a new vector or scalar mediator can contribute with the correct, positive sign to $\deltaamu$. In this article, we focus on a dark photon kinetically mixed with the SM hypercharge~\cite{Holdom:1985ag,Pospelov:2008zw}. Like several other low-scale new physics solutions to $\deltaamu$, the minimal dark photon model is excluded.  Scalars mixed with the Higgs face severe constraints from meson decay, beam dump, and collider experiments, and are ruled out as an explanation of $\Delta a_\mu$ in minimal realizations. Particles with muon-specific couplings are still allowed, including $U(1)_{L_\mu-L_\tau}$ models~\cite{Amaral:2020tga} and leptonic Higgs portal scalars~\cite{Chen:2015vqy,Batell:2016ove,Batell:2017kty}. Parity-violating interactions may also contribute to $\deltaamu$ via one-loop vertex corrections. These come with a negative sign and can co-exist with scalar or vector contributions, which could still remain sufficiently large. Parity-violating gauge bosons as part of a solution to $\deltaamu$ have been discussed in the context of the proton radius puzzle~\cite{Batell:2011qq,Karshenboim:2014tka} and arise naturally in the so-called dark-Z ($Z_d$) models~\cite{Davoudiasl:2012qa,Davoudiasl:2012ag,Davoudiasl:2014kua}. In the case of axion-like-particles, Barr-Zee diagrams~\cite{Barr:1990vd} may dominate the contribution to $\deltaamu$, even though such diagrams are effectively a two-loop effect with heavy charged fermions $f$ loops~\cite{Chang:2000ii}. This contribution is positive and enhanced with respect to the one-loop result by $(m_f/m_\mu)^2$. Models with axion-like-particles coupled to a dark force~\cite{Kaneta:2016wvf,deNiverville:2018hrc} have also been proposed as an explanation of $\deltaamu$ with the added assumption of a direct coupling of $a$ to the muon~\cite{Ge:2021cjz}.




\bibliographystyle{apsrev4-1}
\bibliography{main}{}

\end{document}